\documentclass[12pt,oneside]{article}

\usepackage{apacite}

\usepackage{color}
\usepackage[dvipsnames]{xcolor}

\usepackage{natbib}
\setcitestyle{authoryear,round}

\usepackage{hyperref}
\usepackage{fullpage, url, graphicx, subcaption}
\usepackage{fancyhdr}
\usepackage{mathrsfs}
\usepackage{epsfig}
\usepackage{graphics}
\usepackage{epsf}
\usepackage{amsmath, amsthm, amssymb, multirow, paralist, mathtools}
\usepackage{algorithm}
\usepackage[noend]{algorithmic}

\usepackage{tikz}
\usepackage{caption}
\usepackage{lscape}
\usepackage{array}
\usepackage{epstopdf}

\def\diag{\mathrm{diag}}

\def\T{ {\mathrm{\scriptscriptstyle T}} }

\def\bbR{\mathbb R}

\def\argmin{\mathrm{argmin}}

\def\TV{\mbox{TV}}
\def\HTV{\mathrm{HTV}}
\def\mymu{v}
\newcommand\prox{\mathbf{prox}}

\newtheorem{pro}{Proposition}

\allowdisplaybreaks

\begin{document}

\begin{titlepage}

\begin{center}
{\Large Block-wise Primal-dual Algorithms for Large-scale Doubly Penalized ANOVA Modeling}

\vspace{.15in} Penghui Fu\footnotemark[1] and Zhiqiang Tan\footnotemark[1]

\vspace{.1in}
\today
\end{center}

\footnotetext[1]{Department of Statistics, Rutgers University. Address: 110 Frelinghuysen Road,
Piscataway, NJ 08854. E-mails: penghui.fu@rutgers.edu, ztan@stat.rutgers.edu.
}

\paragraph{Abstract.}

For multivariate nonparametric regression,
doubly penalized ANOVA modeling (DPAM) has recently been proposed, using hierarchical total variations (HTVs) and empirical norms as penalties
on the component functions such as main effects and multi-way interactions in a functional ANOVA decomposition of the underlying regression function.
The two penalties play complementary roles: the HTV penalty promotes sparsity in the selection of basis functions within each component function, whereas the empirical-norm penalty promotes sparsity in the selection of component functions.
We adopt backfitting or block minimization for training DPAM,
and develop two suitable primal-dual algorithms, including both batch and stochastic versions, for updating each component function in single-block optimization.
Existing applications of primal-dual algorithms are intractable in our setting with both HTV and empirical-norm penalties.
Through extensive numerical experiments, we demonstrate the validity and advantage of our stochastic primal-dual algorithms, compared
with their batch versions and a previous active-set algorithm, in large-scale scenarios.

\paragraph{Key words and phrases.} ANOVA modeling; Nonparametric regression; Penalized estimation; Primal-dual algorithms; Stochastic algorithms; Stochastic gradient methods; Total variation.

\end{titlepage}


\section{Introduction} \label{sec:intro}

Consider functional analysis-of-variance (ANOVA) modeling for multivariate nonparametric regression (e.g., \cite{gu2013smoothing}).
Let $Y_i$ and $X_i=(X_{i1}, \ldots, X_{ip})^\T$, $i=1,\ldots, n$,
be a collection of independent observations of a response variable and a covariate vector.
For continuous responses, nonparametric regression can be stated such that
\begin{equation}\label{eqn:model}
  Y_i = f(X_{i1},  \ldots, X_{ip}) + \varepsilon_i,
\end{equation}
where $f(x)=f(x_1,\ldots,x_p)$ is an unknown function,
and $\varepsilon_i$ is a noise with mean zero and a finite variance given $X_i$.
In the framework of functional ANOVA modeling, the multivariate function $f$ is decomposed as
\begin{align} \label{eqn:f-anova}
  f(x_1, \ldots, x_p) &= f_0 + \sum_{1 \leq j_1 \leq p}f_{j_1}(x_{j_1}) + \sum_{1 \leq j_1 < j_2 \leq p}f_{j_1, j_2}(x_{j_1}, x_{j_2}) + \cdots \nonumber \\
  & \quad + \sum_{1\leq j_1 <\cdots < j_K \leq p}f_{j_1,\ldots, j_K}(x_{j_1}, \cdots, x_{j_K}),
\end{align}
where $f_0$ is a constant, $f_{j_1}$'s are univariate functions representing main effects,
$f_{j_1, j_2}$'s are bivariate functions representing two-way interactions, etc, and $K$ is the maximum way of interactions allowed.
The special case of (\ref{eqn:model})--(\ref{eqn:f-anova}) with $K=1$ is known as additive modeling
\citep{stone1986dimensionality, hastie1990generalized}.
For general $K \ge 2$, a notable example is smoothing spline ANOVA modeling \citep{wahba1995smoothing, lin2006component, gu2013smoothing}, where the component functions in (\ref{eqn:f-anova}) are assumed
to lie in tensor-product reproducing kernel Hilbert spaces (RKHSs),
defined from univariate Sobolev-$L_2$ spaces as in smoothing splines.
Alternatively, in \cite{yang2021hierarchical},
a class of hierarchical total variations (HTVs) is introduced to measure roughness of main effects and multi-way interactions
by properly extending the total variation associated with the univariate Sobolev-$L_1$ space.
Compared with smoothing spline ANOVA modeling,
this approach extends univariate regression splines using total variation penalties \citep{mammen1997locally}.

Recently, theory and methods have been expanded for additive and ANOVA modeling to high-dimensional settings, with $p$ close to or greater than $n$.
Examples include \cite{ravikumar2009sparse}, \cite{meier2009high}, \cite{koltchinskii2010sparsity}, \cite{radchenko2010variable}, \cite{raskutti2012minimax}, \cite{petersen2016fused}, \cite{tan2019doubly}, and \cite{yang2018backfitting,yang2021hierarchical} among others.
An important idea from the high-dimensional methods is to employ
empirical $L_2$ norms of component functions as penalties, in addition to functional semi-norms which measure
roughness of the component functions, such as the Sobolev-$L_2$ semi-norm or total variation for univariate functions in the case of additive modeling ($K=1$).
For $K \ge 2$, the incorporation of empirical-norm penalties is relevant even when $p$ is relatively small against $n$, because
the total number of component functions in (\ref{eqn:f-anova}) scales as $p^K$.

In this work, we are interested in doubly penalized ANOVA modeling (DPAM) in \cite{yang2021hierarchical}, using both
the HTVs and empirical $L_2$ norms as penalties on the component functions in (\ref{eqn:f-anova}).
To describe the method, the ANOVA decomposition (\ref{eqn:f-anova}) can be expressed in a more compact notation as
\begin{align}\label{eqn:f-anova2}
  & f(x_1, \ldots, x_p) = f_0 + \sum_{S_1: |S_1|=1} f_{S_1}  + \sum_{S_2: |S_2|=2} f_{S_2} +\cdots  + \sum_{S_K: |S_K|=K} f_{S_K} ,
\end{align}
where $S_k$ is a subset of size $k$ from $\{1,\ldots,p\}$ for $k=1,\ldots,K$, and
$f_{S_k} = f_{j_1,\ldots,j_k} (x_{j_1}, \ldots, x_{j_k})$ if $S_k = \{j_1, \ldots, j_k\}$.
For identifiability,  the component functions are assumed to be uniquely defined as
$ f_{S_k}=(\prod_{j\in S_k}(I-H_j)\prod_{j\notin S_k} H_j) f$,
where $H_j$ is a marginalization operator over $x_j$, for example, defined by averaging over $X_{ij}$'s in the training set.
For $m \ge 1$, the HTV of differentiation order $m$, denoted as $\HTV^m$, is defined inductively in $m$.
For example, for a univariate function $f=f(x_1)$, $\HTV^1(f) = \TV(f)$ and $\HTV^2 (f) = \HTV(D_1 f)$,
where $D_1$ is the differentiation operator in $x_1$
and $\TV$ is the standard total variation as in \cite{mammen1997locally}.
For a bivariate function $f=f(x_1,x_2)$, the HTV with $m=1$ or $2$ is considerably more complicated (even after ignoring scaling constants):
\begin{align*}
\HTV^{1} (f) &= \mathrm{TV} (f_{12}) + \TV (f_1) + \TV (f_2) , \\
\HTV^{2}(f ) &=  \HTV^1 (D_1D_2 f_{12}) + \HTV^1(D_1 f_1) + \HTV^1(D_2f_2),
\end{align*}
where $(f_1,f_2,f_{12})$ are the component functions in the ANOVA decomposition (\ref{eqn:f-anova2}).
Nevertheless, suitable basis functions are derived in \cite{yang2021hierarchical} to achieve
the following properties in a set of multivariate splines, defined as the union of all tensor products of up to $K$ sets of univariate splines, depending on differentiation order $m$ and pre-specified marginal knots over each coordinate.
First, the ANOVA decomposition (\ref{eqn:f-anova2}) for such a multivariate spline $f$ can be represented as
\begin{align}
f=\beta_0 + \sum_{k=1}^K \sum_{S_k:|S_k|=k} \Psi_{S_k}^\T  \beta_{S_k},  \label{eqn:f-beta}
\end{align}
with $f_0=\beta_0$ and $f_{S_k} = \Psi_{S_k}^\T  \beta_{S_k}$,
where $\Psi_{S_k}$ is the basis vector in $(x_{j_1}, \ldots, x_{j_k})$
and $\beta_{S_k}$ is the associated coefficient vector for $S_k = \{j_1, \ldots, j_k\}$ and $k=1, \ldots, K$.
Moreover, the HTV can be transformed into a Lasso representation, with $\HTV^m (f_{S_k}) = \|\Gamma_{S_k}\beta_{S_k}\|_1$:
\begin{align}
\HTV^m(f) = \sum_{k=1}^K \sum_{S_k:|S_k|=k} \|\Gamma_{S_k}\beta_{S_k}\|_1 , \label{eqn:HTV-beta}
\end{align}
where $\|\cdot\|_1$ denotes the $L_1$ norm of a vector, $\rho_k$ is a scaling constant for HTV of $k$-variate component $f_{S_k}$,
and $\Gamma_{S_k}$ is a diagonal matrix with each diagonal element either 0 or $\rho_k$ (with $0$ indicating that the
corresponding element of $\beta_{S_k}$ is not penalized, for example, the coefficient of a fully linear basis in the case of $m=2$).
For $m=1$ or 2, the univariate spline bases are piecewise constant or linear,
and the resulting multivariate spline bases are, respectively, piecewise constant or cross-linear
(i.e., being piecewise linear in each coordinate with all other coordinates fixed).
Readers are referred to \cite{yang2021hierarchical} for further details, although
such details are not required in our subsequent discussion.

With the preceding background, linear DPAM is defined by solving
\begin{align*}
\min_{\beta}\; \frac{1}{2} \|Y-f  \|^2_n+\sum_{k=1}^K\sum_{S_k:|S_k|=k}\Bigl(\HTV^m (f_{S_k})+\lambda_k\|f_{S_k}\|_n\Bigr),
\end{align*}
or, with the representations (\ref{eqn:f-beta}) and (\ref{eqn:HTV-beta}), by solving
\begin{equation} \label{eq:linear-DPAM}
    \min_{\beta}\;  \frac{1}{2}\Bigl\|Y - \beta_0-\sum_{k=1}^K\sum_{S_k:|S_k|=k} \Psi_{S_k}^\T \beta_{S_k}\Bigr\|^2_n+\sum_{k=1}^K\sum_{S_k:|S_k|=k}\Bigl( \|\Gamma_{S_k}\beta_{S_k}\|_1 +\lambda_k\|\Psi_{S_k}^\T \beta_{S_k}\|_n \Bigr),
\end{equation}
where $\beta$ consists of $\beta_0$ and $(\beta_{S_k})_{k,S_k}$,
$\|\cdot\|_n$ denotes the empirical $L_2$ norm (or in short, empirical norm), e.g.,
$\|f \|_n = \{n^{-1} \sum_{i=1}^n f ^2(X_i )\}^{1/2}$, and $\lambda_k$ is a tuning parameter for $k=1,\ldots, K$.
As reflected in the name DPAM, there are two penalty terms involved,
the HTV $\|\Gamma_{S_k}\beta_{S_k}\|_1$ and the empirical norm $\lambda_k\|\Psi_{S_k}^\T \beta_{S_k}\|_n$,
which play different roles in promoting sparsity; see Proposition \ref{pro:soft-threshold} and related discussion.
The HTV penalty promotes sparsity among the elements of $\beta_{S_k}$ (or the selection of basis functions) for each block $S_k$, whereas
the empirical-norm penalty promotes sparsity among the coefficient vectors $\beta_{S_k}$ (or the component functions $f_{S_k}$) across $S_k$.
The use of these two penalties is also instrumental in the theory for doubly penalized estimation in additive or ANOVA modeling \citep{tan2019doubly}.

For binary outcomes, replacing (\ref{eqn:model}) by (nonparametric) logistic regression,
$P( Y_i=1 |X_i) = \{ 1+ \exp(-f(X_i)) \}^{-1}$, and the square loss in (\ref{eq:linear-DPAM})
by the likelihood loss (or cross-entropy) leads to logistic DPAM, which solves
\begin{equation}\label{eq:logistic-DPAM}
    \min_{\beta}\; \frac{1}{n}\sum_{i=1}^n l \Bigl(Y_i, \beta_0 + \sum_{k,S_k} \Psi_{i,S_k} ^\T \beta_{S_k}\Bigr) + \sum_{k, S_k} \Bigl(\|\Gamma_{S_k}\beta_{S_k}\|_1+\lambda_k\|\Psi_{S_k}\beta_{S_k}\|_n  \Bigl),
\end{equation}
where $Y_i\in\{0,1\}$, $\Psi_{i,S_k} = \Psi_{S_k} (X_{i,j_1},\ldots,X_{i,j_k})$ for $S_k=\{j_1,\ldots,j_k\}$, and $l(y,f)=\log(1+\exp(f))-yf$
is the likelihood loss for logistic regression.

We develop new optimization algorithms for solving (\ref{eq:linear-DPAM}) and (\ref{eq:logistic-DPAM}), including stochastic algorithms
adaptive to large-scale scenarios (with large sample size $n$).
Similarly as in the earlier literature \citep{hastie1990generalized},
the top-level idea is backfitting or block minimization: the objective is optimized with respect to one block $\beta_{S_k}$ at a time
while fixing the remaining blocks. This approach is valid
for solving (\ref{eq:linear-DPAM}) and (\ref{eq:logistic-DPAM}) due to the block-wise separability of non-differentiable terms \citep{tseng1988coordinate}.
In \cite{yang2018backfitting,yang2021hierarchical}, a backfitting algorithm, called AS-BDT (active-set block descent and thresholding),
is proposed by exploiting two supportive properties. First, in the subproblem with respect to only $\beta_{S_k}$, the solution can be obtained by
jointly soft-thresholding an (exact) solution to the Lasso subproblem with only the penalty $\|\Gamma_{S_k}\beta_{S_k}\|_1$; see Proposition \ref{pro:soft-threshold}.
Second, the desired Lasso solution can be computed using an active-set descent algorithm \citep{osborne2000new},
which tends to be efficient when the training dataset is relatively small or the Lasso solution is sufficiently sparse within each block.
However, AS-BDT is not designed to efficiently handle large datasets, especially in applications with non-sparse blocks.

As a large-scale alternative to AS-BDT, we investigate primal-dual algorithms, including both batch and stochastic versions,
for solving the single-block subproblems in (\ref{eq:linear-DPAM}) and (\ref{eq:logistic-DPAM}).
(We also study a new majorization-minimization algorithm, which is presented in the Supplement.)
The primal-dual methods have been extensively studied for large-scale convex optimization, as shown in the book \cite{ryu2022large}.
However, existing applications of primal-dual algorithms are intractable in our setting, with both HTV/Lasso and empirical-norm penalties.
In particular, SPDC (stochastic primal-dual coordinate method) in \cite{zhang2017stochastic}
can be seen as a stochastic version of the Chambolle--Pock (CP) algorithm
\citep{esser2010general, chambolle2011first}. But SPDC requires the objective to be split
into an empirical risk term ($f$-term), for example, the square loss in (\ref{eq:linear-DPAM}), and a penalty term ($g$-term)
such that its proximal mapping can be easily evaluated. Such a choice would combine the HTV/Lasso and empirical-norm penalties into a single penalty term,
for which the proximal mapping is numerically intractable to evaluate.

To overcome this difficulty, we formulate a suitable split of the objective into $f$- and $g$-terms in our setting with HTV/Lasso and empirical-norm penalties,
and derive two tractable primal-dual batch algorithms for solving the DPAM single-block problems.
The two algorithms are new applications of, respectively, the CP algorithm and a linearized version of AMA (alternating minimization algorithm) \citep{tseng1991applications}.
Furthermore, we propose two corresponding stochastic primal-dual algorithms, with per-iteration cost about $1/n$ of that in the batch algorithms.
In contrast with SPDC (where exact unbiased updates are feasible with a separable $f$-term), our stochastic algorithms are derived by allowing
approximately unbiased updates, due to non-separability of the $f$-term in our split formulation.
We demonstrate the validity and effectiveness of our stochastic algorithms in extensive numerical experiments, while
leaving formal analysis of convergence to future work.

\section{Primal-dual algorithms: Review and new finding}\label{sec:Primal-dual review}
\subsection{Notation}

We introduce basic concepts and notations, mostly following \cite{ryu2022large}. For a closed, convex and proper (CCP) function $f$, we denote its Fenchel conjugate as $f^*(y)=\sup_{x}\,\{y^\T x-f(x)\}$. If $f$ is CCP then $f^*$ is CCP and $f^{**}=f$. Denote as $\argmin_x\, f(x)$ the set of minimizers of $f$.
For a CCP $f$,  we denote its proximal mapping as
$    \prox_{f}(x) = \underset{u}{\argmin}\, \frac{1}{2}\|u-x\|_2^2+f(u)$.
If $f$ is CCP, then $\prox_{f}(x)$ is uniquely defined on the whole space. An important property of proximal mapping is the Moreau's identity
\begin{equation}\label{eq:Moreau's identity}
    \prox_{\alpha f^*}(x) = x - \alpha\cdot \prox_{f/\alpha}(x/\alpha),
\end{equation}
for $\alpha>0$ and CCP $f$. A differentiable function (not necessarily convex) is called $L$-smooth if its gradient is $L$-Lipschitz continuous. A differentiable function $f$ is called $\mu$-strongly convex if there exists $\mu>0$ such that for all $x$ and $y$,
$    f(y)\geq f(x) +\nabla f(x)^\T(y-x) + (\mu/2)\|y-x\|_2^2$.
It is known that a CCP $f$ is $\mu$-strongly convex if and only if $f^*$ is $(1/\mu)$-smooth.

Consider an optimization problem in the form
\begin{align}
\min_{x}\, f(Ax)+g(x), \label{eq:original problem}
\end{align}
or, equivalently, in its split form,
\begin{equation}\label{eq:generic problem for review}
    \min_{z,x}\, f(z)+g(x), \quad \text{subject to}\quad -z+Ax=0,
\end{equation}
where $x\in\bbR^d$, $z\in\bbR^n$, $A\in\bbR^{n\times d}$, and $f$ and $g$ are CCP.
The (full) Lagrangian associated with problem (\ref{eq:generic problem for review}) is
\begin{equation}\label{eq:generic lagrangian}
    L(z,x;u) =f(z)+g(x)+u^\T(Ax-z),
\end{equation}
with $u\in\bbR^n$. Then the dual problem of (\ref{eq:generic problem for review}) is
\begin{equation}\label{eq:generic dual problem for review}
    \max_{u} -(f^*(u)+g^*(-A^\T u)),
\end{equation}
We assume that total duality holds: a primal solution $(z^*,x^*)$  for (\ref{eq:generic problem for review}) exists (with $z^*= Ax^*$),
a dual solution  $u^*$ for (\ref{eq:generic dual problem for review}) exists, and $(z^*,x^*,u^*)$ is a saddle point of (\ref{eq:generic lagrangian}).
By minimizing (\ref{eq:generic lagrangian}) over $z$, we obtain the primal-dual Lagrangian
\begin{equation}\label{eq:primal-dual lagrangian}
    L(x;u)=g(x)+u^\T Ax-f^*(u).
\end{equation}
The dual problem associated with (\ref{eq:primal-dual lagrangian}) is still (\ref{eq:generic dual problem for review}), but the primal problem becomes the unconstrained problem (\ref{eq:original problem}), in that
$ \max_{u} L(x; u) = f(Ax)+g(x)$ and
$ \min_{x} L(x; u) = -(f^*(u)+g^*(-A^\T u))$.
By construction, $(x^*,u^*)$ is a saddle point of (\ref{eq:primal-dual lagrangian}) if and only if $(Ax^*,x^*,u^*)$ is a saddle point of (\ref{eq:generic lagrangian}). The two forms of Lagrangian are used in different scenarios later.

In the following subsections, we review several primal-dual methods which are used in our work,
and propose a new method, linearized AMA, which corresponds to a dual version of
the PAPC/$\mathrm{PDFP^2O}$ method \citep{chen2013primal,drori2015simple}.

\subsection{ADMM, linearized ADMM and Chambolle--Pock}\label{subsec:ADMM}

\textbf{Alternating direction method of multipliers} (ADMM) has been well studied (e.g., \cite{boyd2011distributed}).
For solving (\ref{eq:generic problem for review}), the ADMM iterations are defined as follows: for $k\ge 0$,
\vspace{-\baselineskip}
\begin{subequations}\label{eq:ADMM}
\begin{align}
    z^{k+1} &= \argmin_{z}\, \left( f(z)-\langle u^k,z \rangle+ (\alpha/2)\|Ax^k-z\|_2^2 \right), \label{subeq:ADMM-z} \\
    x^{k+1} &\in \,\argmin_{x}\, \left(g(x)+\langle A^\T u^k,x \rangle + (\alpha/2)\|Ax-z^{k+1}\|_2^2\right), \label{subeq:ADMM-x}\\
    u^{k+1} &= u^k + \alpha(Ax^{k+1}-z^{k+1}),
\end{align}
\end{subequations}
where $\alpha>0$ is a step size. After completing the square, (\ref{subeq:ADMM-z}) is equivalent to evaluating a proximal mapping of $f$ as shown in (\ref{subeq:linear-ADMM-z}). However, (\ref{subeq:ADMM-x}) in general does not admit a closed-form solution. To address this, \textbf{linearized ADMM} has been proposed with the following iterations for $k\ge 0$:
\begin{subequations}\label{eq:linear-ADMM}
\begin{align}
    z^{k+1} & 
    = \prox_{f/\alpha}(Ax^k+u^k/\alpha), \label{subeq:linear-ADMM-z} \\
    x^{k+1} &= \prox_{\tau g}\left(x^k-\tau A^\T(u^k+\alpha(Ax^k-z^{k+1}))\right), \label{subeq:linear-ADMM-x}\\
    u^{k+1} &= u^k + \alpha(Ax^{k+1}-z^{k+1}),
\end{align}
\end{subequations}
where $\tau >0$ is a step size in addition to $\alpha$.
The iterates $(z^{k+1},x^{k+1},u^{k+1})$ can be shown to converge to a saddle point of the full Lagrangian (\ref{eq:generic lagrangian}) provided
$0< \alpha\tau \|A\|_2^2\leq 1$,
where $\|A \|_2$ denotes the spectral norm of $A$, i.e., the square root of the largest eigenvalue of $A^\T A$.
The method is called linearization because (\ref{subeq:linear-ADMM-x}) can be obtained by linearizing the quadratic $\|Ax-z^{k+1}\|_2^2$ in (\ref{subeq:ADMM-x}) at $x^k$ with an additional regularization $(1/(\alpha\tau))\|x-x^k\|_2^2$, i.e.,
$$
x^{k+1}=\argmin_{x}\, \left( g(x)+ \langle A^\T u^k,x \rangle + \alpha \langle A^\T (Ax^k-z^{k+1}),x \rangle+(1/(2\tau))\|x-x^k\|_2^2 \right).
$$
Moreover, linearized ADMM (\ref{eq:linear-ADMM}) can be transformed to the \textbf{primal-dual hybrid gradient} (PDHG) or \textbf{Chambolle--Pock algorithm} (CP) \citep{esser2010general,chambolle2011first}
with the following iterations for $k \ge 0$:
\begin{subequations}\label{eq:CP}
\begin{align}
    \mymu^{k+1} &= \prox_{\alpha f^*}(\mymu^k + \alpha A(2x^k-x^{k-1})),\label{subeq:CP-v} \\
    x^{k+1}   &= \prox_{\tau g}(x^k-\tau A^\T\mymu^{k+1})\label{subeq:CP-x} ,
\end{align}
\end{subequations}
which depend on both $(\mymu^k, x^k)$ and $x^{k-1}$.
In fact, if $u^0 = v^0 + \alpha A(x^0-x^{-1})$, then the two algorithms can be matched with each other, with the following relationship for $k\ge 0$:
\begin{subequations}
\begin{align}
   & v^{k+1}=\prox_{\alpha f^*}(\alpha A x^k + u^k), \label{subeq:v-from-ADMM} \\
   & u^{k} = v^{k} + \alpha A (x^{k}- x^{k-1}). \label{subeq:u-from-CP}
\end{align}
\end{subequations}
See the Supplement for details.
Under the same condition $0< \alpha\tau \|A\|_2^2\leq 1$, the iterates $(x^{k+1},\mymu^{k+1})$ converge to a saddle point of the primal-dual Lagrangian (\ref{eq:primal-dual lagrangian}).

\subsection{Proximal gradient, AMA and linearized AMA}\label{subsec:AMA}

As a prologue, we introduce the method of \textbf{proximal gradient},
which can be used to derive AMA.
As an extension of gradient descent, the method of proximal gradient is designed to solve composite optimization
in the form $\min_{x}\, \tilde f(x)+\tilde g(x)$, where
$\tilde f$ and $\tilde g$ are CCP, $\tilde f$ is differentiable but $\tilde g$ may be not. The proximal gradient update is
\begin{equation}\label{eq:proximal gradient method}
    x^{k+1}=\prox_{\alpha \tilde g}(x^k-\alpha \nabla \tilde f(x^k)).
\end{equation}
For $L$-smooth $\tilde f$, if $\alpha \in (0,2/L)$, then the iterate $x^{k+1}$ can be shown to converge to a minimizer of $\tilde f(x)+\tilde g(x)$. There are various interpretations for the proximal gradient method \citep{parikh2014proximal}.
In particular, when $\alpha\in(0,1/L]$, the proximal gradient method can be identified as a majorization-minimization (MM) algorithm,
which satisfies a descent property \citep{hunter2004tutorial}.  
By the definition of proximal mapping, the update $x^{k+1}$ in (\ref{eq:proximal gradient method}) is
a minimizer of $\tilde{f}(x,x^k)+\tilde g(x)$, where
$\tilde{f}(x,x^k) = \tilde f(x^k) + \langle x-x^k, \nabla \tilde f(x^k) \rangle + \frac{1}{2\alpha}\|x-x^k\|_2^2$.
When $\alpha\in(0,1/L]$, $\tilde{f}(x,x^k)$ is an upper bound (or a majorizing function) for $\tilde f$, and hence
the update $x^{k+1}$ leads to an MM algorithm.
For $\alpha \in (1/L,2/L)$, the proximal gradient method is no longer an MM algorithm, although convergence can still be established.

Now return to problem (\ref{eq:original problem}). If we further assume that $f$ is $\mu$-strongly convex, then $f^*$ is $(1/\mu)$-smooth.
Applying proximal gradient (\ref{eq:proximal gradient method}) to the dual problem (\ref{eq:generic dual problem for review}), i.e.,
\begin{equation*}
    \min_u\, \underbrace{f^*(u)}_{\tilde{f}(u)} + \underbrace{g^*(-A^\T u)}_{\tilde{g}(u)},
\end{equation*}
we obtain the \textbf{alternating minimization algorithm} (AMA) \citep{tseng1991applications}:
\begin{subequations}\label{eq:AMA}
\begin{align}
    z^{k+1} &= \argmin_{z}\, \left( f(z)-\langle u^k,z \rangle \right)=\nabla f^*(u^k) , \label{subeq:AMA-z}\\
    x^{k+1} &\in \,\argmin_{x}\, \left(g(x)+\langle A^\T u^k,x \rangle + (\alpha/2)\|Ax-z^{k+1}\|_2^2\right), \label{subeq:AMA-x}\\
    u^{k+1} &= u^k + \alpha(Ax^{k+1}-z^{k+1}).
\end{align}
\end{subequations}
See \cite{ryu2022large} for details about deriving AMA from proximal gradient.
Compared with ADMM (\ref{eq:ADMM}), a major difference is that there is no quadratic term (the augmented term) in (\ref{subeq:AMA-z}),
and $z^{k+1}$ takes a simpler form depending on $u^k$ only.

Similarly as (\ref{subeq:ADMM-x}) in ADMM, the step (\ref{subeq:AMA-x}) in AMA is in general difficult to implement.
By linearizing (\ref{subeq:AMA-x}) in the same way as (\ref{subeq:linear-ADMM-x}),
we propose \textbf{linearized AMA}:
\begin{equation}\label{eq:linear-AMA}
\begin{split}
    z^{k+1} &= \argmin_{z}\, \left( f(z)-\langle u^k,z \rangle \right)=\nabla f^*(u^k), \\
    x^{k+1} &= \prox_{\tau g}\left(x^k-\tau A^\T(u^k+\alpha(Ax^k-z^{k+1}))\right), \\
    u^{k+1} &= u^k + \alpha(Ax^{k+1}-z^{k+1}).
\end{split}
\end{equation}
This method represents a new application of the (heuristic) linearization technique (\cite{ryu2022large}, Section 3.5).
In the proof of Proposition \ref{pro:linear-AMA},
we show that (\ref{eq:linear-AMA}) corresponds to PAPC/$\mathrm{PDFP^2O}$ \citep{chen2013primal,drori2015simple} applied to the dual problem (\ref{eq:generic dual problem for review}). Hence, by the relationship of AMA and proximal gradient discussed above,
PAPC/$\mathrm{PDFP^2O}$ can be viewed as a linearized version of proximal gradient.
The convergence of (\ref{eq:linear-AMA}) can be deduced from that of PAPC/$\mathrm{PDFP^2O}$ as in \cite{li2021new}.

\begin{pro}\label{pro:linear-AMA}
Assume $f$ and $g$ are CCP, and $f$ is $\mu$-strongly convex. Assume that total duality holds and let $(z^{k+1},x^{k+1},u^{k+1})$ be the sequence generated in (\ref{eq:linear-AMA}).  If $0<\alpha<2\mu$ and $0<\alpha\tau\|A\|_2^2\leq 4/3$, then $(z^{k+1},x^{k+1},u^{k+1})$ converge to a saddle point of (\ref{eq:generic lagrangian}).
\end{pro}

\section{Single-block optimization for DPAM} \label{sec:single-block}

To implement doubly penalized ANOVA regression using backfitting (or block coordinate descent),
the subproblem of (\ref{eq:linear-DPAM}) or (\ref{eq:logistic-DPAM}) with respect to one block while fixing the rest takes the following form (see Section~\ref{sec:multi-block} for more details):
\begin{equation}\label{eq:single-block}
    \min_{\beta} \, \frac{1}{2n}\|r-X\beta\|_2^2 + \|\Gamma\beta\|_1 + \lambda \|X\beta\|_n,
\end{equation}
where $\beta\in\bbR^d$ is the coefficient vector for a specific block, $X\in\bbR^{n\times d}$ is the basis matrix,
$r\in \bbR^n$ is the residual vector after adjusting for the other blocks, $\Gamma$ is a diagonal matrix of scaling constants for the HTV/Lasso penalty,
and $\lambda\geq 0$ is a tuning parameter for the empirical-norm penalty. For simplicity we omit the subscript of the block index.

In this section, we consider the optimization of (\ref{eq:single-block}) where the basis dimension $d$ is of moderate size, but the sample size $n$ can be large.
The backfitting algorithm in \cite{yang2021hierarchical} relies on the following result.

\begin{pro}[\cite{yang2018backfitting}] \label{pro:soft-threshold}
Let $\tilde{\beta}$ be a minimizer for the Lasso problem
\begin{equation}\label{eq:single-block lasso}
    \min_{\beta} \, \frac{1}{2n}\|r-X\beta\|_2^2 + \|\Gamma\beta\|_1.
\end{equation}
If $\|X\tilde{\beta}\|_n=0$, take $\hat{\beta}=0$. Otherwise, take $\hat{\beta}= (1 - \lambda/\|X\tilde{\beta}\|_n )_+ \tilde{\beta}$,
where $c_+ = c$ for $c\ge 0$ or $0$ for $c <0$. Then $\hat\beta$ is a minimizer for problem (\ref{eq:single-block}).
\end{pro}

Proposition~\ref{pro:soft-threshold} not only makes explicit the different shrinkage effects of the two penalties,
HTV and empirical norm, but also provides a direct approach to solving (\ref{eq:single-block}): first solving the Lasso problem (\ref{eq:single-block lasso}) and then
rescaling (or jointly soft-thresholding) the Lasso solution $\tilde{\beta}$. However, this approach requires determination of the exact Lasso solution
and hence can be inefficient in large-scale applications.
Soft-thresholding an inexact Lasso solution may not decrease the objective value in problem (\ref{eq:single-block})
even when the objective value in (\ref{eq:single-block lasso}) is decreased before soft-thresholding.
It is also unclear how to derive appropriate stochastic algorithms in this approach for handling large datasets.

We propose and study three first-order methods for directly solving (\ref{eq:single-block})
without relying on Proposition~\ref{pro:soft-threshold}.
Both batch and stochastic algorithms are derived for each method.
The first two methods are based on the primal-dual algorithms in Section \ref{sec:Primal-dual review}.
The third method, called concave conjugate (CC), is derived from a different application of Fenchel duality
and can be interpreted as an MM algorithm to tackle the empirical-norm penalty.
From our numerical experiments, the CC method performs worse than or similarly as the primal-dual algorithms.
For space limitation, the CC method is presented in the Supplement.

In addition to the two-operator splitting methods described in Section~\ref{sec:Primal-dual review}, it seems natural to consider three-operator splitting methods
for solving (\ref{eq:single-block}). In Supplement Section~\ref{sec:additional discussions}, we present a (tractable) batch algorithm based on the Condat--V\~u algorithm \citep{condat2013primal,vu2013splitting}. However, its randomization seems to be difficult in our setting.

\subsection{Batch primal-dual algorithms}\label{subsec:PD-single block}

We apply the batch Chambolle--Pock algorithm (\ref{eq:CP}) and linearized AMA (\ref{eq:linear-AMA}) to the single-block problem. To make notations more convenient, we rescale problem (\ref{eq:single-block}) to
\begin{equation}\label{eq:rescaled single-block}
    \min_{\beta} \, \frac{1}{2}\|r-X\beta\|_2^2 + n\|\Gamma\beta\|_1 + \lambda\sqrt{n} \|X\beta\|_2,
\end{equation}
and then formulate problem (\ref{eq:rescaled single-block}) in the form of (\ref{eq:generic problem for review}) as
\begin{equation}\label{eq:generic problem for single-block}
    \min_{z,\beta} \, f(z)+g(\beta), \quad \text{subject to} \quad -z+X\beta=0,
\end{equation}
where
\begin{equation}\label{eq:f of single-block}
    f(z)=(1/2)\|r-z\|_2^2+\lambda\sqrt{n}\|z\|_2,
\end{equation}
for $z\in\bbR^n$, and $g(\beta)=n\|\Gamma\beta\|_1$ for $\beta\in\bbR^d$. The dual problem of (\ref{eq:generic problem for single-block}) is $\max_{u}\, -f^*(u)-g^*(-X^\T u)$, with the associated (full) Lagrangian
\begin{equation}\label{eq:generic lagrangian for single-block}
    L(z,\beta;u)=f(z)+g(\beta)+\langle u,X\beta-z \rangle,
\end{equation}
and the primal-dual Lagrangian
\begin{equation}\label{eq:primal-dual lagrangian for single-block}
    L(\beta;u)=g(\beta)+\langle u,X\beta \rangle - f^*(u).
\end{equation}
Although the choices of $f$ and $g$ are not unique, our choices above are
motivated by the fact that both the error term $\|r-X\beta\|_2^2$ and the empirical norm $\|X\beta\|_2$ depend on $\beta$ only through
the linear predictor $X\beta$.
Under our formulation (\ref{eq:generic problem for single-block}) of problem (\ref{eq:rescaled single-block}), the conjugate functions and
proximal mappings for $f$ and $g$ can be calculated in a tractable manner.
See Section \ref{subsec:stoc-CP} for a comparison with an alternative formulation.

The conjugate functions $f^*$ and $g^*$ can be calculated in a closed form as
\begin{equation}\label{eq:f* of single-block}
    f^*(u)=(1/2)(\|u+r\|_2-\lambda\sqrt{n})_+^2-(1/2)\|r\|_2^2,
\end{equation}
for $u\in\bbR^n$ and $g^*(t)=\sum_{j=1}^d \delta(|t_j|\leq n\Gamma_j)$ for $t \in\bbR^d$,
where $\Gamma_j$ is the $j$th diagonal element of $\Gamma$, and $\delta_S(x)$ denotes an indicator function for a set $S$ such that $\delta_S(x)=0$ if $x\in S$, and $\infty$ otherwise.
For scalars $x\in\bbR$ and $\gamma\geq 0$, denote the soft-thresholding operation
as $\mathcal{S}(x,\gamma) = (1-\gamma/|x|)_+ x$. If $x=0$, then set $\mathcal{S}(x,\gamma)=0$.
For vectors $x$ and $r$ of the same dimensions, we still use $\mathcal{S}(x,r)$ to denote the vector obtained
by applying soft-thresholding element-wise, i.e., $\mathcal{S}(x,r) =[(1-r_i/|x_i|)_+x_i]_i$. For a vector $x$ and a scalar $\gamma\geq 0$, we denote the joint soft-thresholding as $\mathcal{T}(x,\gamma) = (1-\gamma/\|x\|_2)_+ x.$ If $x=0$, then set $\mathcal{T}(x,\gamma)=0$.
The result in Proposition \ref{pro:soft-threshold} can be stated as $X\hat\beta = \mathcal{T} ( X\tilde\beta, \lambda \sqrt{n} )$.
With the preceding notation, it can be directly calculated that
\begin{equation} \label{eq:prox of f of single-block}
    \prox_{\alpha f}(z)=\frac{\mathcal{T}(z+\alpha r, \alpha \lambda\sqrt{n})}{1+\alpha},
\end{equation}
and $\prox_{\tau g}(\beta)=\mathcal{S}(\beta,\tau n \cdot\diag(\Gamma))$.

Applying the linearized ADMM (\ref{eq:linear-ADMM}) to problem (\ref{eq:generic problem for single-block}), we obtain
\begin{subequations}\label{eq:batch-linear-ADMM}
\begin{align}
    z^{k+1} 
    &= \prox_{f/\alpha}(X\beta^k+u^k/\alpha) \\
    &=\mathcal{T}(\alpha X\beta^k+u^k+r,\lambda\sqrt{n})/(1+\alpha), \label{subeq:batch-l-admm-z-closed form} \\
    \beta^{k+1} &=\prox_{\tau g/n}\left(\beta^k-\frac{\tau}{n} X^\T(u^k+\alpha(X\beta^k-z^{k+1}))\right) \\
    &=\mathcal{S}\left(\beta^k-\frac{\tau}{n} X^\T(u^k+\alpha(X\beta^k-z^{k+1})),\tau\cdot\diag(\Gamma)\right), \\
    u^{k+1} &= u^k+\alpha(X\beta^{k+1}-z^{k+1}) .
\end{align}
\end{subequations}
Moreover, application of the CP algorithm (\ref{eq:CP}) yields Algorithm~\ref{alg:batch-CP}, where line (\ref{subeq:batch CP-v-closed form}) follows from (\ref{eq:prox of f of single-block}) and Moreau's identity (\ref{eq:Moreau's identity}).
In fact, Algorithm \ref{alg:batch-CP} can be equivalently transformed from (\ref{eq:batch-linear-ADMM}) as discussed in Section \ref{subsec:ADMM}.
Both algorithms are stated, because we find it somewhat more direct to derive stochastic algorithms from the CP algorithm,
whereas the relationship of the CP algorithm with linearized ADMM (\ref{eq:batch-linear-ADMM}) helps us find a simple criterion to
declare a zero solution of $\beta$, which is discussed below.

\begin{algorithm} [t]
\caption{Single-block Chambolle--Pock}\label{alg:batch-CP}
\begin{algorithmic}
\STATE {\bf Input} Initial values $(\beta^{-1}=\beta^0,u^0)$, number of batch steps $B$, step sizes $\tau$ and $\alpha$.
\FOR{$k=0,1,\ldots,B-1$}
\STATE \vspace{-.35in} 
\begin{subequations}
\begin{align}
    \mymu^{k+1} &= \prox_{\alpha f^*}(\mymu^k + \alpha X(2\beta^k-\beta^{k-1}))\label{subeq:batch CP-v}   \\
    &= \mymu^k + \alpha X(2\beta^k-\beta^{k-1}) - \frac{\alpha}{1+\alpha}\mathcal{T}(\mymu^k + \alpha X(2\beta^k-\beta^{k-1})+r,\lambda\sqrt{n}), \label{subeq:batch CP-v-closed form} \\
    \beta^{k+1} &= \prox_{\tau g/n}(\beta^k-\frac{\tau}{n} X^\T\mymu^{k+1})\label{subeq:batch CP-beta} \\
    &= \mathcal{S}\left(\beta^k-\frac{\tau}{n} X^\T \mymu^{k+1},\tau\cdot \diag(\Gamma)\right). \label{subeq:batch CP-beta-closed form}
\end{align}
\end{subequations}
\ENDFOR
\end{algorithmic}
\end{algorithm}

Next, we notice that $f$ in (\ref{eq:f of single-block}) is $1$-strongly convex and $f^*$ in (\ref{eq:f* of single-block}) is $1$-smooth, with the gradient $\nabla f^*(u)=\mathcal{T}(u+r,\lambda\sqrt{n})$.
Hence linearized AMA (\ref{eq:linear-AMA}) can also be applied to problem (\ref{eq:generic problem for single-block}), leading to Algorithm~\ref{alg:batch-linearized-AMA}, which differs from (\ref{eq:batch-linear-ADMM}) only in the $z^{k+1}$-step.

\begin{algorithm} [t]
\caption{Single-block linearized AMA}\label{alg:batch-linearized-AMA}
\begin{algorithmic}
\STATE {\bf Input} Initial values $(\beta^0,u^0)$, number of batch steps $B$,
step sizes $\tau$ and $\alpha$.
\FOR{$k=0,1,\ldots, B-1$}
\STATE  \vspace{-.35in}  
\begin{subequations}
\begin{align}
    z^{k+1} &= \argmin_{z}\, \left( f(z)-\langle u^k,z \rangle \right)=\nabla f^*(u^k)  \\
    &=\mathcal{T}(r+u^k,\lambda\sqrt{n}), \label{subeq:batch AMA-z}   \\
    \beta^{k+1} &=\prox_{\tau g/n}\left(\beta^k-\frac{\tau}{n} X^\T(u^k+\alpha(X\beta^k-z^{k+1}))\right) \\
    &=\mathcal{S}\left(\beta^k-\frac{\tau}{n} X^\T(u^k+\alpha(X\beta^k-z^{k+1})),\tau\cdot\diag(\Gamma)\right), \label{subeq:batch AMA-beta} \\
    u^{k+1} &= u^k+\alpha(X\beta^{k+1}-z^{k+1}).
\end{align}
\end{subequations}
\ENDFOR
\end{algorithmic}
\end{algorithm}

Convergence results for Algorithms~\ref{alg:batch-CP} and \ref{alg:batch-linearized-AMA} can be directly obtained from \cite{chambolle2011first} and Proposition~\ref{pro:linear-AMA} respectively.

\begin{pro}\label{pro:convergence of batch primal-dual}
For Algorithm~\ref{alg:batch-CP}, if $\alpha>0$, $\tau>0$ and $\alpha\tau\|X\|_2^2\leq n$, then $(\beta^k,u^k)$ converges to a saddle point of (\ref{eq:primal-dual lagrangian for single-block}). For Algorithm~\ref{alg:batch-linearized-AMA}, if $0<\alpha<2$, $\tau>0$ and $\alpha\tau\|X\|_2^2\leq 4n/3$, then $(z^k,\beta^k,u^k)$ converges to a saddle point of (\ref{eq:generic lagrangian for single-block}).
\end{pro}

To conclude this section, we discuss how the sparsity of $\beta$ can be reached in Algorithms \ref{alg:batch-CP} and \ref{alg:batch-linearized-AMA}.
As shown by Proposition \ref{pro:soft-threshold}, a solution $\hat\beta$ to problem (\ref{eq:single-block}) may
exhibit two types of sparsity.
One is element-wise sparsity: $\hat\beta$ may be a sparse vector (with some elements being 0), induced by the Lasso penalty $\| \Gamma \beta\|_1$.
The other is group sparsity: $\hat\beta$ may be an entirely zero vector, as a result of the empirical-norm penalty $\| X\beta\|_n$.
For both Algorithms \ref{alg:batch-CP} and \ref{alg:batch-linearized-AMA}, each iterate $\beta^{k+1}$ in (\ref{subeq:batch CP-beta-closed form}) or (\ref{subeq:batch AMA-beta}) is obtained
using the element-wise soft-thresholding operator $\mathcal S$.
On one hand, such $\beta$ iterates may directly achieve the element-wise sparsity, with some elements being 0.
On the other hand,  the group sparsity may unlikely be satisfied by any iterate $\beta^{k+1}$, because
the element-wise soft-thresholding operator $\mathcal S$ does not typically produce a zero vector,
especially when not all elements of $\beta$ are penalized (with some diagonal elements of $\Gamma$ being $0$) as in the case of piecewise cross-linear basis functions.

The preceding phenomenon can be attributed to the splitting of the two penalties in (\ref{eq:generic problem for single-block}):
$\| \Gamma \beta\|_1$ is assigned to the function $g$,
and $\| X\beta\|_2$ is assigned to the function $f$ through the slack variable $z = X\beta$.
From this perspective, a zero solution for $\beta$ can be more properly detected by checking whether $z^{k+1}=0$, instead of $\beta^{k+1}=0$,
although the iterates $( \beta^{k+1}, z^{k+1})$ converge to $(\hat\beta, X\hat\beta)$,
and $\hat\beta=0$ if and only if $X \hat\beta=0$ for $X$ of rank $d$.
In fact, for Algorithm \ref{alg:batch-linearized-AMA}, $z^{k+1}$ is determined as (\ref{subeq:batch AMA-z}) using the joint soft-thresholding operator $\mathcal T$,
which may likely produce a zero vector.
For Algorithm~\ref{alg:batch-CP}, by the relationship (\ref{subeq:u-from-CP}),
the corresponding $z^{k+1}$ from (\ref{subeq:batch-l-admm-z-closed form}) in linearized ADMM
can be rewritten in terms of $\mathcal T$ as
$z^{k+1} = \mathcal{T}(\alpha X( 2 \beta^k- \beta^{k-1}) + v^k+r,\lambda\sqrt{n})/(1+\alpha)$.
In our implementation, we reset $\beta$ to 0 after the final iteration if $z^{k+1}=0$ or equivalently
\begin{align}
& \text{Algorithm \ref{alg:batch-CP}:} \quad \|\alpha X(2\beta^k-\beta^{k-1})+v^k+r\|_n\leq \lambda, \label{eq:condition for CP}\\
& \text{Algorithm \ref{alg:batch-linearized-AMA}:} \quad \|r+u^k\|_n\leq \lambda. \label{eq:condition for AMA}
\end{align}
By this scheme, a zero solution for $\beta$ can be effectively recovered from Algorithms \ref{alg:batch-CP} and \ref{alg:batch-linearized-AMA},
even though the $\beta$ iterates themselves may not yield a zero vector.

\subsection{Stochastic primal-dual algorithms}\label{subsec:stoc-PD}

For large datasets (with large $n$), it is desirable to develop stochastic primal-dual algorithms.
Typically, a batch optimization method, such as gradient descent or Algorithms~\ref{alg:batch-CP} and \ref{alg:batch-linearized-AMA}, has
a per-iteration cost of  $O(nd)$, because the full data matrix (or basis matrix) $X$ needs to be scanned.
Stochastic optimization methods, on the other hand, act only on a single row of $X$, thus lowering the per-iteration cost to $O(d)$,
free of $n$. A potential advantage of stochastic methods
is that their overall efficiency, when measured in the number of batch steps (with each batch step consisting of $n$ iterations) over the full data matrix $X$,
can still be superior over that of batch methods, measured in the number of iterations.

We develop stochastic versions of Algorithms~\ref{alg:batch-CP} and \ref{alg:batch-linearized-AMA} based on two general principles: replacing the batch gradient with an unbiased or approximately unbiased stochastic gradient for the update of $\beta$, and performing randomized coordinate updates for the dual variable $u$ or $v$.
The formal analysis of convergence is left for future work.

\subsubsection{Stochastic Chambolle--Pock}\label{subsec:stoc-CP}

As a background, we briefly discuss SPDC (stochastic primal-dual coordinate method) proposed by \cite{zhang2017stochastic}, as a stochastic version of
the CP algorithm (\ref{eq:CP}). The method deals with minimizing an objective function in the form
\begin{equation}\label{eq:ERM}
 \min_{\beta}\,  \frac{1}{n}\sum_{i=1}^n f_i(X_{i,\cdot}^\T \beta)+\tilde{g}(\beta),
\end{equation}
where $X_{i,\cdot} \in \bbR^d$ denotes the transpose of $i$th row of $X$, each $f_i$ is convex and smooth, and $\tilde g$ is a regularizer
(for example, the Lasso penalty) such that its proximal mapping can be easily evaluated.
After rescaling by $n$, (\ref{eq:ERM}) can be put in the form of (\ref{eq:generic problem for single-block}), where
$f(z)=\sum_{i=1}^n f_i(z_i)$ is coordinate-wise separable with $z = (z_1,\ldots, z_n)^\T$ and $g(\beta)=n\tilde g(\beta)$. Consequently,
the conjugate of $f$ is separable:
$f^*(v)=\sum_{i=1}^n f_i^*(v_i)$, where $v=(v_1,\ldots,v_n)^\T $ and $f^*_i$ is the conjugate of $f_i$.
The dual update in the CP algorithm (\ref{eq:CP}) also becomes separable:
the $i$th element, $\mymu^{k+1}_i$, of $\mymu^{k+1}$ in (\ref{subeq:CP-v}) can be obtained as
\begin{equation}
    \mymu^{k+1}_i = \prox_{\alpha f_i^*}(\mymu^k_i + \alpha X_{i,\cdot}^\T(2\beta^k-\beta^{k-1})). \label{eq:SPDC-i}
\end{equation}
In other words, each element $\mymu^{k+1}_i$ can be updated with cost $O(d)$,  
independently of the other elements in $\mymu^{k+1}$.
Such separability is exploited by SPDC to achieve two properties of unbiasedness, given the history up to $(v^k,\beta^k)$.
First, one coordinate of $\mymu^k$ (for example $i$th) is randomly selected and then updated to $\mymu^{k+1}_i$,
the $i$th element in the full update $\mymu^{k+1}$ in (\ref{eq:SPDC-i}).
The resulting update of $\mymu$, different from $\mymu^k$ by one coordinate, is unbiased for the full update $\mymu^{k+1}$.
Second, a stochastic gradient, $G^{k+1}$, is created to replace the batch gradient $X^\T \mymu^{k+1}/n$ in the update $\beta^{k+1}$ in (\ref{subeq:CP-x}),
such that $G^{k+1}$ is unbiased for $X^\T \mymu^{k+1}/n$ with the full update $\mymu^{k+1}$.
The choice of $G^{k+1}$ in SPDC is defined as
$G^{k+1} =  X_{i,\cdot}( \mymu_i^{k+1}-\mymu_i^k) + \frac{1}{n}X^\T \mymu^k$,
corresponding to the SAGA method \citep{defazio2014saga} discussed below.

Our problem (\ref{eq:single-block}) can be put in the form of (\ref{eq:ERM}), with $f_i (t_i) = (y_i-t_i)^2/2$
and $\tilde g(\beta) = \|\Gamma\beta\|_1 + \lambda \|X\beta\|_n$, for which the proximal mapping is difficult to evaluate.
Hence our problem does not fit into the setting of SPDC.
Moreover, our formulation (\ref{eq:generic problem for single-block}) of problem (\ref{eq:single-block}) differs substantially from
(\ref{eq:ERM}) in that the function $f$ in (\ref{eq:f of single-block}) is not separable due to the inclusion of the empirical norm.
In our CP algorithm (Algorithm~\ref{alg:batch-CP}),
evaluating one coordinate of $\mymu^{k+1}$ in (\ref{subeq:batch CP-v-closed form}) is as difficult as evaluating the full $\mymu^{k+1}$, costing $O(nd)$.
Therefore, we need to generalize related ideas in SPDC to derive a stochastic CP algorithm with non-separable $f$.

To approximate the batch update $\mymu^{k+1}$, we apply randomized coordinate minimization,
i.e., evaluating the proximal mapping in (\ref{subeq:batch CP-v}) for a randomly selected coordinate while fixing the remaining coordinates.
This operation is distinct from that of evaluating one randomly selected coordinate of the full update $\mymu^{k+1}$, although
the two operations coincide in the special case of separable $f$ as in SPDC.
We show that the coordinate minimization can be done with cost $O(d)$ provided we additionally maintain $\|\mymu^k+r\|_2^2$.

\begin{pro}\label{pro:cd of proximal}
For $b\in\bbR^n$ and $\alpha\geq 0$, let
\begin{align}
\prox_{\alpha f^*}(b,i,\mymu^k_{-i}) = \underset{v_i}{\argmin} \, \frac{1}{2} \|\mymu-b\|_2^2+\alpha f^*(\mymu), \label{eq:CD-prox}
\end{align}
where  $f^*$ is defined in (\ref{eq:f* of single-block}), and $\mymu_i$ is the $i$th coordinate of $\mymu$, while the remaining coordinates of $\mymu$ are fixed at
$\mymu^k_{-i}=(\mymu^k_1,\ldots,\mymu^k_{i-1},\mymu^k_{i+1},\ldots,\mymu^k_n)$.
Then
$\prox_{\alpha f^*}(b,i,\mymu^k_{-i}) = -r_i$ if $b_i+r_i=0$; otherwise $\prox_{\alpha f^*}(b,i,\mymu^k_{-i}) = c(b_i+r_i)-r_i$, with
\begin{equation*}
    c = \begin{cases}
    1, & \text{if } (b_i+r_i)^2+\|\mymu^k+r\|^2_{-i}\leq n\lambda^2, \\
    \left(1+\frac{\alpha\lambda\sqrt{n}}{|b_i+r_i|}\right)/(1+\alpha), & \text{if } \|\mymu^k+r\|^2_{-i}=0 \text{ and } |b_i+r_i|>\lambda\sqrt{n}, \\
    c^*, & otherwise,
    \end{cases}
\end{equation*}
where $\|\mymu^k+r\|^2_{-i} = \|\mymu^k+r\|^2_2-(\mymu^k_i+r_i)^2$ and $c^*$ is the root of equation
$$
\left(1+\alpha - \frac{\alpha\lambda\sqrt{n}}{\sqrt{c^2(b_i+r_i)^2+\|\mymu^k+r\|^2_{-i}}} \right)c=1 .
$$
Such $c^*$ uniquely lies in $(0,1)$.
\end{pro}

Given the history up to $(\mymu^k, \beta^k)$, we randomly select one coordinate (for example $i$th) and
define the stochastic update $\mymu^{k+1}$ as follows:
\begin{equation}\label{eq:random cd-stoc-CP}
    \mymu^{k+1}_i = \prox_{\alpha f^*}(\mymu^k + \alpha X^\T(2\beta^k-\beta^{k-1}),i,\mymu^k_{-i}) \quad \text{or} \quad
     \mymu_l^{k+1} = \mymu_l^{k} \text{ for } l\neq i,
\end{equation}
while keeping the remaining elements as in $\mymu^k$.
Note that $\mymu^{k+1}$ here differs from the batch update $\mymu^{k+1}$ in (\ref{subeq:batch CP-v}),
which will henceforth be denoted as $v^{k+1}_{\text{B}}$.
From Proposition~\ref{pro:cd of proximal},
we observe that the coordinate proximal mapping (\ref{eq:CD-prox}) costs $O(1)$ to compute, once $b_i$ and $\|\mymu^k+r\|^2_{-i}$ are determined.
For our Algorithm \ref{alg:stoc-CP}, the former is evaluated as $\mymu^k_i + \alpha X_{i,\cdot}^\T(2\beta^k-\beta^{k-1})$ with cost $O(d)$,
and the latter costs $O(1)$ to compute as we maintain $\|\mymu^k+r\|_2^2$.
Therefore, the proposed update $\mymu^{k+1}$ can be computed with cost $O(d)$.

For the primal update $\beta^{k+1}$, the general idea is to replace the batch gradient $G^{k+1}_{\text{B}}=X^\T v^{k+1}_{\text{B}}/n$
by a stochastic approximation $G^{k+1}$, and define
\begin{equation}\label{eq:beta-update-stoc-CP}
    \beta^{k+1} =\prox_{\tau g/n}(\beta^k-\tau G^{k+1}) = \mathcal{S}(\beta^k-\tau G^{k+1},\tau\cdot \diag(\Gamma)),
\end{equation}
where $v^{k+1}_{\text{B}}$ denotes the batch update $v^{k+1}$ in (\ref{subeq:batch CP-v}), to be distinguished from the partial update $v^{k+1}$ defined above.
There are at least three choices for $G^{k+1}$, corresponding to three related stochastic gradient methods.
Given the approximation $v_i^{k+1}$,
the first is the standard stochastic gradient method (SG) \citep{bottou2018optimization}:
$G^{k+1} = X_{i,\cdot}v_i^{k+1}$.
The second is based on the stochastic
average gradient (SAG) method \citep{roux2012stochastic}:
\begin{equation}\label{eq:stoc-CP-SAG}
    G^{k+1} = \frac{1}{n}X_{i,\cdot}(v_i^{k+1}-v_i^k) + \frac{1}{n}\sum_{i=1}^n X_{i,\cdot}v_i^k=\frac{1}{n} X_{i,\cdot}(v_i^{k+1}-v_i^k) + \underbrace{\frac{1}{n}X^\T v^k}_{w^k}.
\end{equation}
The third is based on SAGA \citep{defazio2014saga}:
\begin{equation}\label{eq:stoc-CP-SAGA}
    G^{k+1} = X_{i,\cdot}(v_i^{k+1}-v_i^k) + \frac{1}{n}\sum_{i=1}^n X_{i,\cdot}v_i^k=X_{i,\cdot}( v_i^{k+1}-v_i^k) + \underbrace{\frac{1}{n}X^\T v^k}_{w^k}.
\end{equation}
For (\ref{eq:stoc-CP-SAG}) and (\ref{eq:stoc-CP-SAGA}), the variable $w^k$ can be updated as
$w^{k+1}=w^k+ X_{i,\cdot}(v_i^{k+1}-v_i^k)/n$,
so that there is no need to re-calculate the average in each iteration.

There is an extensive literature on the three methods and other variants in the standard setting (\ref{eq:ERM}) or similar ones.
Both SAG and SAGA are designed for variance reduction, and SAG often yields smaller variance than SAGA.
Moreover, given the history,
the stochastic gradients in SG and SAGA are known to be unbiased, but that in SAG is biased, in standard settings.
Following SPDC, we adopt (\ref{eq:stoc-CP-SAGA}) based on SAGA
as the stochastic gradient in our update of $\beta$. The resulting algorithm is summarized in Algorithm~\ref{alg:stoc-CP}.

Convergence analysis of the proposed stochastic algorithm remains to be studied.
As discussed earlier in this section, due to separable $f$ in problem (\ref{eq:ERM}),
the closely related SPDC method enjoys two properties of (exact) unbiasedness
for the updates of $\mymu$ and $\beta$, when compared with the corresponding batch CP algorithm.
Such a simple relationship no longer holds between the stochastic Algorithm~\ref{alg:stoc-CP}
and batch Algorithm~\ref{alg:batch-CP} in our setting, due to non-separable $f$.
Given up to $k$th iteration, our stochastic update $\mymu^{k+1}$ or gradient $G^{k+1}$
is not exactly unbiased for the batch version $\mymu^{k+1}_{\text{B}}$ or $G^{k+1}_{\text{B}}$,
because randomized coordinate minimization is used
instead of evaluating a randomly selected coordinate of the batch update $\mymu^{k+1}_{\text{B}}$
(which is as costly as evaluating $\mymu^{k+1}_{\text{B}}$ itself with non-separable $f$).
Further study is needed to tackle these complications for theoretical analysis.

\begin{algorithm}[t!] 
\caption{Stochastic Chambolle--Pock}\label{alg:stoc-CP}
\begin{algorithmic}
\STATE {\bf Input} Initial $(\beta^0,\mymu^0)$, number of batch steps $B$, step sizes $\alpha$ and $\tau$.
\STATE {\bf Initialize} $w^0=X^\T \mymu^0/n$, and $L^2=\|\mymu^0+r\|_2^2$.
\FOR{$k=0,1,\ldots, nB-1$}
\STATE Pick $i$ uniformly from $\{1,\ldots,n\}$, and perform the following updates: \vspace{-.1in}
\begin{align*}
    & \text{$\mymu^{k+1}$, $G^{k+1}$, and $\beta^{k+1}$ by (\ref{eq:random cd-stoc-CP}) with $L^2=\|\mymu^k+r\|_2^2$, (\ref{eq:stoc-CP-SAGA}), and (\ref{eq:beta-update-stoc-CP}),}  \\
    & w^{k+1} = w^k + X_{i,\cdot}(\mymu^{k+1}_i-\mymu^{k}_i)/n, \\
    & L^2 \gets L^2 - (\mymu^k_i+r_i)^2 + (\mymu^{k+1}_i+r_i)^2.
\end{align*}
\ENDFOR
\end{algorithmic}
\end{algorithm}

\subsubsection{Stochastic linearized AMA}\label{subsec:stoc-linear-AMA}
We develop a stochastic version of our linearized AMA algorithm (Algorithm~\ref{alg:batch-linearized-AMA}).
For technical convenience, we reorder Algorithm~\ref{alg:batch-linearized-AMA} such that the dual variable is updated first,
and the primal update is moved to the end of each iteration, i.e.,
\begin{subequations}\label{eq:re-ordered linear AMA}
\begin{align}
    u^{k+1} &= u^k + \alpha(X\beta^k-\nabla f^*(u^k)) \label{subeq:re-ordered AMA-u}\\
    \beta^{k+1} &= \prox_{\tau g/n}\left[\beta^k-\frac{\tau}{n} X^\T\left(u^{k+1}+\alpha (X\beta^k-\nabla f^*(u^{k+1}))\right)\right].\label{subeq:re-ordered AMA-beta}
\end{align}
\end{subequations}
Note that the auxiliary variable $z^{k}=\nabla f^*(u^k)$ and $z^{k+1}=\nabla f^*(u^{k+1})$ are absorbed in the preceding updates such that only primal and dual variables are left.

To approximate the dual update (\ref{subeq:re-ordered AMA-u}), it is desirable to perform a randomized coordinate update as follows.
We randomly select one coordinate (for example $i$th) and define the stochastic update $u^{k+1}$ as follows:
\begin{align} \label{eq:random update-stoc-AMA}
u_i^{k+1} &= u_i^k+\alpha (X_{i,\cdot}^\T\beta^k -\nabla_i f^*(u^k)) \quad \text{or}\quad u_l^{k+1} = u_l^k \text{ for } l\neq i,
\end{align}
where $\nabla_i f^*(u^k)$ denotes the $i$th element of $\nabla f^*(u^k)$, i.e., $\partial f^*(u^k)/\partial u_i$.
Note that $u^{k+1}$ here differs from the batch update $u^{k+1}$ in (\ref{subeq:re-ordered AMA-u}),
which will henceforth be denoted as $u^{k+1}_{\text{B}}$.
While evaluating $X_{i,\cdot}^\T\beta^k$ costs $O(d)$,
evaluating one coordinate of $\nabla f^*(u)$ may in general cost $O(n)$ for non-separable $f$.
However, unlike stochastic Chambolle--Pock in Section~\ref{subsec:stoc-CP} where coordinate minimization is employed to handle the non-separability, evaluating $\nabla_i f^*(u^k)$ turns out to cost only $O(1)$ in the current setting, due to the special form of $f^*$.
In fact, by the definition of $f^*$ in (\ref{eq:f* of single-block}), $\nabla f^*(u^k)$ takes the form of joint soft-thresholding:
$$
\nabla f^*(u^k) = \mathcal{T}(r+u^k,\lambda\sqrt{n})=\left(1-\frac{\lambda\sqrt{n}}{\|r+u^k\|_2}\right)_+\cdot (r+u^k).
$$
To evaluate $\nabla_i f^*(u^k)$, it suffices to compute the scalar factor above, which costs $O(1)$ if we additionally maintain $\|r+u^k\|_2$
as shown in Algorithm \ref{alg:stoc-linear-AMA}.
Therefore, the proposed update $u^{k+1}$ can be computed with cost $O(d)$.

To approximate the primal update (\ref{subeq:re-ordered AMA-beta}), the general strategy is to construct a stochastic estimate ${G}^{k+1}$ for the batch gradient
\begin{equation}\label{eq:batch gradient of single block AMA}
   G^{k+1}_{\text{B}} = \frac{1}{n}X^\T u^{k+1}_{\text{B}} + \frac{\alpha}{n}X^\T(X\beta^k-\nabla f^*(u^{k+1}_{\text{B}})),
\end{equation}
where $u^{k+1}_{\text{B}}$ denotes the batch update $u^{k+1}$ in (\ref{subeq:re-ordered AMA-u}), to be distinguished from the partial update $u^{k+1}$
defined above (which differs from $u^k$ by one coordinate).
Note that (\ref{eq:batch gradient of single block AMA}) is more complicated than the batch gradient $X^\T v^{k+1}_{\text{B}}/n$ in the CP algorithm.
The additional second term involves $\nabla f^*(u^{k+1}_{\text{B}})$, which depends on the unknown full update $u^{k+1}_{\text{B}}$ in a nonlinear manner.
To proceed, the only feasible approach seems to be replacing the batch update $u^{k+1}_{\text{B}}$ by the partial update $u^{k+1}$
and approximating the second term by
$\alpha(X_{i,\cdot}^\T\beta^k-\nabla_i f^*( u^{k+1}))\cdot X_{i,\cdot}$.
For the first term in (\ref{eq:batch gradient of single block AMA}), in line with the approximation of the second term, it is natural to also replace
the batch update $u^{k+1}_{\text{B}}$ by the partial update $u^{k+1}$.
Combining the preceding choices leads to the stochastic gradient
\begin{align} \label{eq:stocGrad-AMA-SAG}
G^{k+1} = w^k +\frac{1}{n}X_{i,\cdot}(u_i^{k+1}-u_i^k) + \alpha X_{i,\cdot}( X_{i,\cdot}^\T\beta^k-\nabla_i f^*( u^{k+1})),
\end{align}
where $u^{k+1}$ is the partial update defined in (\ref{eq:random update-stoc-AMA}), and $w^k=X^\T u^k /n$ can be updated as
$w^{k+1}=w^k+ X_{i,\cdot}(u_i^{k+1}-u_i^k)/n$.
The primal update of $\beta$ is then defined as
\begin{align} \label{eq:beta-update-stoc-AMA}
\beta^{k+1} =\prox_{\tau g/n}(\beta^k-\tau G^{k+1} )= \mathcal{S}(\beta^k-\tau G^{k+1} , \tau\cdot\diag(\Gamma)) ,
\end{align}
The resulting algorithm is summarized in  Algorithm~\ref{alg:stoc-linear-AMA}.
The stochastic gradient $G^{k+1}$ above can be seen to be in spirit of SAG.
Alternatively, SAGA can also be employed to approximate the first term in (\ref{eq:batch gradient of single block AMA}).
The corresponding stochastic gradient is
\begin{align} \label{eq:stocGrad-AMA-SAGA}
G^{k+1} = w^{k} + X_{i,\cdot}(u_i^{k+1}-u_i^k) + \alpha X_{i,\cdot}( X_{i,\cdot}^\T\beta^k-\nabla_i f^*(u^{k+1} )) ,
\end{align}
which leads to the SAGA option in Algorithm~\ref{alg:stoc-linear-AMA}.

Convergence analysis of Algorithm~\ref{alg:stoc-linear-AMA} remains to be studied,
although with slightly different complications than in Algorithm \ref{alg:stoc-CP}.
Given up to $k$th iteration, the stochastic update $u^{k+1}$ is unbiased for the batch update $u^{k+1}_{\text{B}}$
in spite of non-separable $f$ in our problem.
However, the stochastic gradient $G^{k+1}$ is not exactly unbiased for the batch gradient $G^{k+1}_{\text{B}}$,
due to the nonlinear dependency of $\nabla f^*(u^{k+1}_{\text{B}})$ on $u^{k+1}_{\text{B}}$ as discussed earlier.

\vspace{.1in}
\begin{algorithm} 
\caption{Single-block stochastic linearized AMA}\label{alg:stoc-linear-AMA}
\begin{algorithmic}
\STATE {\bf Input} Initial $(\beta^0,u^0)$, number of batch steps $B$, step sizes $\alpha$ and $\tau$.
\STATE {\bf Initialize} $w^0=X^\T u^0/n$, and $L^2=\|u^0+r\|_2^2$.
\FOR{$k=0,1,\ldots, nB-1$}
\STATE Pick $i$ uniformly from $\{1,\ldots,n\}$, and perform the following updates:\vspace{-.1in}
\begin{flalign*}
    &\quad \text{$u^{k+1}$ by (\ref{eq:random update-stoc-AMA}) with $\nabla_i f^*(u^k) = (1-\lambda\sqrt{n}/{L})_+\cdot (r_i+u_i^k)$,} &\\
    &\quad L^2 \gets L^2 - (r_i+u_i^k)^2 + (r_i+u_i^{k+1})^2, &\\
    &\quad \nabla_i f^*(u^{k+1}) = (1-\lambda\sqrt{n}/{L})_+\cdot (r_i+u_i^{k+1}), &\\
    &\quad \textbf{if }\text{AMA-SAG }\textbf{then} &\\
    &\quad \text{\quad $G^{k+1}$ and $\beta^{k+1}$ by  (\ref{eq:stocGrad-AMA-SAG}) and (\ref{eq:beta-update-stoc-AMA}),} &\\
    &\quad \textbf{if }\text{AMA-SAGA }\textbf{then} &\\
    &\quad \text{\quad $G^{k+1}$ and $\beta^{k+1}$ by (\ref{eq:stocGrad-AMA-SAGA}) and (\ref{eq:beta-update-stoc-AMA}),} &\\
    &\quad w^{k+1} = w^k + X_{i,\cdot}(u_i^{k+1}-u_i^k)/n.
\end{flalign*} \vspace{-.45in}
\ENDFOR
\end{algorithmic}
\end{algorithm}

\section{Multi-block optimization for DPAM}\label{sec:multi-block}
\subsection{Training and predictions}
We use the backfitting algorithm in \cite{yang2018backfitting, yang2021hierarchical} to solve the multi-block problems (\ref{eq:linear-DPAM}) and (\ref{eq:logistic-DPAM}). For completeness, we briefly introduce the method here. The top-level idea is updating one selected block while fixing the rest
and then cycling over all blocks.
To solve the single-block problem, methods in Section~\ref{sec:single-block} can be used.

Let $\Psi^\dag_{S_k}$ be the basis matrix formed from the basis vector $\Psi_{S_k}$, i.e., the $i$th row of $\Psi^\dag_{S_k}$ is the transpose of $\Psi_{S_k}(X_i)$ for $i=1,\ldots,n$. For the linear regression problem (\ref{eq:linear-DPAM}), to avoid interference between $\beta_0$ and other coefficients,
we solve a slightly modified problem
\begin{equation}
    \min_{(\beta_{S_k})_{k,S_k}}\, \frac{1}{2}\Bigl\|\tilde{Y}-\sum_{k,S_k} \tilde\Psi^\dag_{S_k}\beta_{S_k}\Bigr\|^2_n+\sum_{k,S_k} \Bigl(\|\Gamma_{S_k}\beta_{S_k}\|_1+\lambda_k\|\tilde\Psi^\dag_{S_k}\beta_{S_k}\|_n  \Bigl), \label{eq:linear-DPAM-centered}
\end{equation}
where $\tilde{Y}=Y-\bar{Y}$ and $\tilde\Psi^\dag_{S_k}=\Psi^\dag_{S_k}-\bar{\Psi}^\dag_{S_k}$ are empirically centered versions of $Y$ and $\Psi^\dag_{S_k}$.
See \cite{yang2021hierarchical}, Section 3.3, for a formal justification, which shows that the modified problem is equivalent to
the original problem (\ref{eq:linear-DPAM}) with an intercept introduced within each block in addition to $\beta_0$.
For backfitting, the subproblem of (\ref{eq:linear-DPAM-centered})
with respect to the block $\beta_{S_k}$ while fixing the rest at the current estimates $(\hat\beta_{S_j})_{j,S_j}$ is
\begin{equation*}
   \min_{\beta_{S_k}} \,  \frac{1}{2n}\|r-\tilde\Psi^\dag_{S_k}\beta_{S_k}\|_2^2 + \|\Gamma_{S_k}\beta_{S_k}\|_1 + \lambda_k\|\tilde\Psi^\dag_{S_k}\beta_{S_k}\|_n,
\end{equation*}
where $r=Y-\sum_{S_j\neq S_k}\tilde\Psi^\dag_{S_j}\hat\beta_{S_j}$. This is in the form of the single-block problem studied in Section~\ref{sec:single-block},
with $X = \tilde\Psi^\dag_{S_k}$, $\beta = \beta_{S_k}$, and $\Gamma= \Gamma_{S_k}$.

For the prediction given new data matrix $X^{\text{new}}$, we construct the basis matrices $(\Psi_{S_k}^{\text{new}})_{k,S_k}$
accordingly (for example, using marginal knots fixed at univariate quantiles from the training data). Then the predicted response vector is
$Y^{\text{pred}}=\bar{Y}+\sum_{k,S_k}( \Psi_{S_k}^{\text{new}} -\bar{\Psi}^\dag_{S_k})\hat{\beta}_{S_k}$.
Note that $\bar{Y}$ and $\bar{\Psi}^\dag_{S_k}$ are the means of the training data.

For the logistic regression problem (\ref{eq:logistic-DPAM}), we keep
the original response vector and the overall intercept $\beta_0$ and solve the modified problem:
\begin{equation*}
    \min_{\beta_0,(\beta_{S_k})_{k,S_k}} \, \frac{1}{n}\sum_{i=1}^n l(Y_i, f_i)+ \sum_{k,S_k} \Bigl(\|\Gamma_{S_k}\beta_{S_k}\|_1+\lambda_k\|\tilde\Psi^\dag_{S_k}\beta_{S_k}\|_n  \Bigl),
\end{equation*}
where $f=(f_1,\ldots,f_n)^\T=\beta_0+\sum_{k,S_k}\tilde\Psi^\dag_{S_k}\beta_{S_k}$ and $\tilde{\Psi}^\dag_{S_k}$ are empirically centered basis matrices.
For training, we replace the logistic loss by its second order Taylor expansion at the current estimates $\hat\beta_0$ and $(\hat\beta_{S_k})_{k,S_k}$, and further replace the Hessian matrix $H=\hat{p}(1-\hat{p})$ by a constant upper bound $1/4$ to obtain a majorization of the logistic loss.
For backfitting, the objective in the sub-problem with respect to $\beta_0$ and $\beta_{S_k}$ reduces to
\begin{equation*}
    \min_{\beta_0,\beta_{S_k}}\,\frac{1}{2n}\|r-\beta_0-\tilde\Psi^\dag_{S_k}\beta_{S_k}\|_2^2 + 4\|\Gamma_{S_k}\beta_{S_k}\|_1 + 4\lambda_k\|\tilde\Psi^\dag_{S_k}\beta_{S_k}\|_n,
\end{equation*}
where $r=\hat\beta_0+\tilde\Psi^\dag_{S_k}\hat\beta_{S_k}+4(Y-\hat{p})$, $\hat{p}=1/(1+\exp(-\hat{f}))$ and $\hat{f}=\hat{\beta}_0+\sum_{j,S_j}{\Psi}^\dag_{S_j}\hat\beta_{S_j}$. Then the intercept can be directly updated as the mean of $r$, denoted as $\bar{r}$. The sub-problem with respect to only $\beta_{S_k}$ becomes
$$
\min_{\beta_{S_k}}\,\frac{1}{2n}\|\tilde{r}-\tilde\Psi^\dag_{S_k}\beta_{S_k}\|_2^2 + 4\|\Gamma_{S_k}\beta_{S_k}\|_1 + 4\lambda_k\|\tilde\Psi^\dag_{S_k}\beta_{S_k}\|_n,
$$
where $\tilde{r}=r-\bar{r}$. This is the standard form of single-block optimization in Section~\ref{sec:single-block}.

For the prediction or classification given new data matrices $X^{\text{new}}$, we construct the basis matrices $(\Psi_{S_k}^{\text{new}})_{k,S_k}$ similarly as in the linear modeling. Then the predicted $f$ is $f^{\text{pred}}=\hat{\beta}_0+\sum_{k,S_k}(\Psi_{S_k}^{\text{new}}-\bar{\Psi}^\dag_{S_k})\hat\beta_{S_k}$, and the predicted probability is $p^{\text{pred}}=1/(1+\exp(-f^{\text{pred}}))$.

\subsection{Comparison with AS-BDT}\label{subsec:comparison with AS-BDT and hybrid}

In \cite{yang2018backfitting, yang2021hierarchical}, the single-block problem is
solved by first solving the Lasso problem (\ref{eq:single-block lasso}) using an active-set descent algorithm \citep{osborne2000new}
and then jointly soft-thresholding the Lasso solution. The resulting backfitting algorithm is called AS-BDT (active-set block descent and thresholding).
As the active-set algorithm is efficient for solving sparse Lasso problems,
AS-BDT tends to perform well if the average size of nonzero coefficients per block (corresponding to active basis functions) is small.
Moreover, the active-set information, including the signs of nonzero coefficients and the associated Cholesky decompositions, can be passed from the previous cycle
to speed up the active-set algorithm within each block and hence achieve computational savings for AS-BDT.
However, AS-BDT may not be suitable for handling large datasets, especially with non-sparse blocks.

Compared with AS-BDT,
the proposed methods (particularly stochastic primal-dual algorithms) are more adaptive to large-sample scenarios. They are expected to achieve near optimal
objective values at relatively lower costs. On the other hand, the primal-dual algorithms are not designed to take advantage of structural information from
the previous cycle of backfitting to achieve acceleration as in AS-BDT.

To exploit gains from different methods, a hybrid algorithm can be considered by combining stochastic primal-dual and the active-set algorithms sequentially.
At the beginning of backfitting, stochastic primal-dual algorithms can be used to obtain a decent decrease of the training loss.
Then the active-set method can be used to fine-tune the solutions if necessary.
Alternatively, acceleration can also be achieved by cycling over a subset of nonzero blocks and adjusting the subset iteratively,
as discussed in \cite{radchenko2010variable}.
Our numerical experiments are focused on the direct implementation of AS-BDT and the proposed methods.
We leave investigation of hybrid methods to future work.

\section{Numerical experiments}\label{sec:experiments}

We conduct numerical experiments to evaluate various algorithms for training DPAMs. We first consider the single-block optimization and then move on to the multi-block linear and logistic regressions, with both simulated data and real data.

\subsection{Single-block experiments}
Consider the following regression function, motivated from \cite{lin2006component}, Section 7, but with more complex nonlinear interactions:
\begin{equation}\label{eq:f in single-block simulation}
    \begin{split}
        f(x) = \left(\sum_{i=1}^7 \tilde g_i(x_i)\right) + \tilde g_1(x_3x_4) + \tilde g_2\left(\frac{x_1+x_3}{2}\right) + \tilde g_3(x_1x_2) \\
        + \tilde g_4(x_4x_5) + \tilde g_5\left(\frac{x_4+x_6}{2}\right) + \tilde g_6\left(\frac{x_5+x_2}{2}\right) + \tilde g_7(x_6x_7),
    \end{split}
\end{equation}
where $\{g_i,\,i=1,\dots,7\}$ are functions defined on the interval $[0,1]$ as \vspace{-.1in}
\begin{eqnarray*}
 && g_1(x) = x , \quad g_2(x) = (2x-1)^2  , \quad g_3(x) = \frac{1}{1+x}  ,\\
 && g_4(x) = 0.1\sin(2\pi x) + 0.2 \cos(2\pi x) + 0.3 \sin^2(2\pi x) + 0.4 \cos^3(2\pi x) + 0.5 \sin^3(2\pi x),\\
 && g_5(x) = \frac{\sin(2\pi x)}{2-\sin(2\pi x)}, \quad g_6(x) = \frac{\sin(4\pi x)}{2+\sin(2\pi x)}, \quad g_7(x) = \frac{\cos(4\pi x)}{2+\cos(2\pi x)},
\end{eqnarray*}
and $\{\tilde g_i,\,i=1,\dots,7\}$ are centered versions $\{\tilde g_i(x) = g_i(x) - \int_{0}^1 g_i(x) \mathrm{d}x\}$. The inputs $X_i=(X_{i,1},\ldots,X_{i,7})$ for $i=1,\ldots,n$ are i.i.d. uniformly distributed on $[0,1]^7$. A normal noise with standard deviation $0.5138$ is added to give a signal-to-noise ratio of 3:1. That is, the response variables $\{Y_1,\ldots,Y_n\}$ are generated by $Y_i=f(X_i)+\epsilon_i$ where $\epsilon_i\sim N(0,0.5138^2)$.

We consider the linear DPAM (\ref{eq:linear-DPAM}) for $Y$ on $X$ with differentiation order $m=2$ (i.e., piecewise cross-linear basis functions), the marginalization $H$ being the average operator on the training set (which are used in all our experiments), and $11$ marginal knots defined as the quantiles by 10\% in the training set for each covariate (resulting in $10$ univariate basis functions for each covariate). In particular, we consider the subproblem of fitting
$Y$ on the basis matrix $\Psi^\dag_{4,5}$, associated with two-way interactions from $x_4$ and $x_5$:
\begin{equation}\label{eq:single-block simulation}
    \min_{\beta_{4,5}}\,\frac{1}{2n}\|\tilde
    Y-\tilde\Psi^\dag_{4,5}\beta_{4,5}\|_2^2 + \|\Gamma_{4,5}\beta_{4,5}\|_1+\lambda\|\tilde\Psi^\dag_{4,5}\beta_{4,5}\|_n,
\end{equation}
where $\tilde Y$ and $\tilde \Psi^\dag_{4,5}$ are empirically centered versions of $Y$ and $\Psi^\dag_{4,5}$ respectively. Based on \cite{yang2021hierarchical}, $\Gamma_{4,5}=\rho\cdot\text{diag}(0,1,\ldots,1)$ is defined to represent the HTV/Lasso penalty with differentiation order $m=2$. The basis matrix $\tilde\Psi^\dag_{4,5}$ includes $10^2=100$ basis functions and $\beta_{4,5}\in\bbR^{100}$. Focusing on the large-scale problem, we take $n=50000$.

We design experiments with different $\rho$ and $\lambda$ to evaluate our algorithms, using Proposition~\ref{pro:soft-threshold}.
For any $\rho$, we first compute $\tilde{\beta}_{4,5}(\rho)$ as a solution to the Lasso problem, with the empirical-norm penalty
removed from (\ref{eq:single-block simulation}). Let $\lambda_{0}(\rho)=\|\tilde\Psi^\dag_{4,5}\tilde{\beta}_{4,5}\|_n$.
Then for any $\lambda\geq\lambda_0(\rho)$, a solution $\hat\beta_{4,5}$ to problem (\ref{eq:single-block simulation})
is $0$. For $\lambda < \lambda_0(\rho)$, a solution $\hat\beta_{4,5}$  can be obtained by jointly shrinking $\tilde{\beta}_{4,5}(\rho)$.
We let $\rho$ vary in $\{2^{-15},2^{-18},2^{-21}\}$, and for each $\rho$, we take $\lambda$ to be either $\lambda_0(\rho)/4$ or $2\lambda_0(\rho)$, corresponding to a nonzero or completely zero solution $\hat\beta_{4,5}$.  As $\rho$ decreases in $\{2^{-15},2^{-18},2^{-21}\}$, the number of nonzero scalar coefficients in the Lasso solution $\tilde{\beta}_{4,5}$ increases from $20$ to $37$ and further to $62$ out of $100$.

In the single-block experiments, we compare the following algorithms. The AS-BDT algorithm in \cite{yang2018backfitting} is used to compute exact solutions.
\begin{itemize}
    \item \textbf{Batch algorithms}
    \begin{itemize}
        \item \textbf{CP:} The Chambolle--Pock algorithm (Algorithm~\ref{alg:batch-CP}).

        \item \textbf{AMA:} The linearized AMA (Algorithm~\ref{alg:batch-linearized-AMA}).
        \item \textbf{CC:} The Concave conjugate algorithm with perturbation (Algorithm~\ref{alg:batch-CC} in the Supplement). Choose $\delta=10^{-6}$.
    \end{itemize}
    \item \textbf{Stochastic algorithms}
    \begin{itemize}
        \item \textbf{Stoc-CP:} Algorithm~\ref{alg:stoc-CP}. Each step operates on a single row of $\tilde\Psi^\dag_{4,5}$. Therefore we count $n$ consecutive steps as one batch step. Each step needs to solve a non-linear univariate equation on a fixed interval $(0,1)$.
        \item \textbf{Stoc-AMA-SAG and Stoc-AMA-SAGA:} Algorithm~\ref{alg:stoc-linear-AMA}. Similarly as in Stoc-CP, we count $n$ steps as one batch step.
        \item \textbf{Stoc-CC:} Algorithm~\ref{alg:stoc-CC} in the Supplement. One outer iteration with cost $O(nd)$ is counted as one batch step.
    \end{itemize}
\end{itemize}

Step sizes for batch and stochastic algorithms are tuned manually, as would be done in practice.
See Supplement Section~\ref{subsec:tuning single-block} for detailed information.
The algorithms are evaluated by their performances in reducing the optimality gap where the optimal objective value is computed from AS-BDT.
The results are shown in Figure~\ref{fig: single-block}.
All algorithms start from an initial $\beta^0=0$, and primal-dual algorithms additionally set the initial dual variable as $u^0=\tilde\Psi^\dag_{4,5}\beta^0-\tilde Y=-\tilde Y$. For the stochastic algorithms, we plot the mean performance as well as the minimum and maximum optimality gaps across 10 repeated runs with different random seeds.
We observe several trends across the experiments:

\begin{figure}[h]
    \centering
    \begin{subfigure}{0.32\textwidth}
       \includegraphics[width=\textwidth]{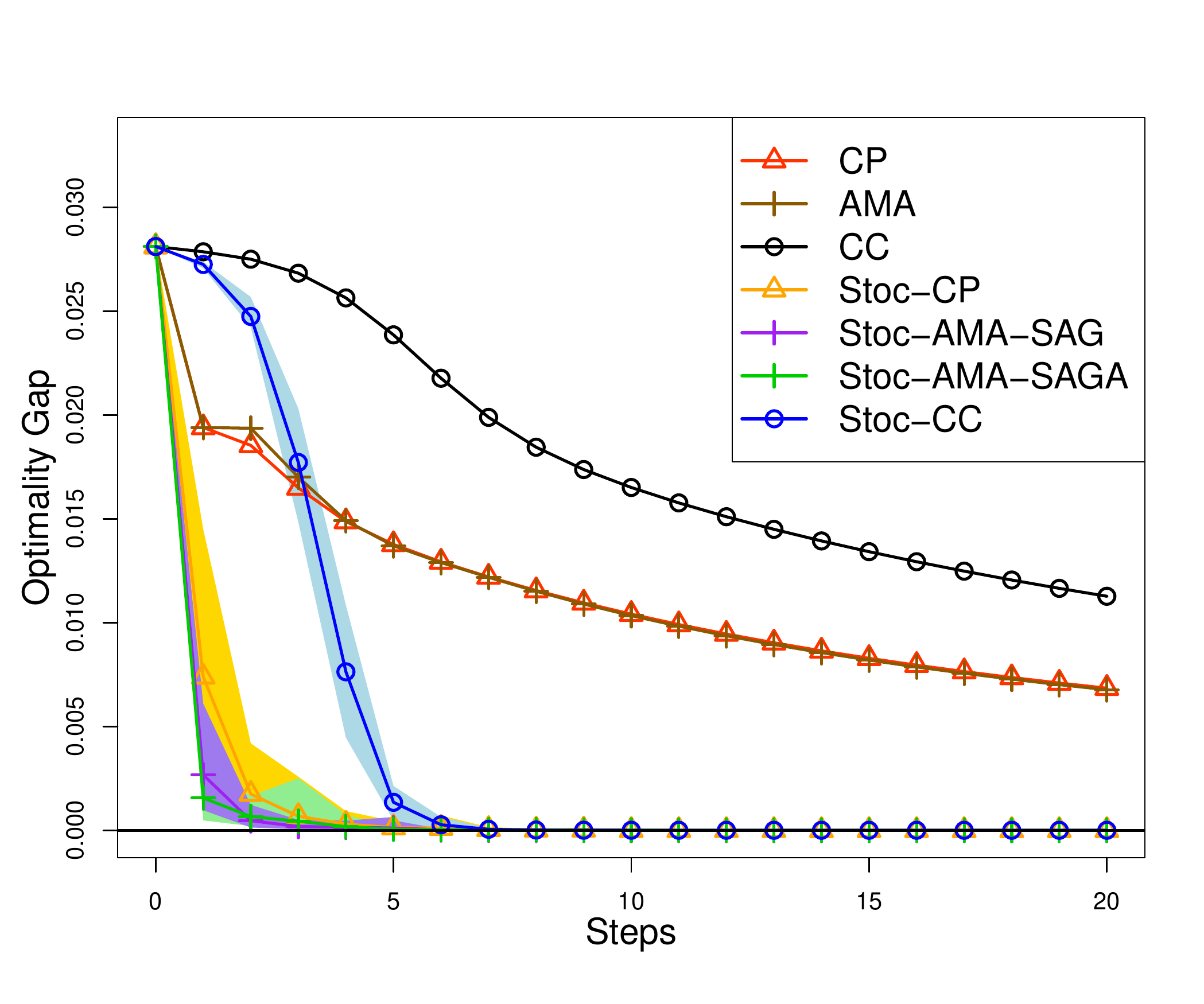}
       \vspace{-0.45in}
       \caption{$\rho=2^{-15}$, $\lambda=\lambda_0/4$}
    \end{subfigure}
    \hfill
    \begin{subfigure}{0.32\textwidth}
       \includegraphics[width=\textwidth]{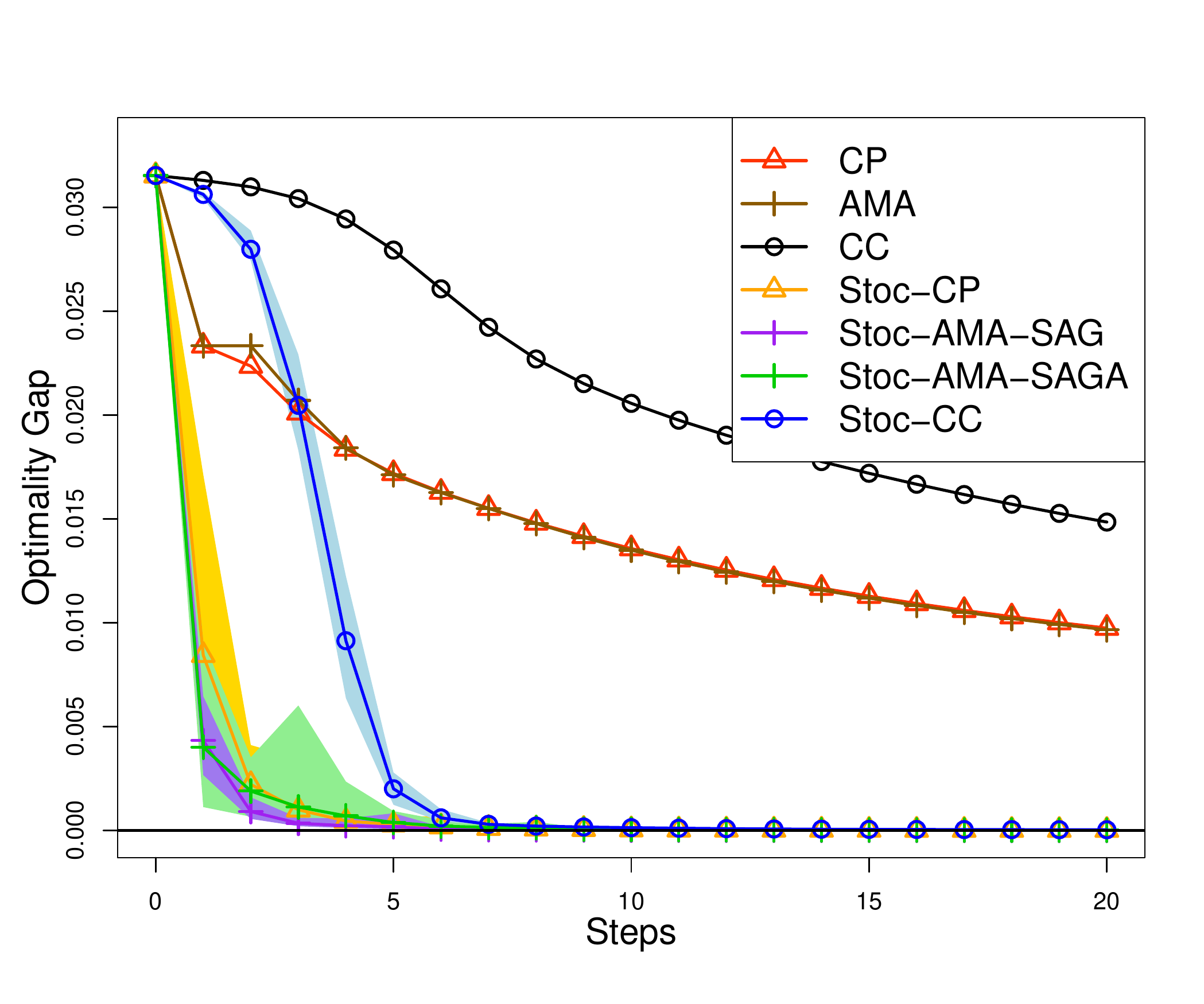}
       \vspace{-0.45in}
       \caption{$\rho=2^{-18}$, $\lambda=\lambda_0/4$}
    \end{subfigure}
    \hfill
    \begin{subfigure}{0.32\textwidth}
       \includegraphics[width=\textwidth]{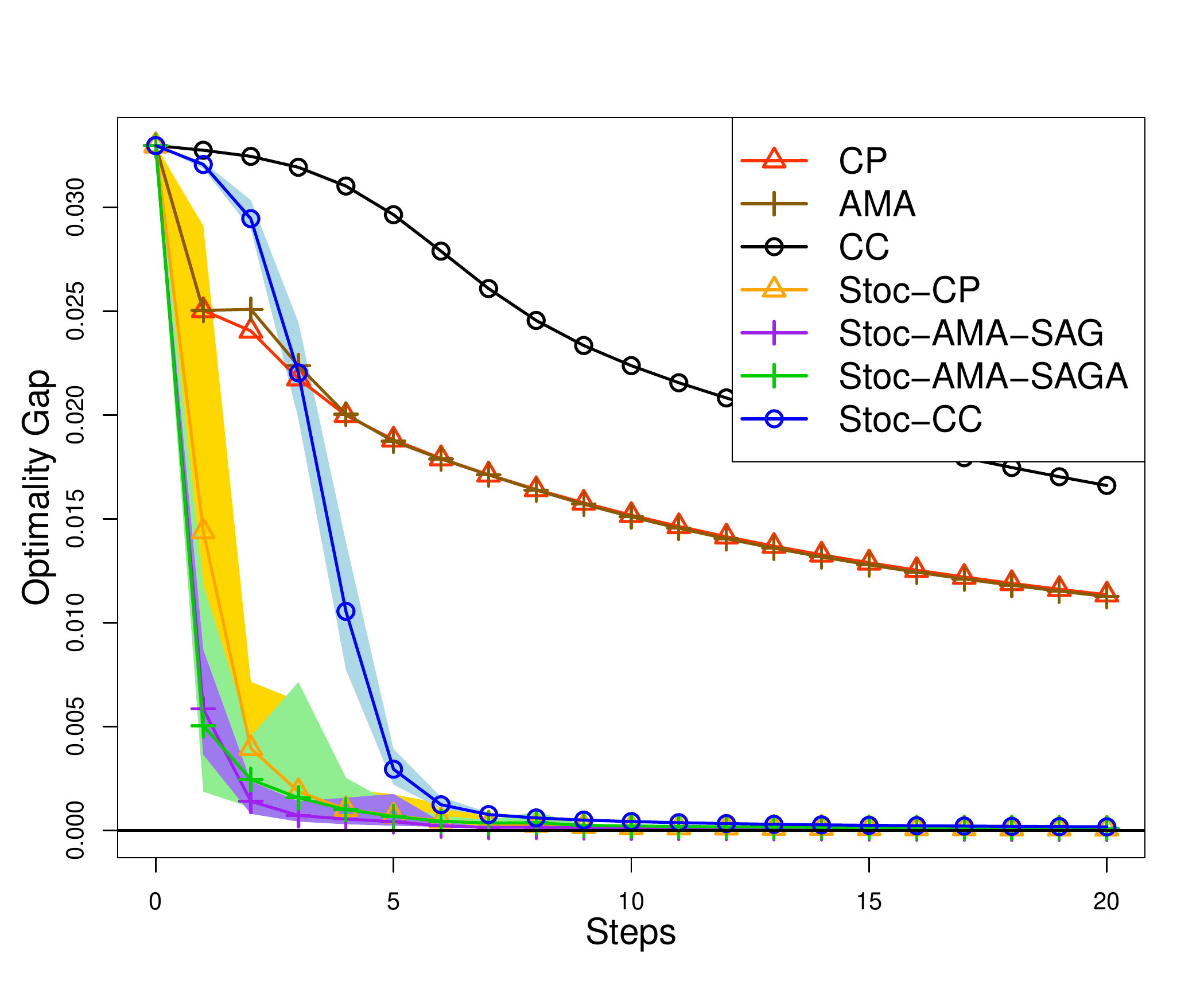}
       \vspace{-0.45in}
       \caption{$\rho=2^{-21}$, $\lambda=\lambda_0/4$}
    \end{subfigure}

    \begin{subfigure}{0.32\textwidth}
       \includegraphics[width=\textwidth]{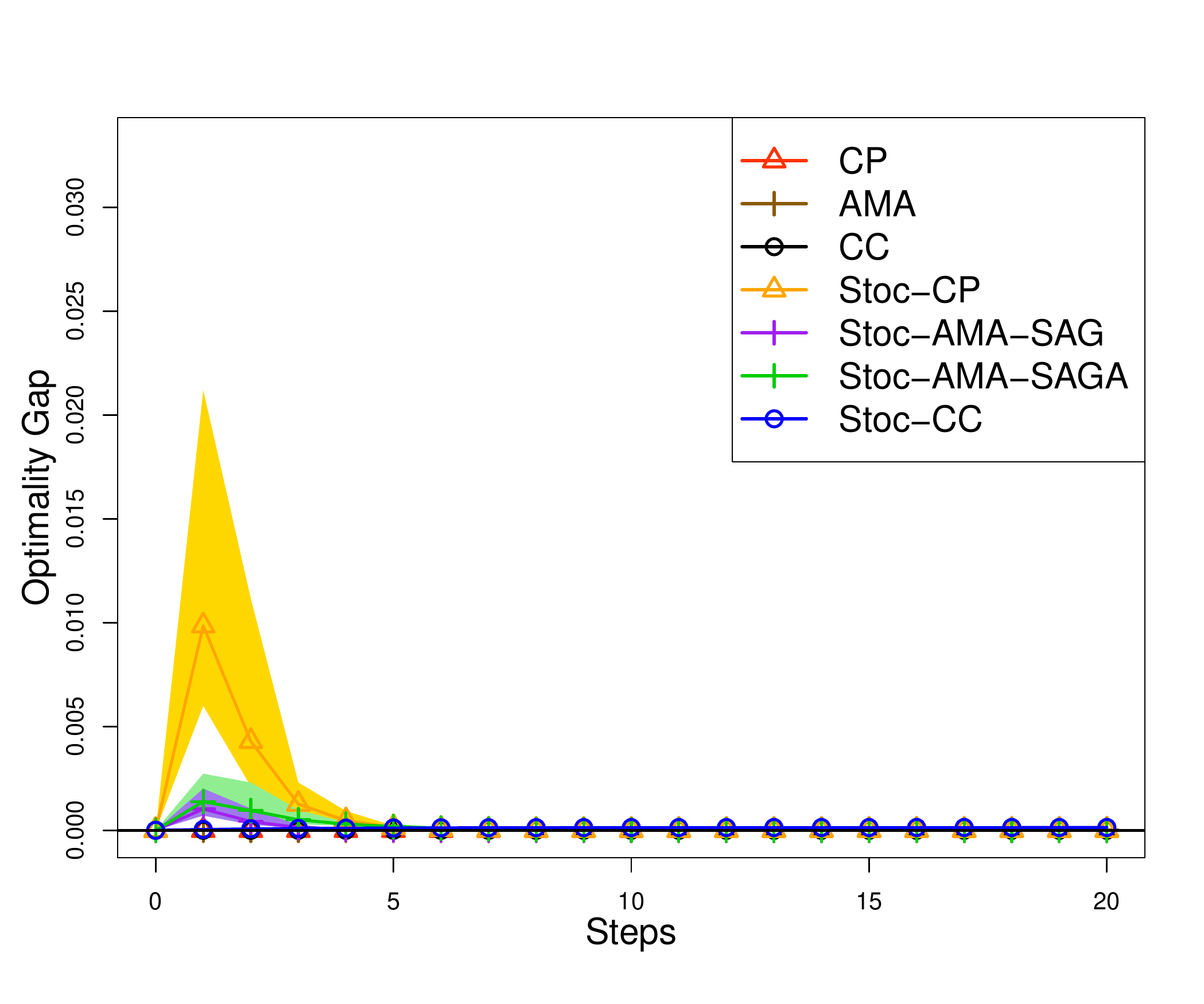}
       \vspace{-0.45in}
       \caption{$\rho=2^{-15}$, $\lambda=2\lambda_0$}
    \end{subfigure}
    \hfill
    \begin{subfigure}{0.32\textwidth}
       \includegraphics[width=\textwidth]{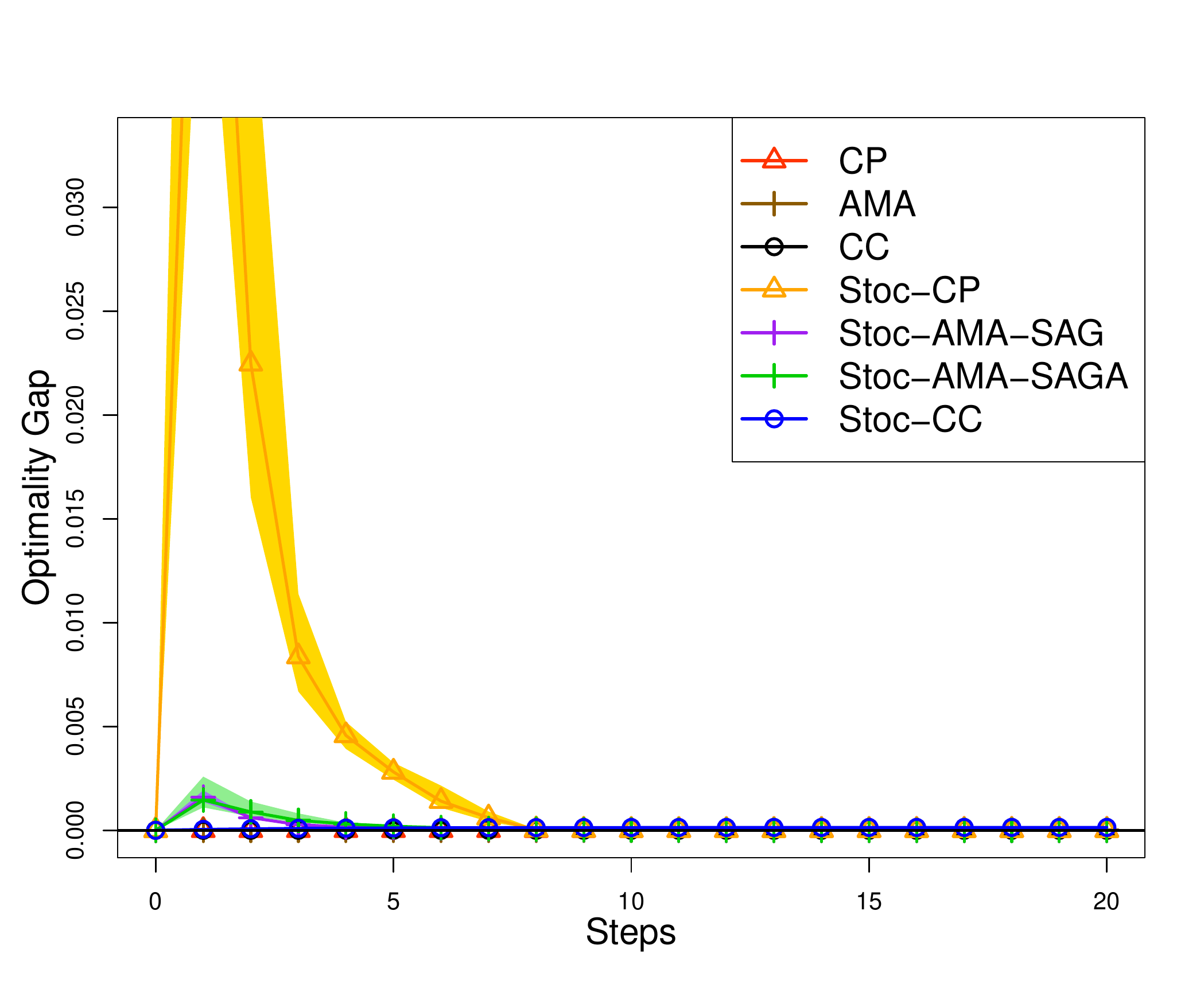}
       \vspace{-0.45in}
       \caption{$\rho=2^{-18}$, $\lambda=2\lambda_0$}
    \end{subfigure}
    \hfill
    \begin{subfigure}{0.32\textwidth}
       \includegraphics[width=\textwidth]{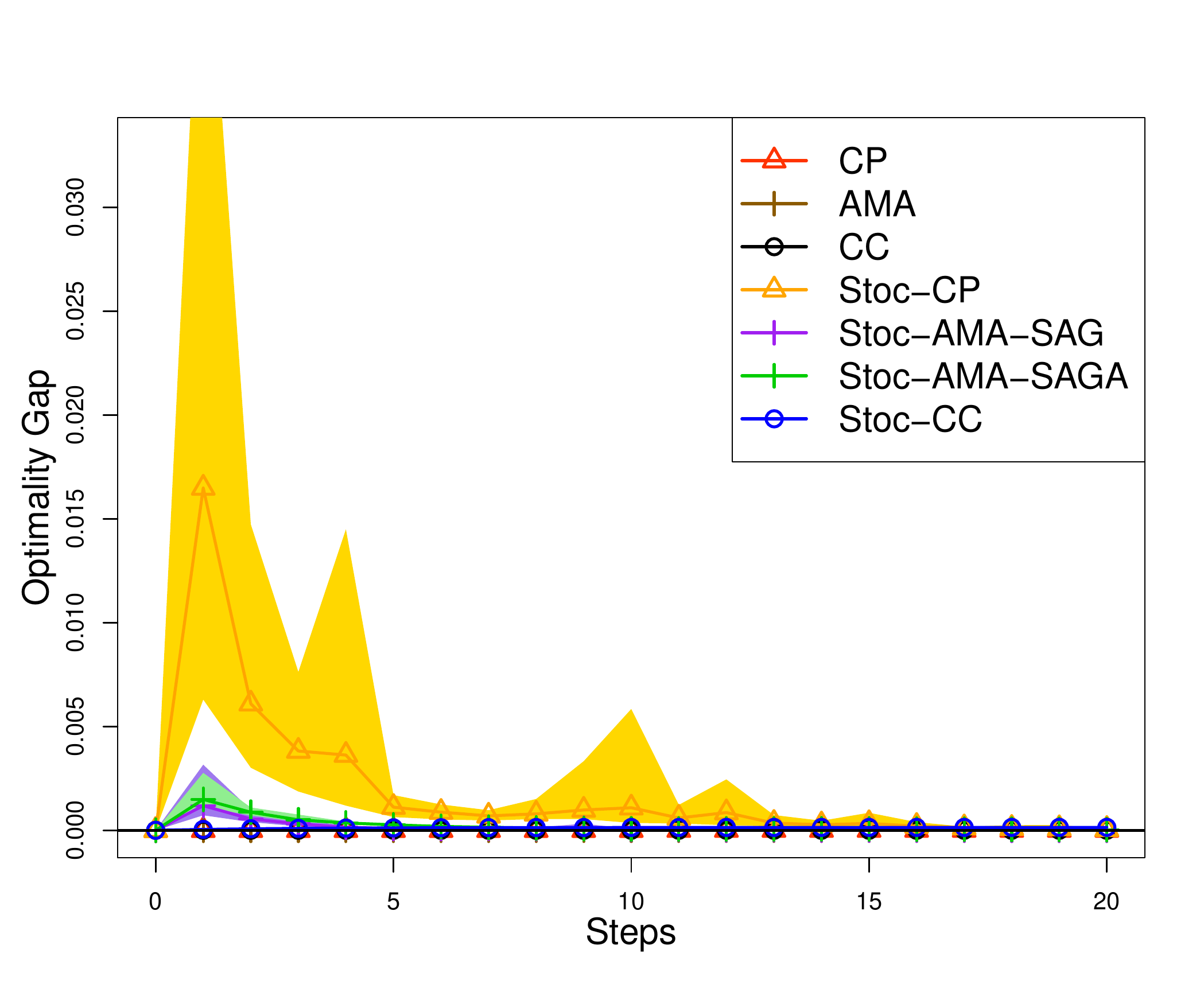}
       \vspace{-0.45in}
       \caption{$\rho=2^{-21}$, $\lambda=2\lambda_0$}
    \end{subfigure}

    \caption{Single-block experiments with varying $\rho$ and $\lambda$. The $0$ gap is indicated by the black solid line.}
    \label{fig: single-block}
\end{figure}

\begin{itemize}
    \item \textbf{Effect of $\lambda$:} 
    When $\lambda=\lambda_0/4$, the initial value $\beta^0 =0 $ is sub-optimal, and all algorithms exhibit a decreasing pattern. When  $\lambda=2\lambda_0$,
    the initial value $\beta^0=0$ is optimal. In this case, all stochastic primal-dual algorithms follow a non-monotone pattern, whereas the batch primal-dual algorithms stay close to $0$ because after tuning, nearly optimal step sizes ($\tau$ close to $0$ and large $\alpha$) are applied.

    \item \textbf{Effect of $\rho$:} With $\lambda=\lambda_0/4$ fixed, as $\rho$ decreases from $2^{-15}$ to $2^{-18}$ and further to $2^{-21}$, while the exact solution becomes denser, both the batch and stochastic algorithms remain relatively stable in their performances.

    \item \textbf{Batch vs Stochastic:} All stochastic algorithms are substantially faster than their deterministic counterparts in the case of $\lambda=\lambda_0/4$.

    \item \textbf{CP vs AMA:} CP and AMA, batch and stochastic versions, lead to similar performances when $\lambda=\lambda_0/4$, although they are derived in technically different manners.

    \item \textbf{Primal-dual vs CC:} When $\lambda=\lambda_0/4$, the primal-dual algorithms, batch and stochastic versions, perform better than CC.
    When $\lambda=2\lambda_0$ and the initial $\beta^0=0$ is optimal, CC benefits from the monotonicity, albeit in terms of the perturbed objective.
\end{itemize}

\subsection{Multi-block experiments}

We compare several algorithms for training multi-block DPAMs \citep{yang2018backfitting,yang2021hierarchical},
in two experiments for linear regression (\ref{eq:linear-DPAM}) on simulated data
and two experiments for logistic regression (\ref{eq:logistic-DPAM}) on both simulated and real data.
The experiment for linear regression using the phase shift model \citep{friedman1991multivariate} is presented in the Supplement,
where DPAM with up to three-way interactions (as well as two-way interactions) is trained.
Throughout, we use constant penalty parameters $\rho$ and $\lambda$ for the HTV/Lasso and empirical-norm penalties.

We evaluate the algorithms in terms of their performances in decreasing the training loss after the same number of epochs over data and basis blocks.
Because the size $d_{S_k}$ of a block $S_k$ varies from block to block, we count one {\it cycle} over all blocks, with one {\it scan} of the full dataset in each block, as one single {\it epoch}.
Specifically, if an algorithm scans the full dataset for $T$ times within the block $S_k$, the corresponding number of epochs is counted as $T d_{S_k}/\sum_{S_l}d_{S_l}$.
All algorithms start from an initial value $0$ within each block.
\begin{itemize}
    \item \textbf{Batch algorithms}
    \begin{itemize}
        \item \textbf{AS-BDT:} In each block, one scan is counted when the active set is adjusted (enlarged or reduced) or when the final active set is identified. The number of scans may vary from block to block, depending on the sparsity of the solution.

        \item \textbf{CC, CP and AMA:} Each batch step is counted as one scan. For simplicity, a fixed number, $6$, of batch steps,
        are performed across all blocks in each backfitting cycle.
        For all primal-dual algorithms (including their stochastic versions), the coefficients are reset to 0 using conditions (\ref{eq:condition for CP}) and (\ref{eq:condition for AMA}) within each block, to promote sparsity and improve performances.
    \end{itemize}
    \item \textbf{Stochastic algorithms}
    \begin{itemize}
        \item \textbf{Stoc-CC, Stoc-CP and Stoc-AMA-SAG:} One
        batch step which consists of $n$ consecutive steps is defined as one scan, as in the single-block experiment.
        The same number of batch steps is fixed across all blocks, either $3$ for linear regression or $5$ for logistic regression. For Stoc-CC (as well as CC),
        only one majorization ($B=1$) is performed within each block.
    \end{itemize}
\end{itemize}

Step sizes are tuned manually; see detailed information in Supplement Section~\ref{subsec:tuning multi-block}.

\subsubsection{Synthetic linear regression} \label{sec:simu-linear}

\begin{figure}[!t]
    \centering
    \begin{subfigure}{0.32\textwidth}
       \includegraphics[width=\textwidth]{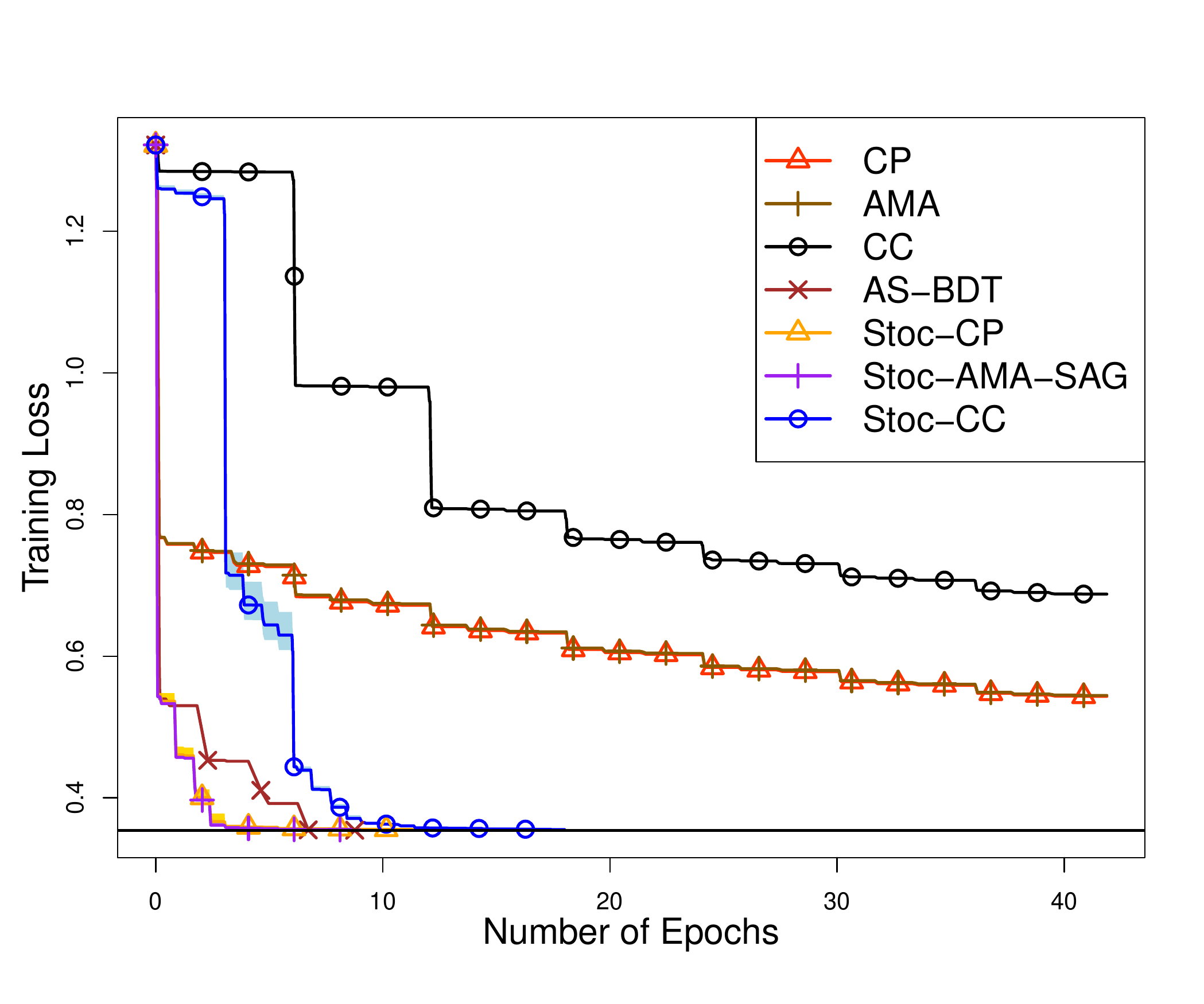}
       \vspace{-.45in}
       \caption{$\rho=2^{-16}$, $\lambda=\|\tilde Y\|_n/2^{6}$}
    \end{subfigure}
    \hfill
    \begin{subfigure}{0.32\textwidth}
       \includegraphics[width=\textwidth]{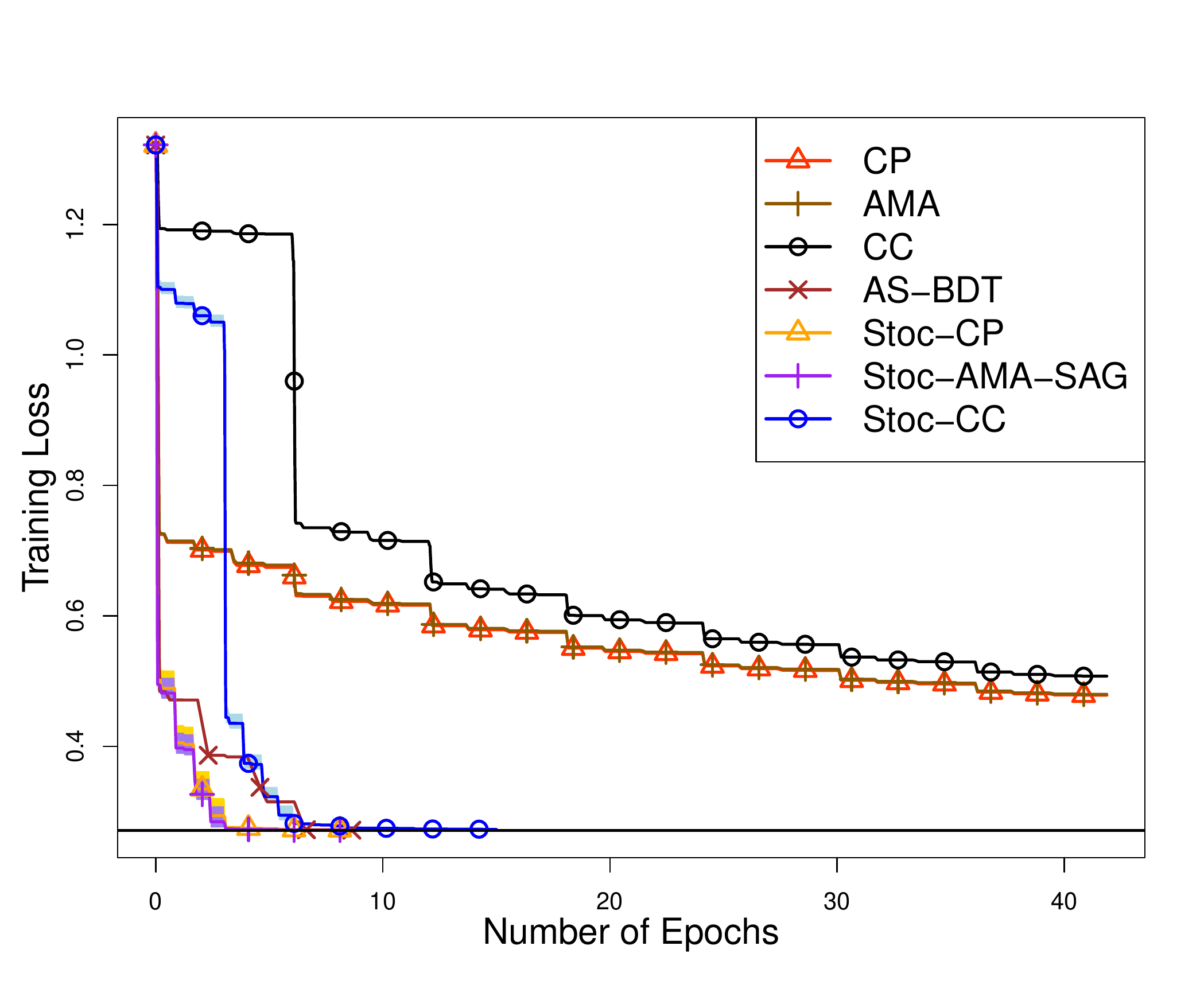}
       \vspace{-.45in}
       \caption{$\rho=2^{-16}$, $\lambda=\|\tilde Y\|_n/2^{8}$}
    \end{subfigure}
    \hfill
    \begin{subfigure}{0.32\textwidth}
       \includegraphics[width=\textwidth]{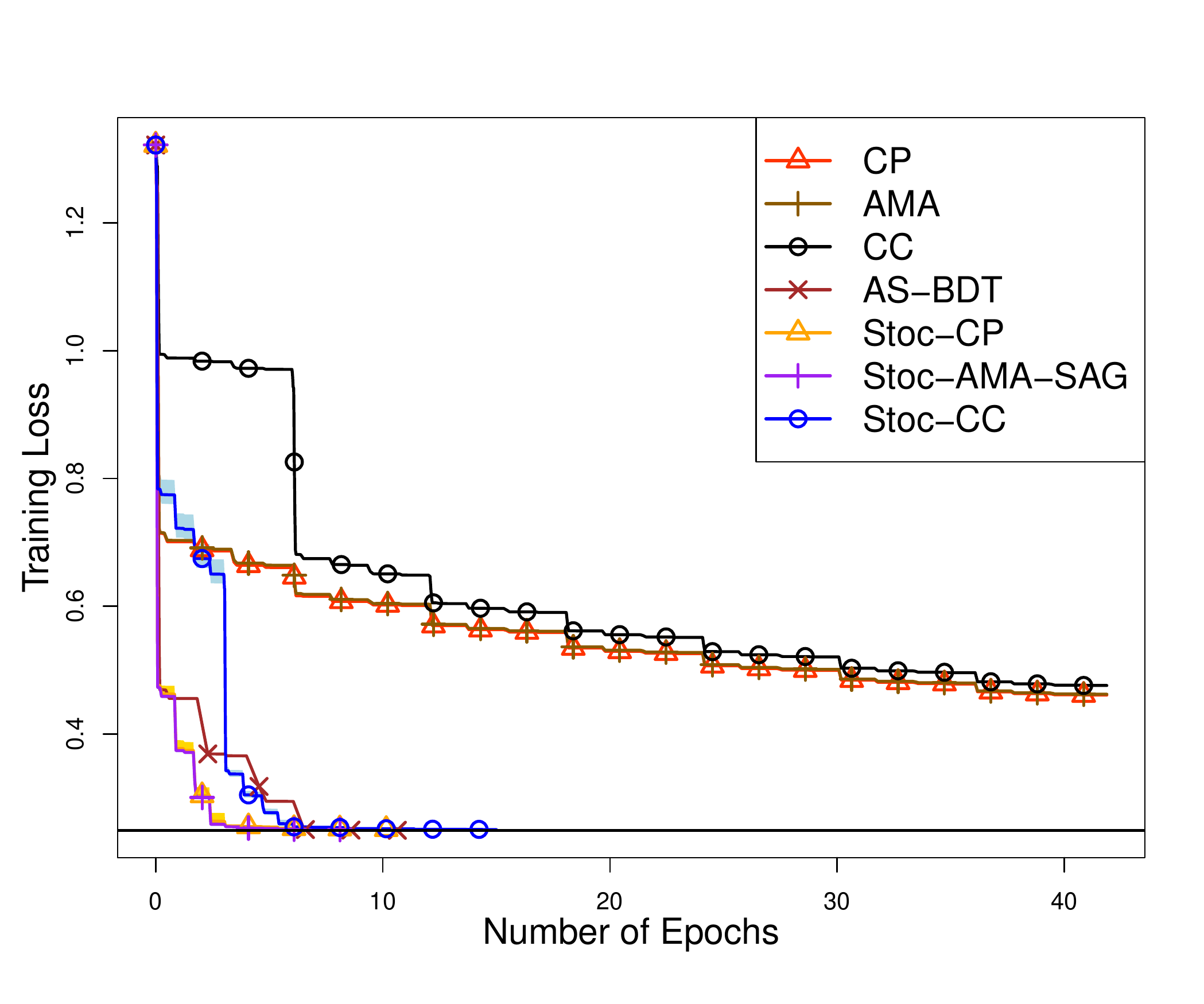}
       \vspace{-.45in}
       \caption{$\rho=2^{-16}$, $\lambda=\|\tilde Y\|_n/2^{10}$}
    \end{subfigure}

    \centering
    \begin{subfigure}{0.32\textwidth}
       \includegraphics[width=\textwidth]{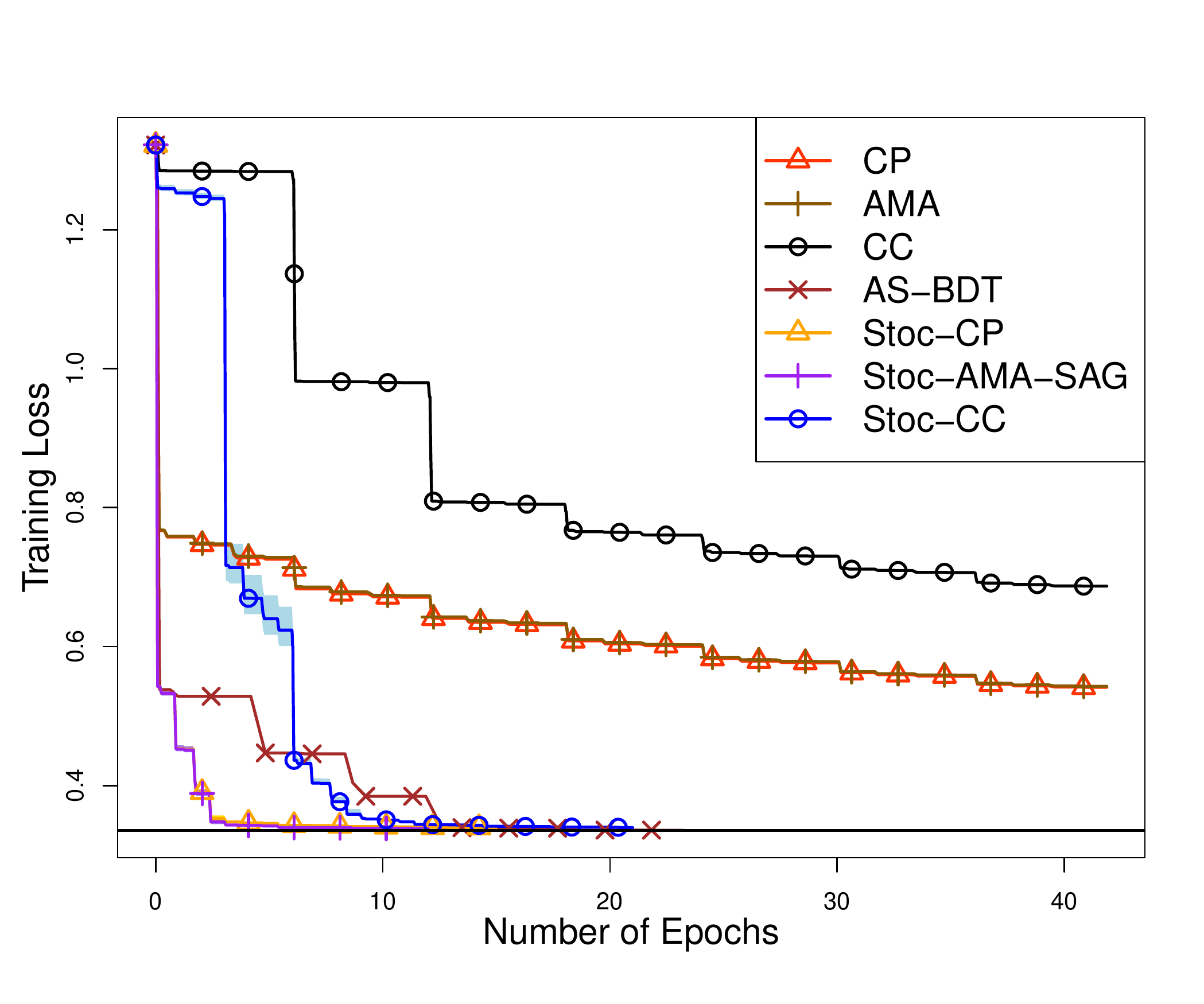}
       \vspace{-.45in}
       \caption{$\rho=2^{-19}$, $\lambda=\|\tilde Y\|_n/2^{6}$}
    \end{subfigure}
    \hfill
     \begin{subfigure}{0.32\textwidth}
       \includegraphics[width=\textwidth]{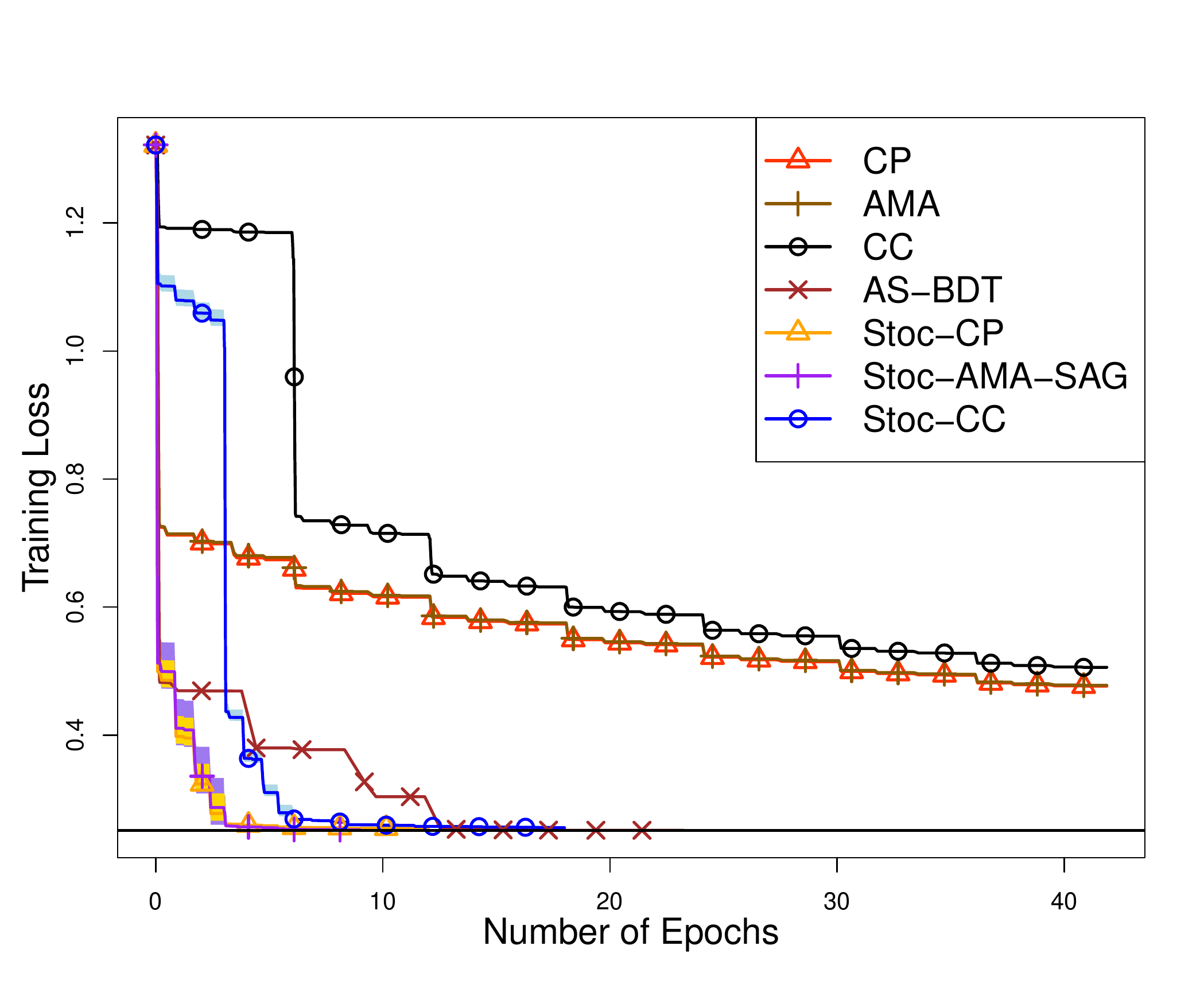}
       \vspace{-.45in}
       \caption{$\rho=2^{-19}$, $\lambda=\|\tilde Y\|_n/2^{8}$}
    \end{subfigure}
    \hfill
    \begin{subfigure}{0.32\textwidth}
       \includegraphics[width=\textwidth]{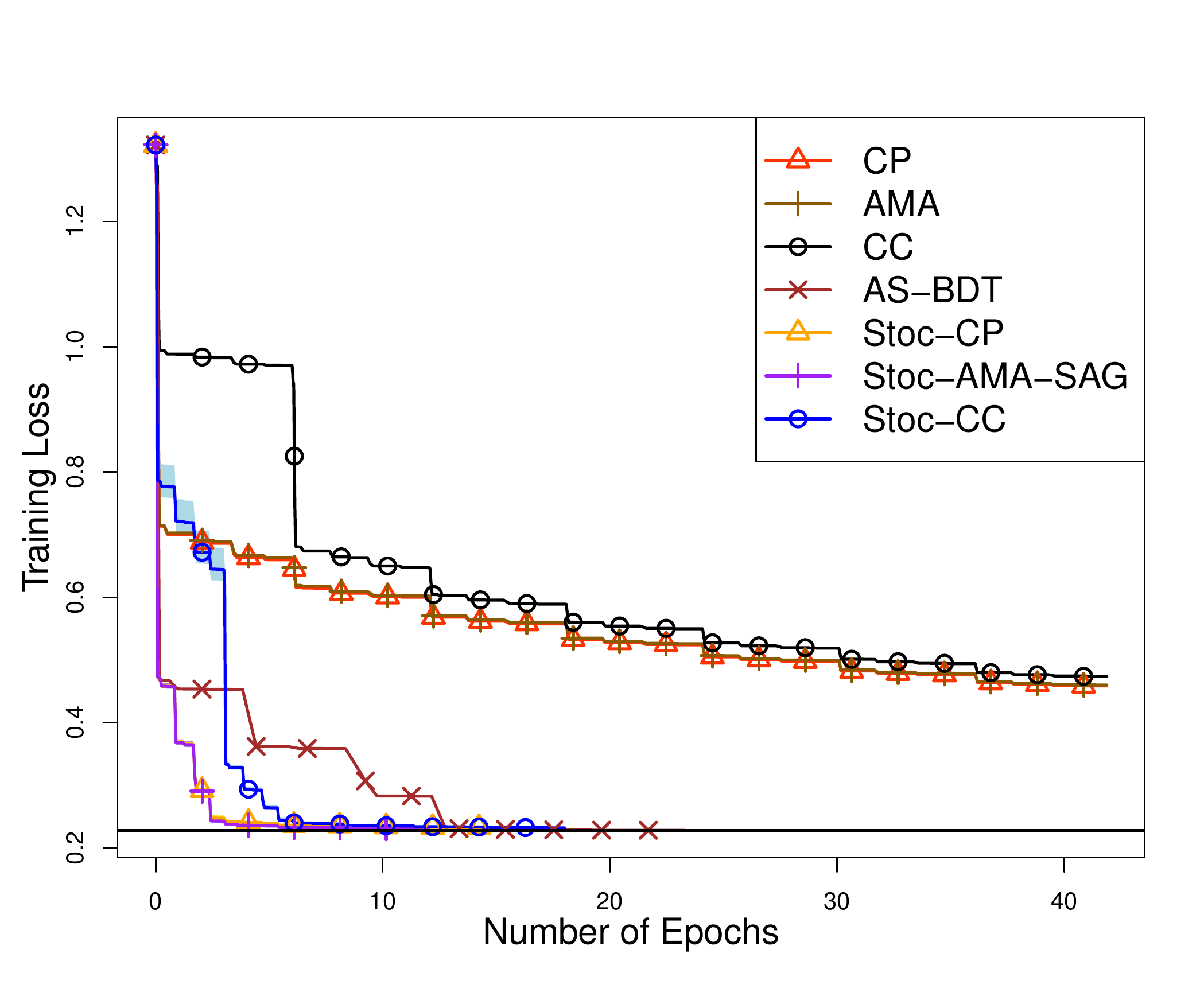}
       \vspace{-.45in}
       \caption{$\rho=2^{-19}$, $\lambda=\|\tilde Y\|_n/2^{10}$}
    \end{subfigure}

    \centering
    \begin{subfigure}{0.32\textwidth}
       \includegraphics[width=\textwidth]{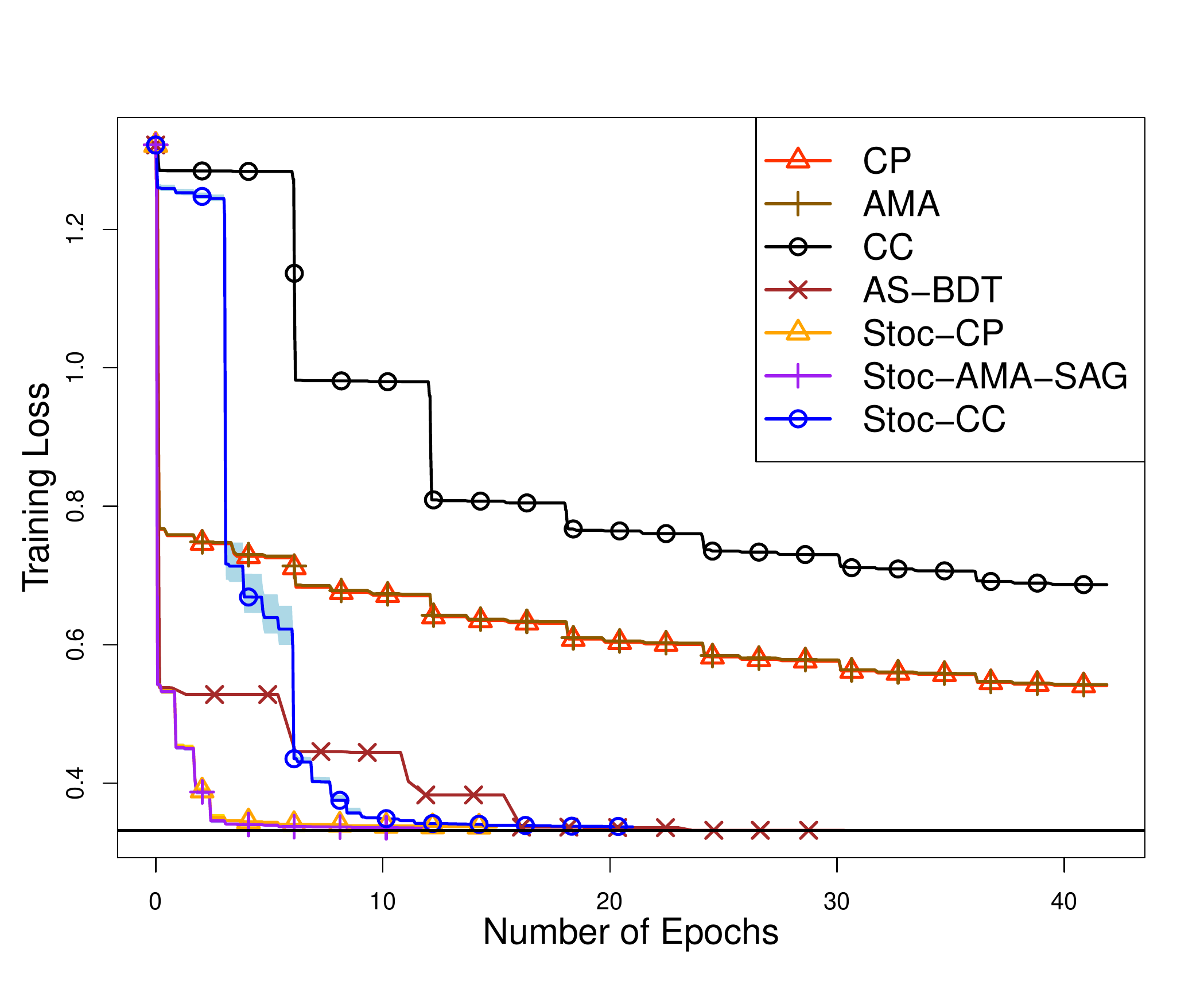}
       \vspace{-.45in}
       \caption{$\rho=2^{-22}$, $\lambda=\|\tilde Y\|_n/2^{6}$}
    \end{subfigure}
    \hfill
    \begin{subfigure}{0.32\textwidth}
       \includegraphics[width=\textwidth]{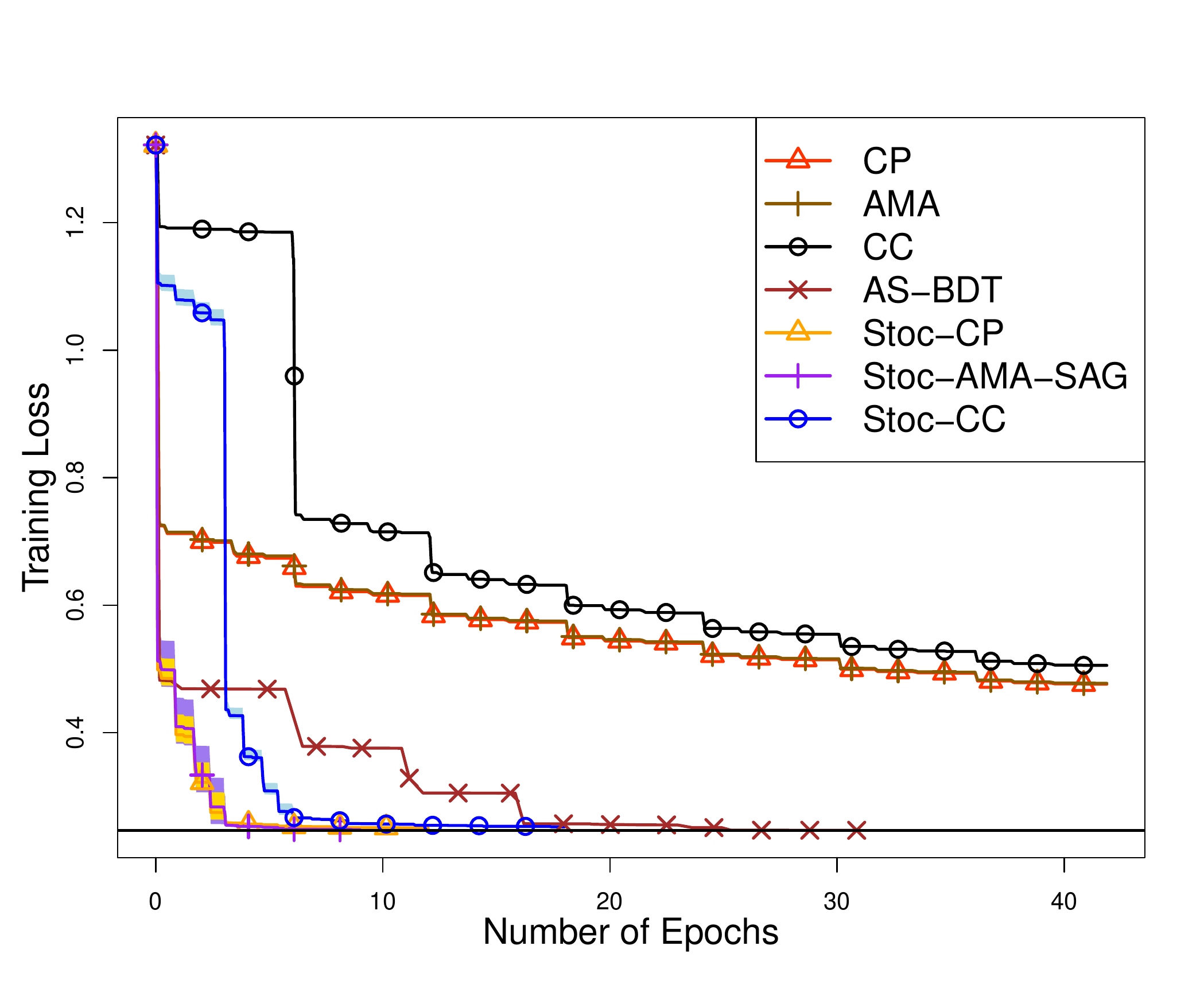}
       \vspace{-.45in}
       \caption{$\rho=2^{-22}$, $\lambda=\|\tilde Y\|_n/2^{8}$}
    \end{subfigure}
    \hfill
    \begin{subfigure}{0.32\textwidth}
       \includegraphics[width=\textwidth]{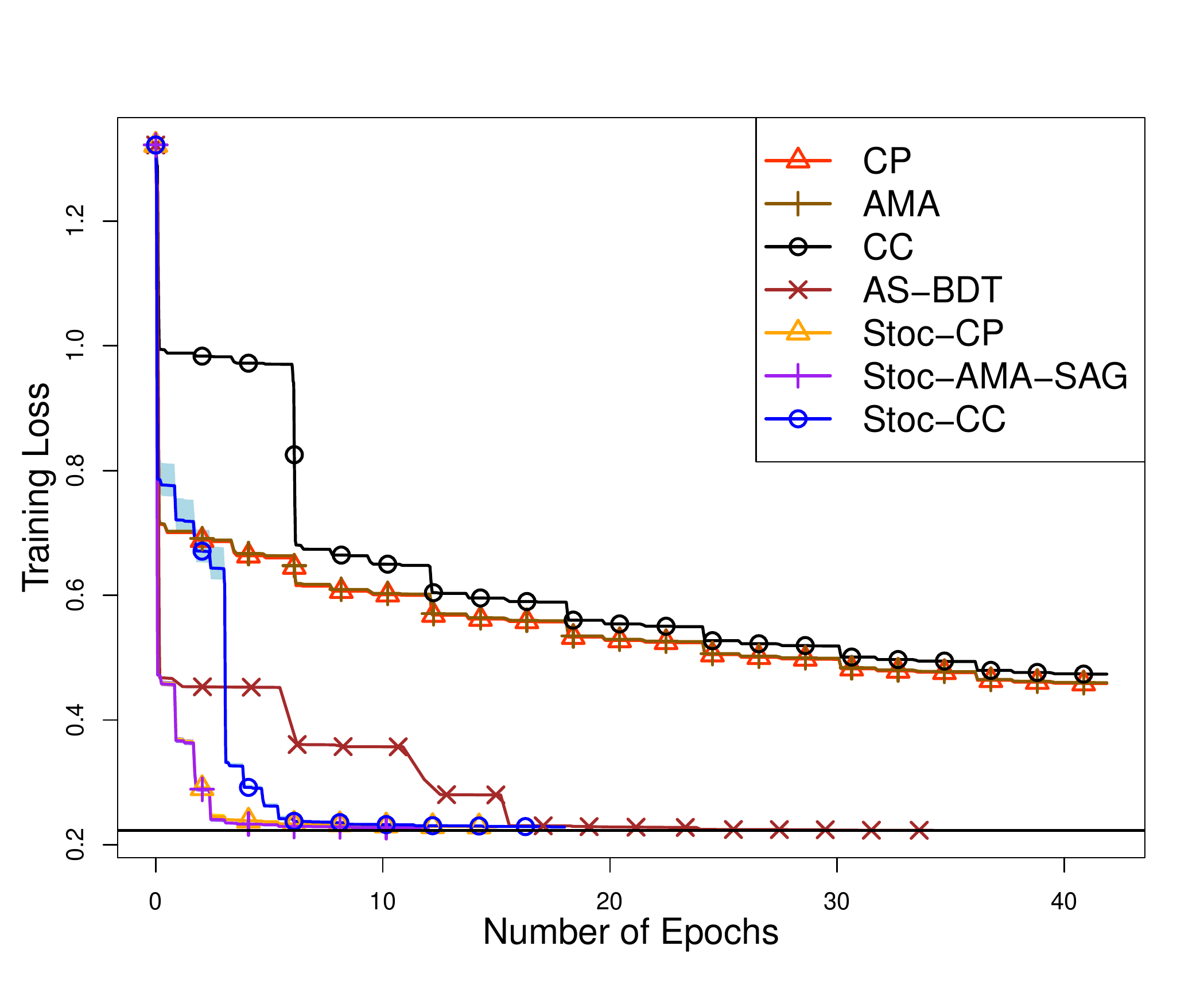}
       \vspace{-.45in}
       \caption{$\rho=2^{-22}$, $\lambda=\|\tilde Y\|_n/2^{10}$}
    \end{subfigure}
    \caption{DPAM training in synthetic linear regression. The minimal training loss achieved from AS-BDT is given by the black solid line. For the stochastic algorithms, we plot the mean as well as the minimum and maximum training losses across 10 repeated runs.}
    \label{fig: synthetic linear}
\end{figure}

For $n=50000$, we generate $X_i=(X_{i,1},X_{i,2},\ldots,X_{i,10})$ uniformly on $[0,1]^{10}$ and $Y_i=f(X_i)+\epsilon_i$, $i=1,\ldots,n$, where $\epsilon_i\sim N(0,0.5138^2)$ and $f$ is defined as (\ref{eq:f in single-block simulation}). Note that $X_8$, $X_9$ and $X_{10}$ are spurious variables. We apply the linear DPAM (\ref{eq:linear-DPAM}),
with differentiation order $m=2$ (i.e., piecewise cross-linear basis functions), interaction order $K=2$ (main effects and two-way interactions), and
$6$ marginal knots from data quantiles by 20\% ($5$ basis functions for each main effect).
The performances of various algorithms are reported in Figure~\ref{fig: synthetic linear}, under different choices of $\rho\in\{2^{-16},2^{-19},2^{-22}\}$ and $\lambda\in\{\|\tilde Y\|_n/2^6,\|\tilde Y\|_n/2^{8},\|\tilde Y\|_n/2^{10}\}$, where $\tilde Y$ is the centered version of $Y$. These penalty parameters $\rho$ and $\lambda$ are chosen, to demonstrate a range of sparsity levels and mean squared errors (MSEs) calculated on a validation set.
For solutions from AS-BDT (after convergence declared), the sparsity levels and MSEs under different tuning parameters are summarized in Table~\ref{tab:sparsity and MSE of linear}. The sparsity levels from stochastic primal-dual algorithms are reported in Supplement Table~\ref{tab:sparsity for synthetic linear}. A tolerance of $10^{-3}$ in the objective value is checked after each cycle over all blocks to declare convergence.

\begin{table}[!ht]
    \centering
    \caption{Sparsity and MSEs for synthetic linear regression.}
    \begin{tabular}{|l|*{2}{c}{c|}*{2}{c}{c|}*{3}{c}|}
     \hline
     $\log_2(\rho^{-1})$ & \multicolumn{3}{c|}{16} & \multicolumn{3}{c|}{19} & \multicolumn{3}{c|}{22} \\
     \hline $\log_2 (\lambda^{-1})$ & 6 & 8 & 10 & 6 & 8 & 10 & 6 & 8 & 10\\
     \hline
     \# nonzero blocks & 14 & 29 & 54 & 14 & 54 & 54 & 14 & 54 & 55 \\
     \# nonzero coefficients & 119 & 195 & 335 & 169 & 670 & 656 & 185 & 841 & 885 \\
     MSEs & 0.462 & 0.451 & 0.45 & 0.447 & 0.439 & 0.440 & 0.446 & 0.439 & 0.442 \\
     \hline
    \end{tabular}\\[.1in]
    \parbox{1\textwidth}{\small Note: For $K=2$, there are $55$ blocks from which $10$ are main effects and $45$ are two-way interactions. There are a total of $1175$ scalar coefficients. For the underlying regression function (\ref{eq:f in single-block simulation}), $14$ of the $55$ components are true. The MSEs are evaluated on a validation set of $n=50000$ data points.}
    \label{tab:sparsity and MSE of linear}
\end{table}

\begin{table}[!ht]
    \centering
    \caption{Sparsity, cross-entropy and misclassification rates for synthetic logistic regression.}
    \begin{tabular}{|l|*{2}{c}{c|}*{2}{c}{c|}*{3}{c}|}
     \hline
     $\log_2(\rho^{-1})$ & \multicolumn{3}{c|}{16} & \multicolumn{3}{c|}{19} & \multicolumn{3}{c|}{22} \\
     \hline $\log_2(\lambda^{-1})$ & 6 & 8 & 10 & 6 & 8 & 10 & 6 & 8 & 10\\
     \hline
     \# nonzero blocks & 13 & 46 & 55 & 18 & 55 & 55 & 39 & 55 & 55\\
     \# nonzero coefficients & 91 & 226 & 258 & 205 & 658 & 655 & 721 & 1011 & 1020 \\
     cross-entropy & 0.541 & 0.537 & 0.536 & 0.535 & 0.533 & 0.534 & 0.534 & 0.535 & 0.537 \\
     misclassification (\%) & 26.92 & 26.8 & 26.85 & 26.62 & 26.69 & 26.81 & 26.58 & 26.87 & 27.03 \\
     \hline
    \end{tabular}\\[.1in]
    \parbox{1\textwidth}{\small Note: For $K=2$, there are $55$ blocks from which $10$ are main effects and $45$ are two-way interactions. There are a total of $1175$ scalar coefficients. For the underlying regression function (\ref{eq:f in single-block simulation}), $14$ of the $55$ components are true. The cross-entropy and misclassification rates are evaluated on a validation set of $n=50000$ data points.}
    \label{tab: sparsity, ce and mr of synthetic logistic}
\end{table}

\subsubsection{Synthetic logistic regression} \label{sec:simu-logistic}

\begin{figure}[!t]
    \centering
    \begin{subfigure}{0.32\textwidth}
       \includegraphics[width=\textwidth]{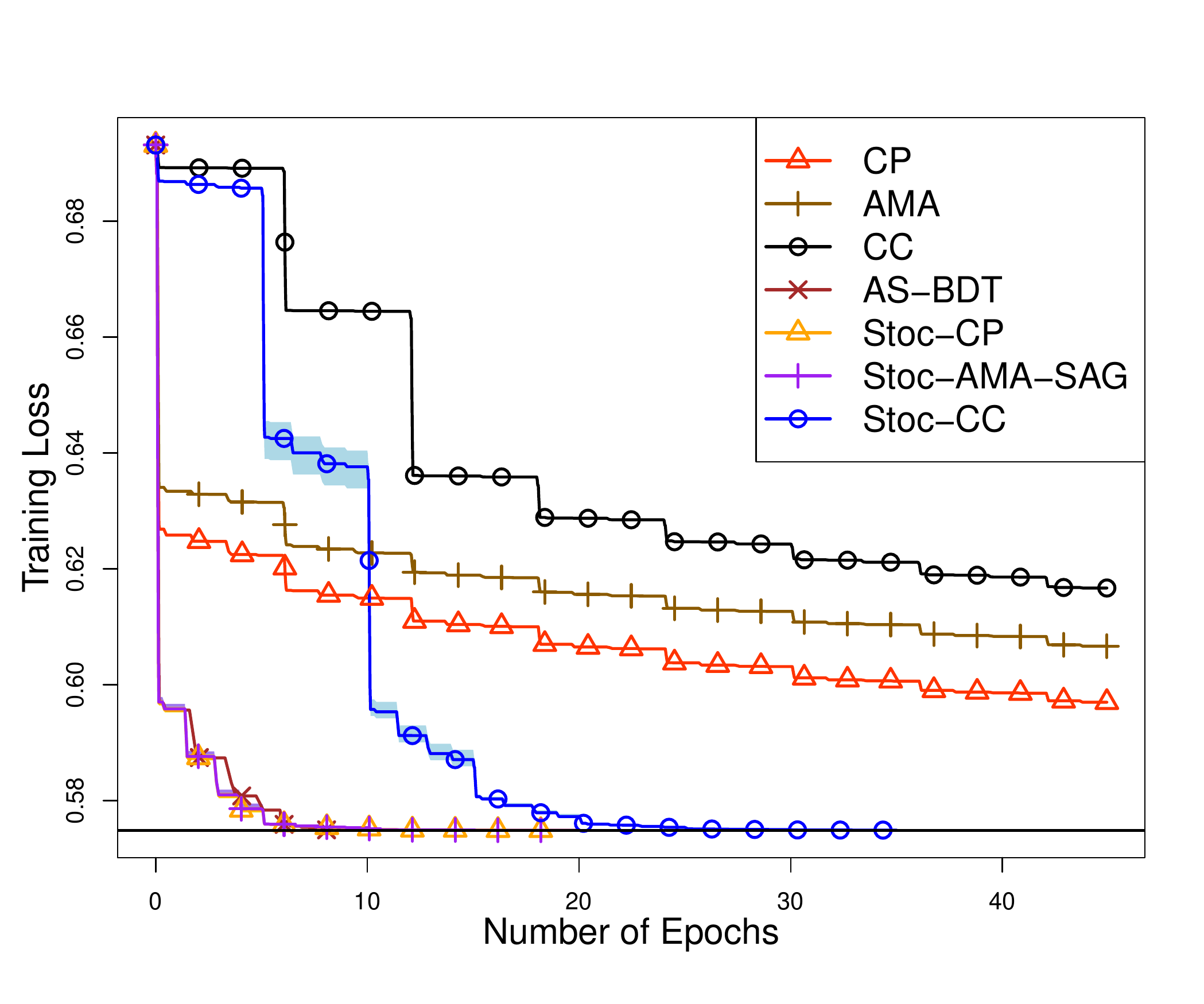}
       \vspace{-.45in}
       \caption{$\rho=2^{-16}$, $\lambda=\|\tilde Y\|_n/2^{6}$}
    \end{subfigure}
    \hfill
    \begin{subfigure}{0.32\textwidth}
       \includegraphics[width=\textwidth]{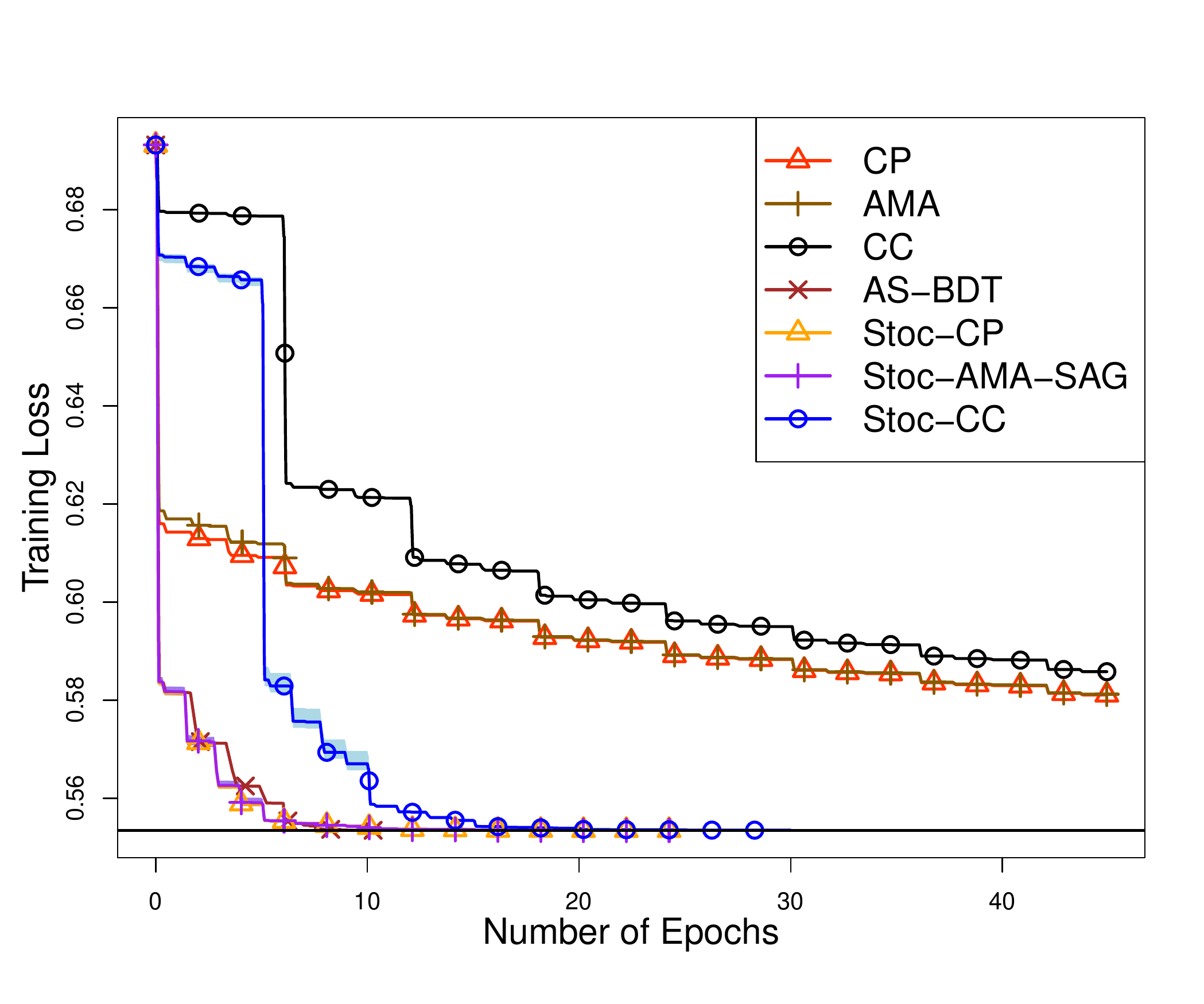}
       \vspace{-.45in}
       \caption{$\rho=2^{-16}$, $\lambda=\|\tilde Y\|_n/2^{8}$}
    \end{subfigure}
    \hfill
    \begin{subfigure}{0.32\textwidth}
       \includegraphics[width=\textwidth]{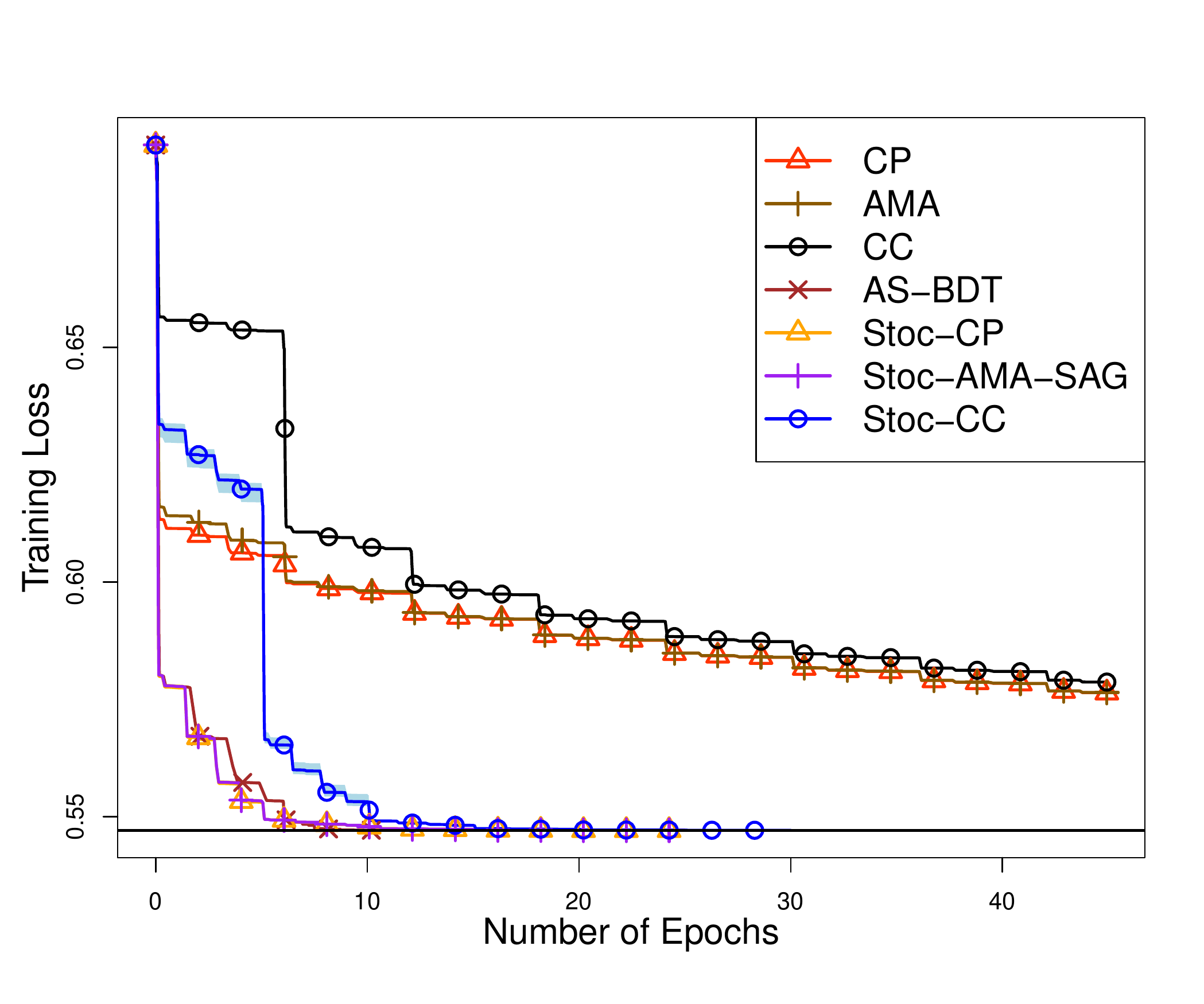}
       \vspace{-.45in}
       \caption{$\rho=2^{-16}$, $\lambda=\|\tilde Y\|_n/2^{10}$}
    \end{subfigure}

    \centering
    \begin{subfigure}{0.32\textwidth}
       \includegraphics[width=\textwidth]{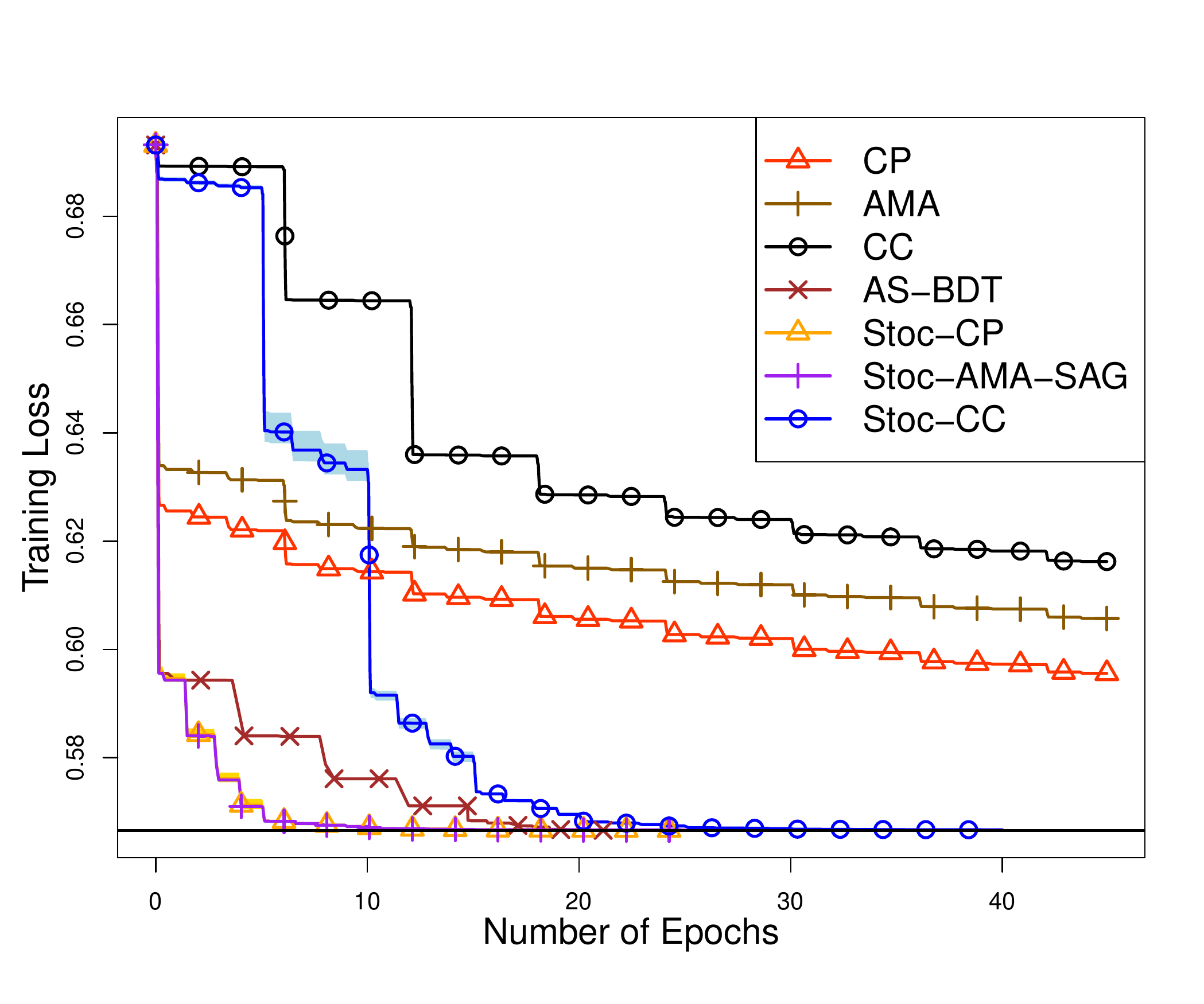}
       \vspace{-.45in}
       \caption{$\rho=2^{-19}$, $\lambda=\|\tilde Y\|_n/2^{6}$}
    \end{subfigure}
    \hfill
    \begin{subfigure}{0.32\textwidth}
       \includegraphics[width=\textwidth]{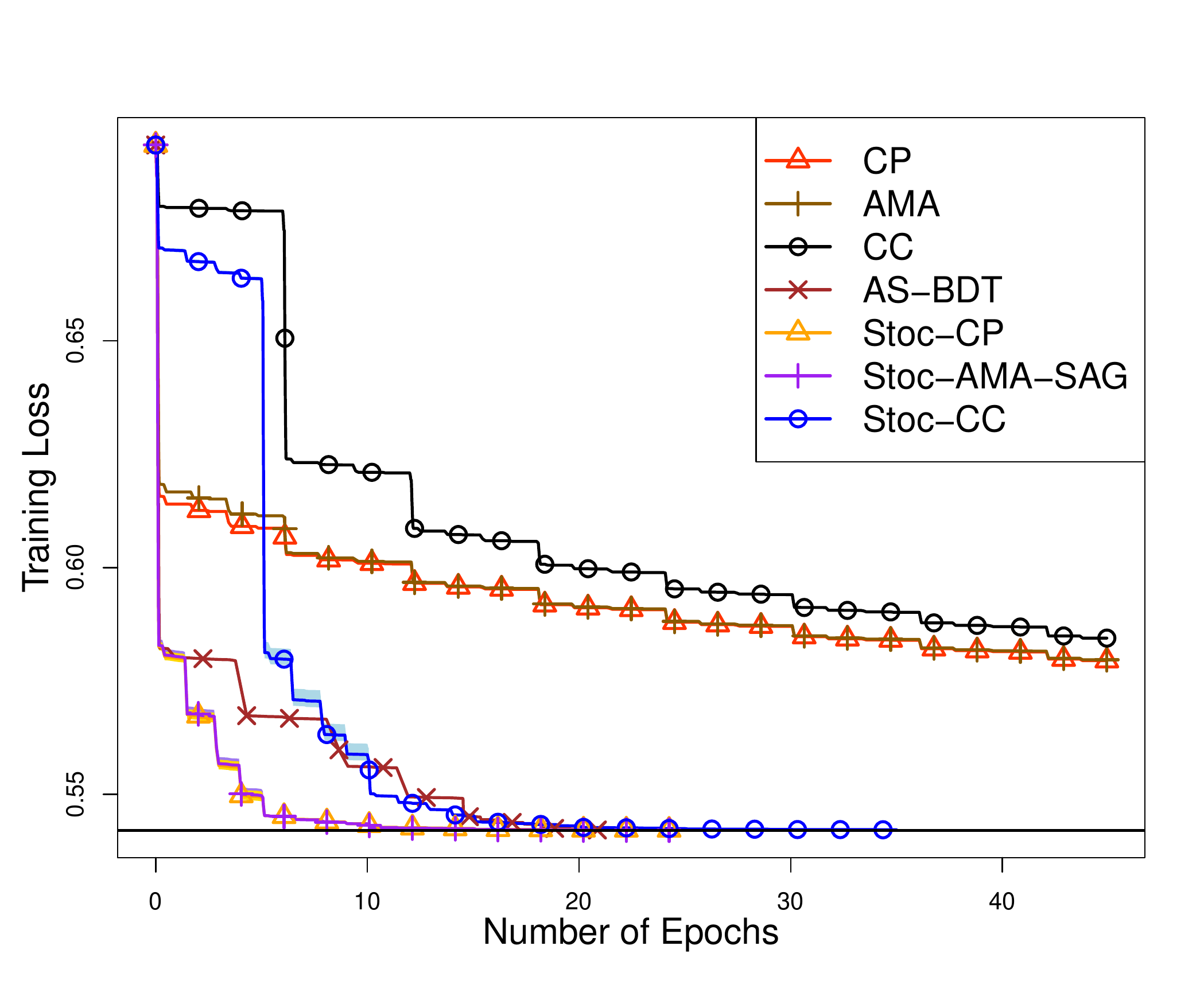}
       \vspace{-.45in}
       \caption{$\rho=2^{-19}$, $\lambda=\|\tilde Y\|_n/2^{8}$}
    \end{subfigure}
    \hfill
    \begin{subfigure}{0.32\textwidth}
       \includegraphics[width=\textwidth]{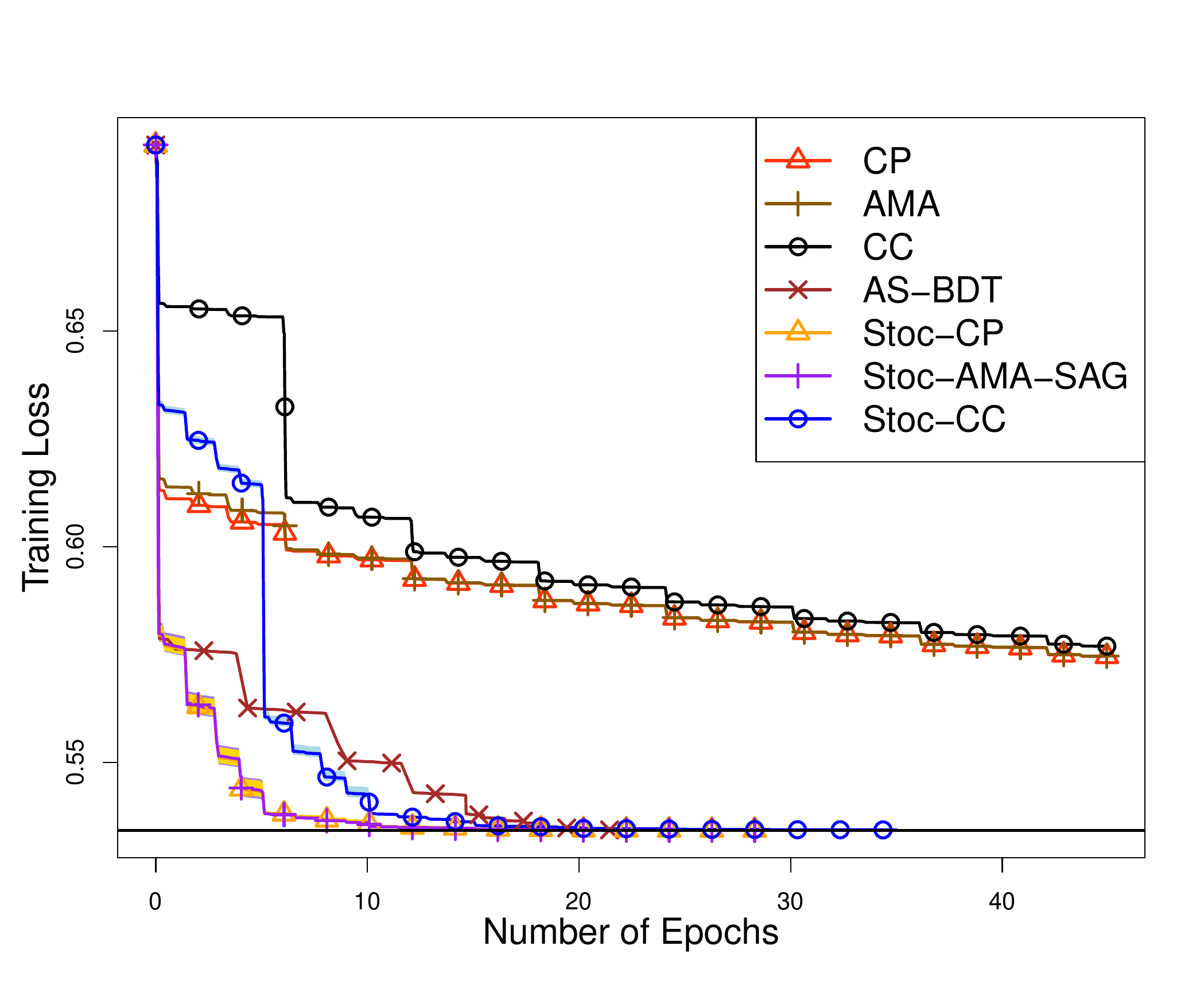}
       \vspace{-.45in}
       \caption{$\rho=2^{-19}$, $\lambda=\|\tilde Y\|_n/2^{10}$}
    \end{subfigure}

    \centering
    \begin{subfigure}{0.32\textwidth}
       \includegraphics[width=\textwidth]{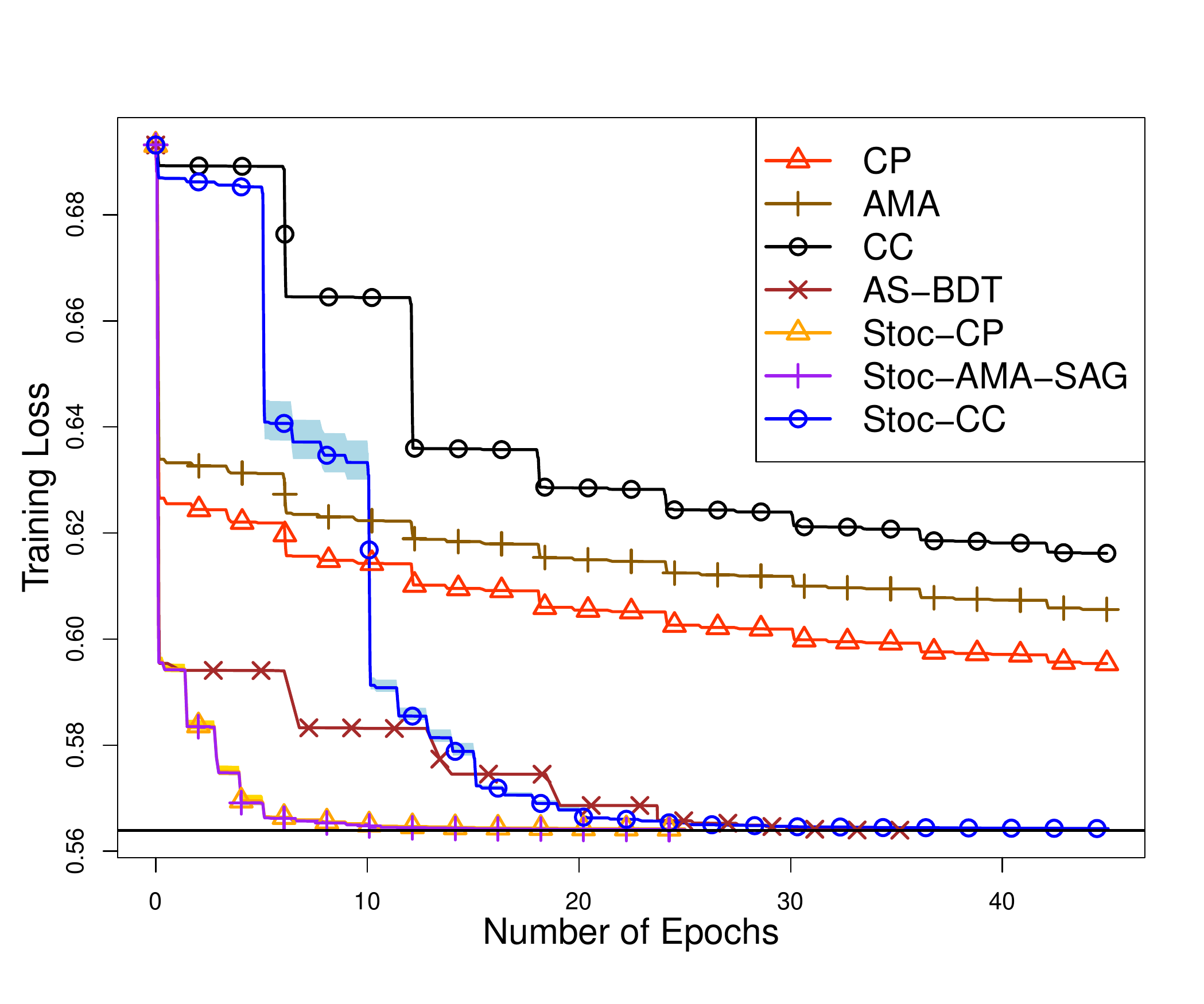}
       \vspace{-.45in}
       \caption{$\rho=2^{-22}$, $\lambda=\|\tilde Y\|_n/2^{6}$}
    \end{subfigure}
    \hfill
    \begin{subfigure}{0.32\textwidth}
       \includegraphics[width=\textwidth]{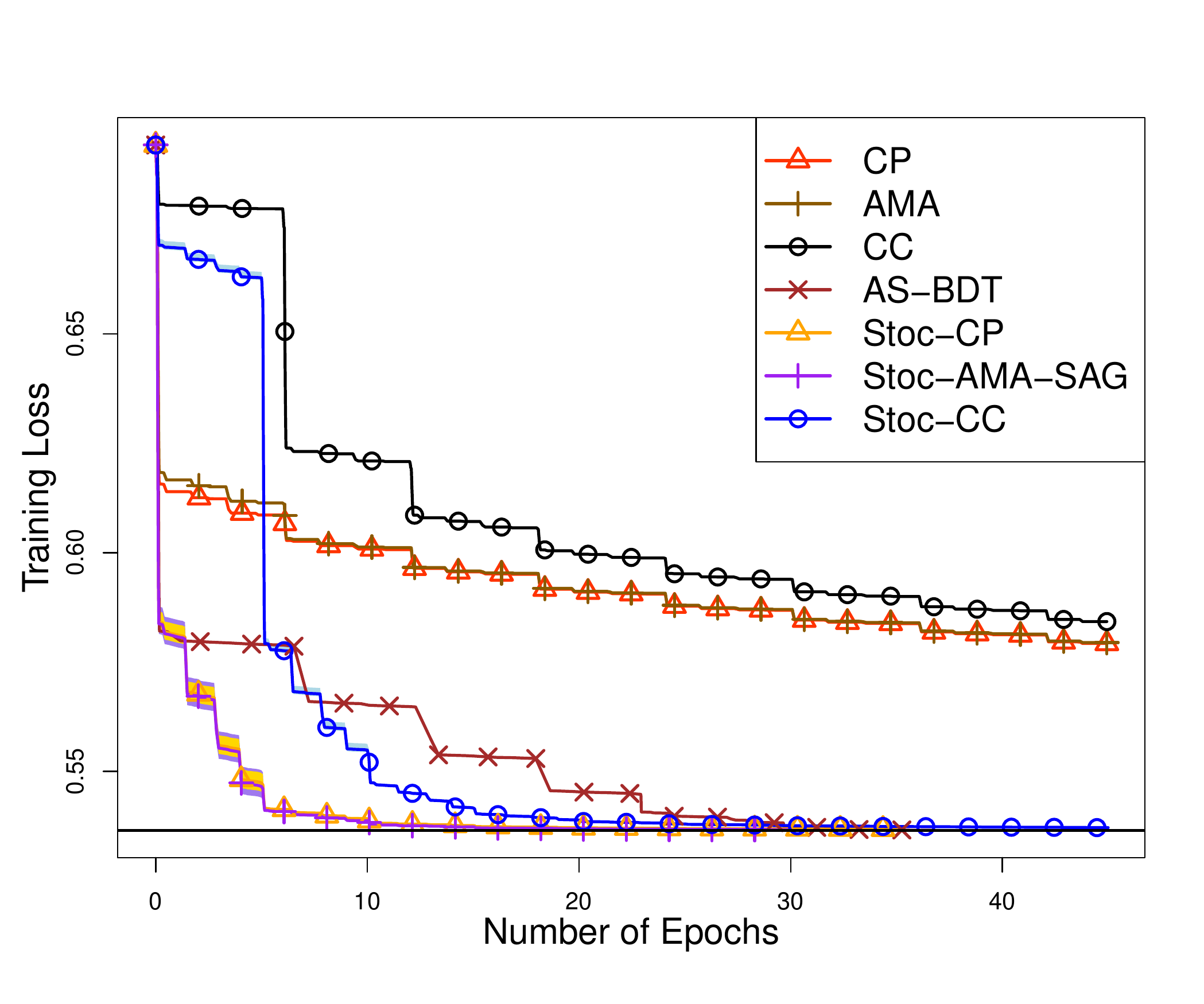}
       \vspace{-.45in}
       \caption{$\rho=2^{-22}$, $\lambda=\|\tilde Y\|_n/2^{8}$}
    \end{subfigure}
    \hfill
    \begin{subfigure}{0.32\textwidth}
       \includegraphics[width=\textwidth]{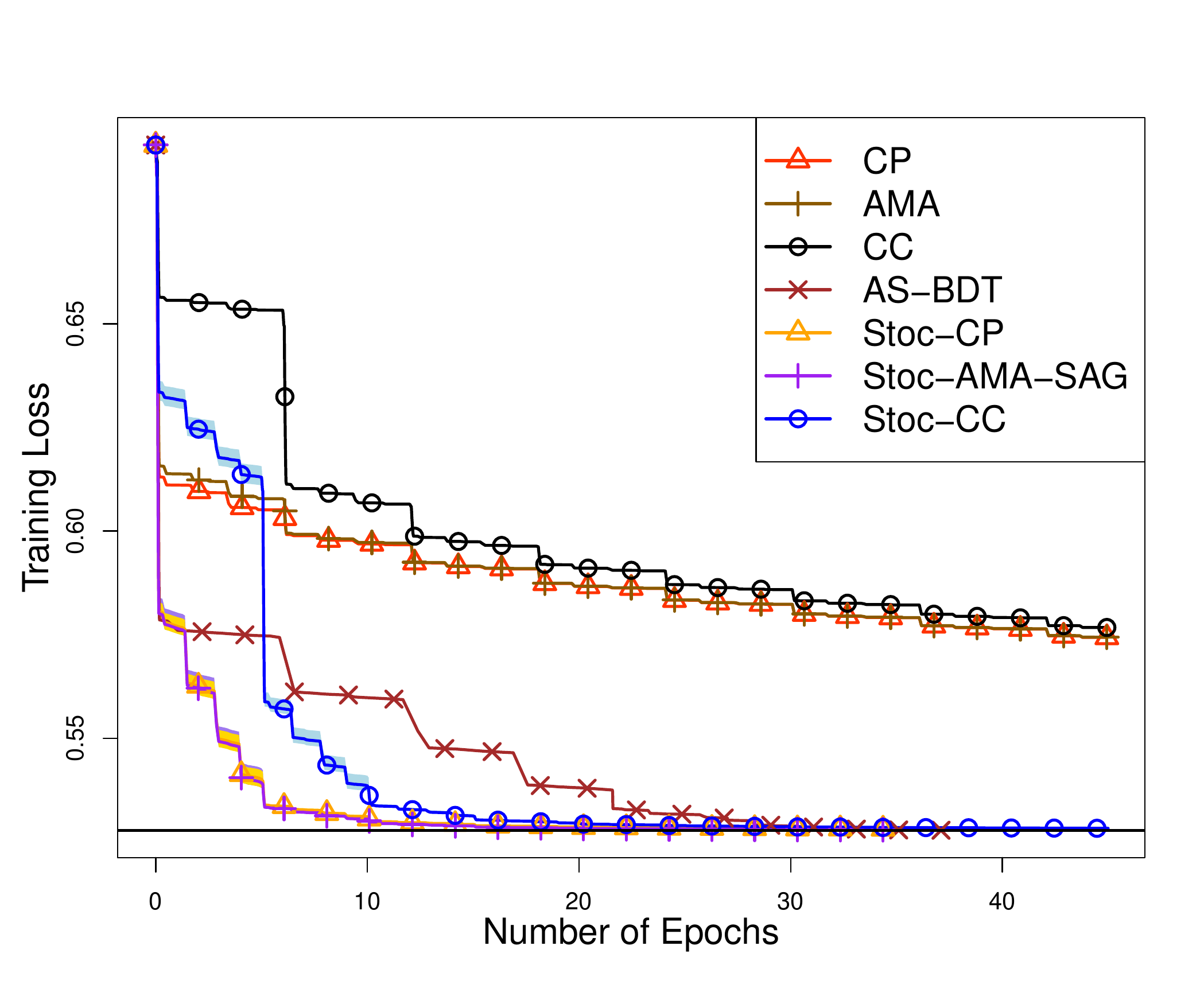}
       \vspace{-.45in}
       \caption{$\rho=2^{-22}$, $\lambda=\|\tilde Y\|_n/2^{10}$}
    \end{subfigure}
    \caption{DPAM training in synthetic logistic regression. The minimal training loss achieved from AS-BDT is given by the black solid line. For the stochastic algorithms, we plot the mean as well as the minimum and maximum training losses across 10 repeated runs.}
    \label{fig: synthetic logistic}
\end{figure}

The logistic regression model is extended from the linear regression model in Section \ref{sec:simu-linear}. For $n=50000$, we generate $X_i=(X_{i,1},X_{i,2},\ldots,X_{i,10})$ uniform on $[0,1]^{10}$ and $Y_i$ as Bernoulli with success probability $\mathrm{expit}(f(X_i))$, $i=1,\ldots,n$, where $\mathrm{expit}(f)= 1/(1+\exp (-f))$, and $f(X)$ is defined as (\ref{eq:f in single-block simulation}). Note that $(X_{8},X_9, X_{10})$ are spurious inputs.
We apply the logistic DPAM (\ref{eq:logistic-DPAM}),
with differentiation order $m=2$ (i.e., piecewise cross-linear basis functions), interaction order $K=2$ (main effects and two-way interactions), and
$6$ marginal knots  ($5$ basis functions for each main effect), similarly as in Section \ref{sec:simu-linear}.
The performances of various algorithms are reported in Figure~\ref{fig: synthetic logistic} under different choices of $\rho\in\{2^{-16},2^{-19},2^{-22}\}$ and $\lambda\in\{\|\tilde Y\|_n/2^6,\|\tilde Y\|_n/2^{8},\|\tilde Y\|_n/2^{10}\}$, where $\tilde Y$ is the centered version of $Y$.
For solutions from AS-BDT (after convergence declared), the sparsity levels, the cross-entropy and the misclassification rates under different tuning parameters are summarized in Table~\ref{tab: sparsity, ce and mr of synthetic logistic}. The sparsity levels from stochastic primal-dual algorithms are reported in Supplement Table~\ref{tab:sparsity for synthetic logistic}.
A tolerance of $10^{-4}$ in the objective value (average log-likelihood) is checked after each backfitting cycle over all blocks to declare convergence.

\subsubsection{Logistic regression on real data}\label{sec:simu-logistic-real}

We evaluate the performances of various algorithms on a real dataset, run-or-walk, where the aim is to classify whether a person is running or walking based on sensor data collected from an iOS device. The dataset is available from Kaggle \url{https://www.kaggle.com/datasets/vmalyi/run-or-walk}. The run-or-walk dataset has $88588$ data points, $6$ explanatory variables and $1$ binary response variable. The $6$ explanatory variables ($acc_x$, $acc_y$, $acc_z$, $gyro_x$, $gyro_y$ and $gyro_z$) are collected from the accelerometer and gyroscope in $3$ dimensions. The binary variable $Y$ indicates whether the person is running (1) or walking (0).

\begin{figure}[!t]
    \centering
    \begin{subfigure}{0.32\textwidth}
       \includegraphics[width=\textwidth]{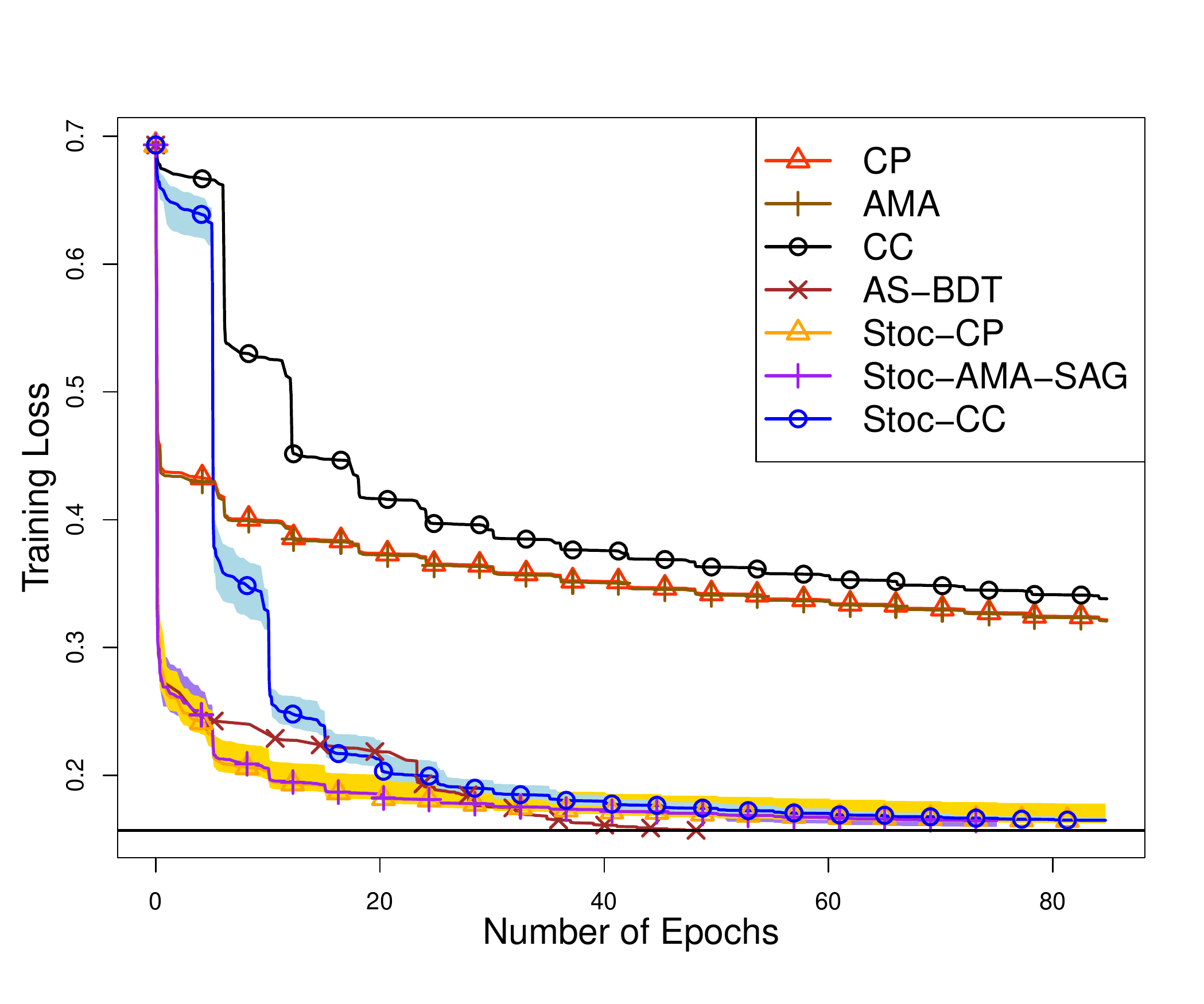}
       \vspace{-.45in}
       \caption{$\rho=2^{-18}$, $\lambda=\|\tilde Y\|_n/2^{6}$}
    \end{subfigure}
    \hfill
    \begin{subfigure}{0.32\textwidth}
       \includegraphics[width=\textwidth]{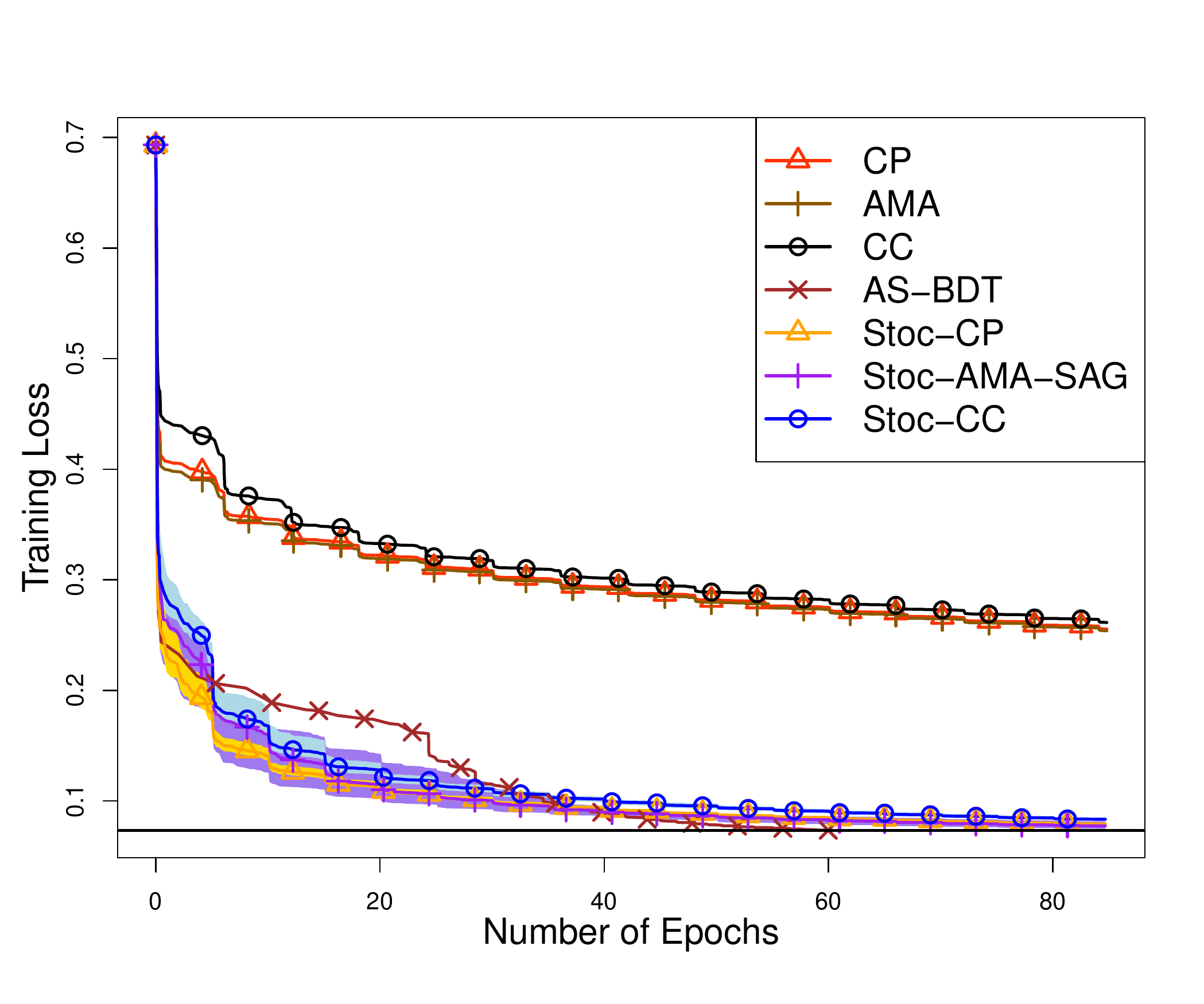}
       \vspace{-.45in}
       \caption{$\rho=2^{-18}$, $\lambda=\|\tilde Y\|_n/2^{13}$}
    \end{subfigure}
    \hfill
    \begin{subfigure}{0.32\textwidth}
       \includegraphics[width=\textwidth]{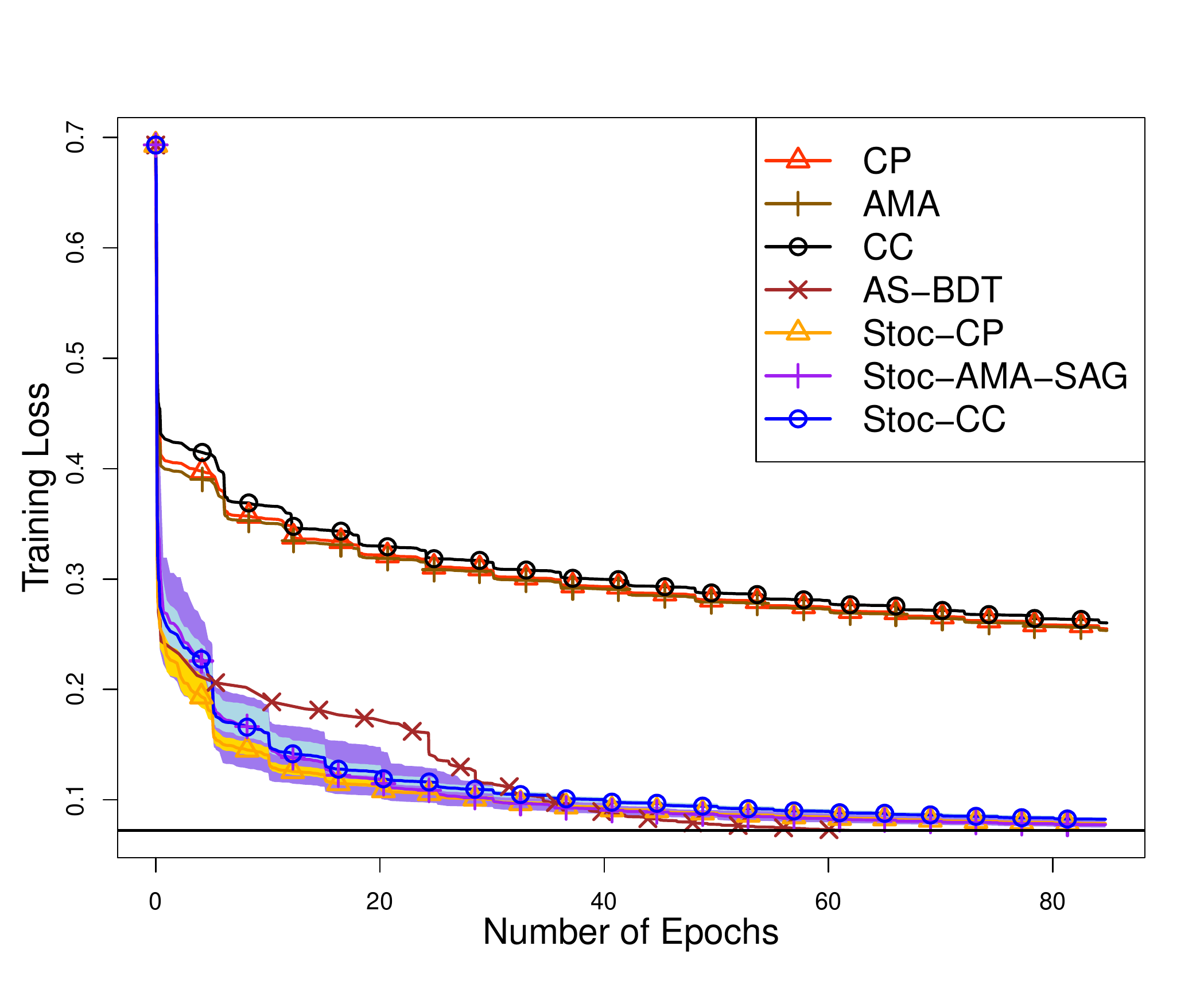}
       \vspace{-.45in}
       \caption{$\rho=2^{-18}$, $\lambda=\|\tilde Y\|_n/2^{15}$}
    \end{subfigure}

    \centering
    \begin{subfigure}{0.32\textwidth}
       \includegraphics[width=\textwidth]{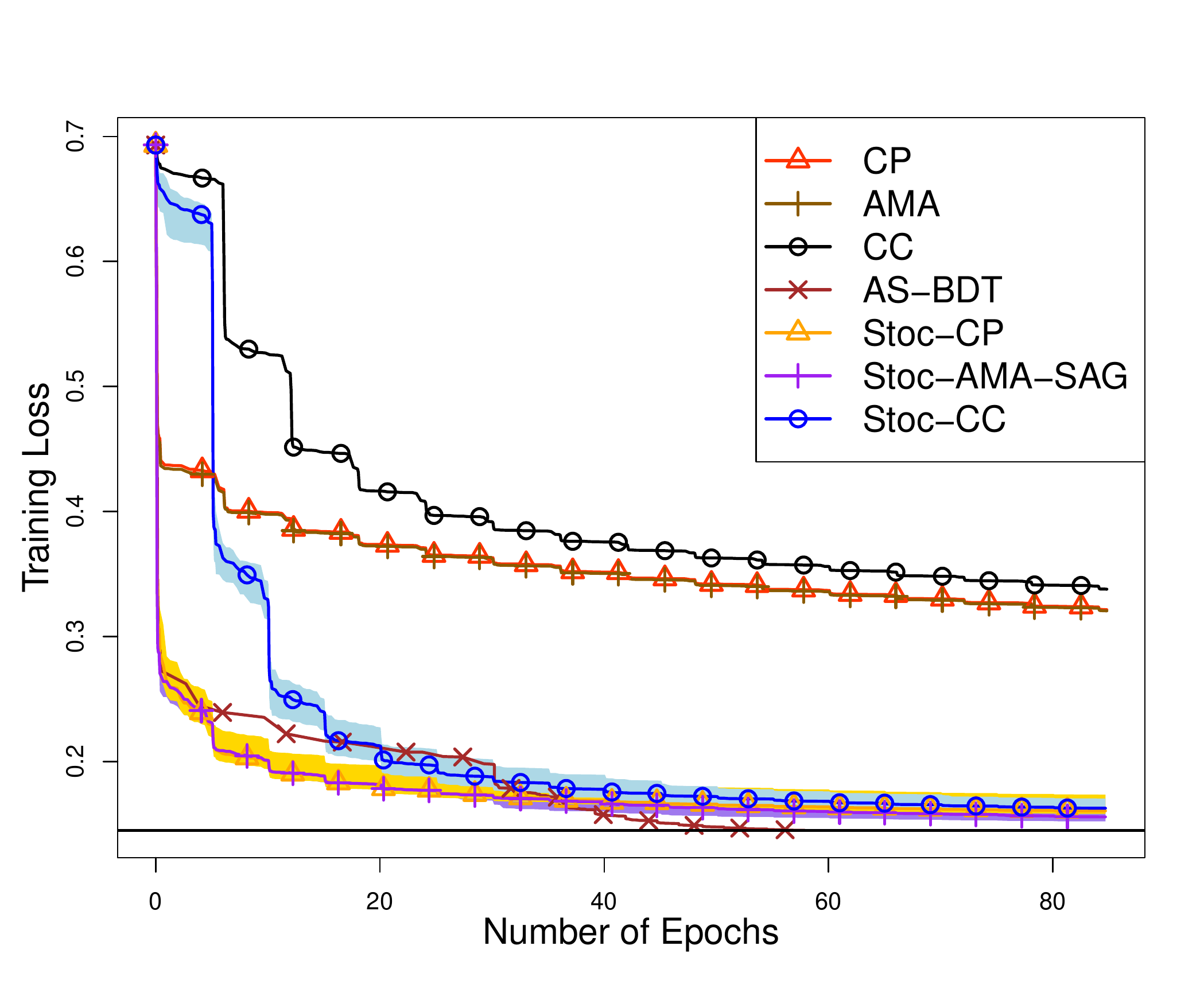}
       \vspace{-.45in}
       \caption{$\rho=2^{-23}$, $\lambda=\|\tilde Y\|_n/2^{6}$}
    \end{subfigure}
    \hfill
    \begin{subfigure}{0.32\textwidth}
       \includegraphics[width=\textwidth]{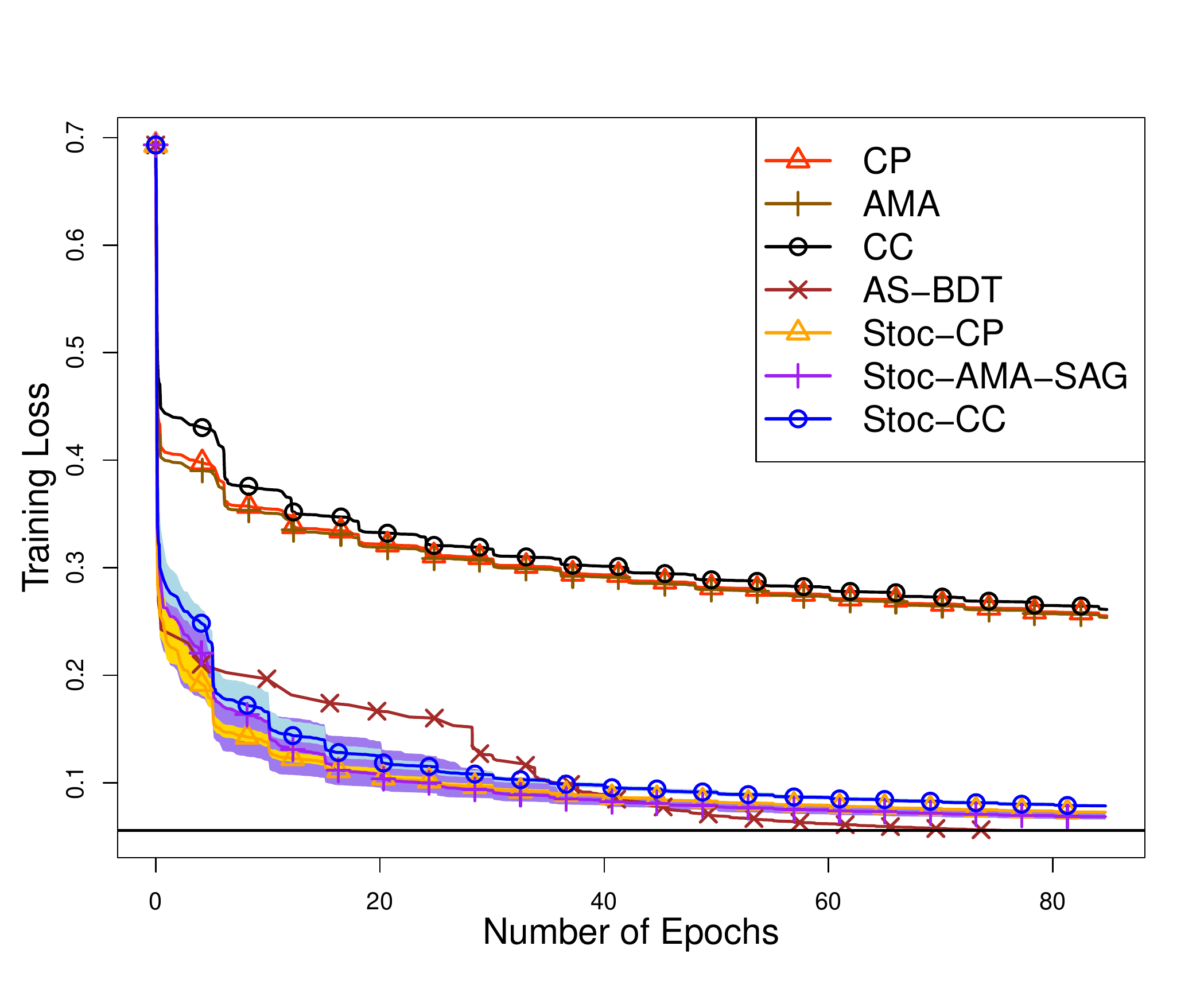}
       \vspace{-.45in}
       \caption{$\rho=2^{-23}$, $\lambda=\|\tilde Y\|_n/2^{13}$}
    \end{subfigure}
    \hfill
    \begin{subfigure}{0.32\textwidth}
       \includegraphics[width=\textwidth]{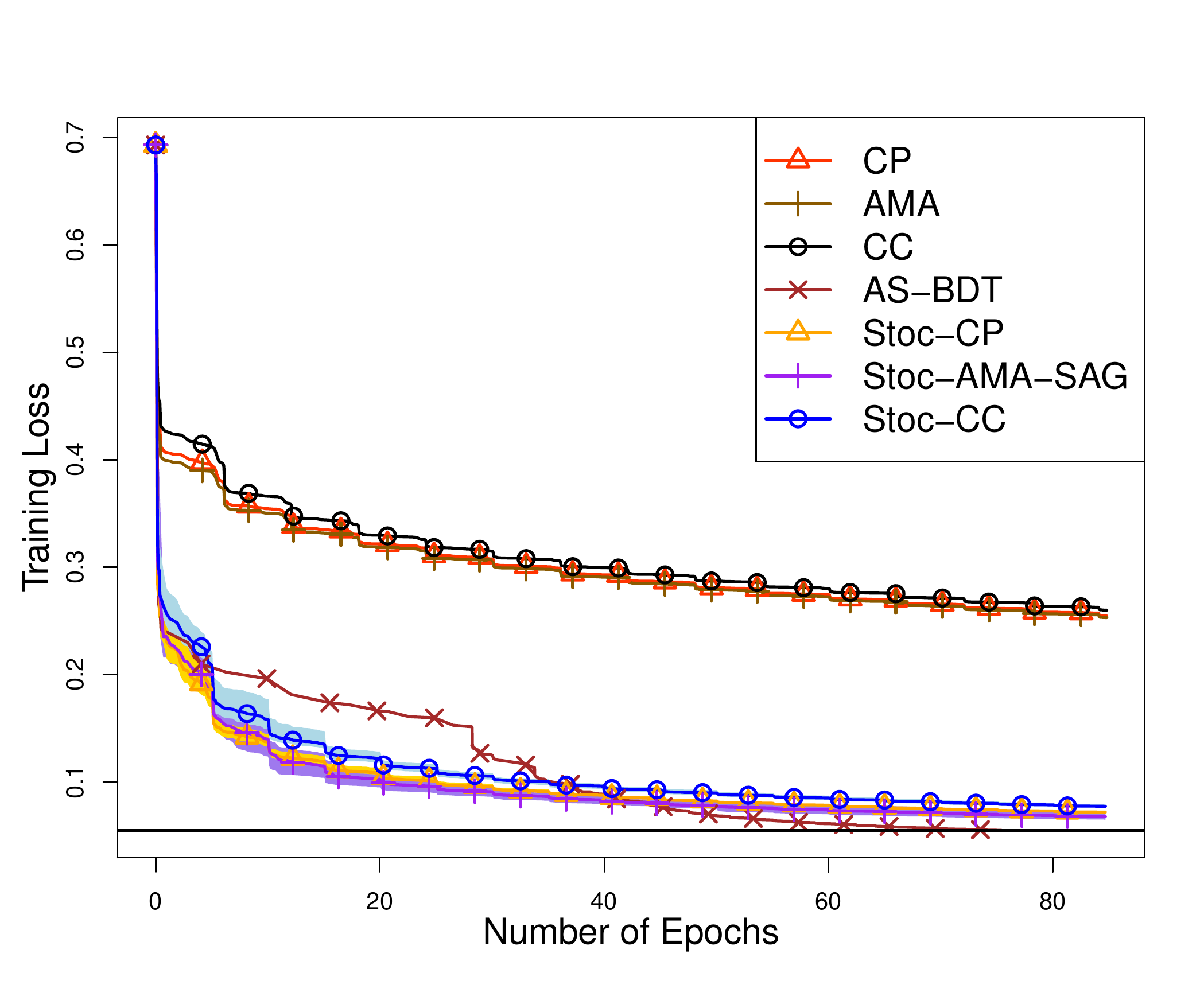}
       \vspace{-.45in}
       \caption{$\rho=2^{-23}$, $\lambda=\|\tilde Y\|_n/2^{15}$}
    \end{subfigure}

    \centering
    \begin{subfigure}{0.32\textwidth}
       \includegraphics[width=\textwidth]{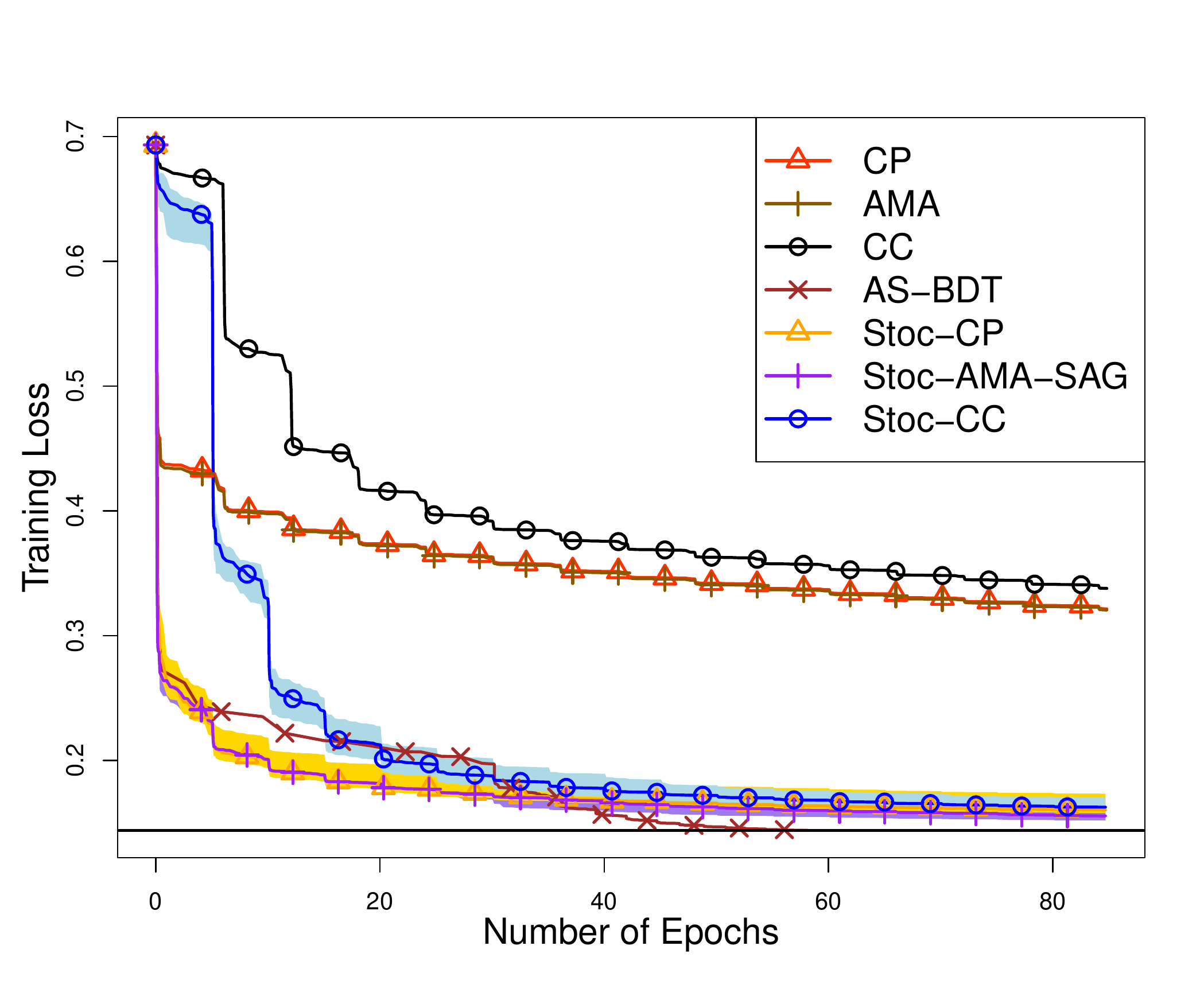}
       \vspace{-.45in}
       \caption{$\rho=2^{-25}$, $\lambda=\|\tilde Y\|_n/2^{6}$}
    \end{subfigure}
    \hfill
    \begin{subfigure}{0.32\textwidth}
       \includegraphics[width=\textwidth]{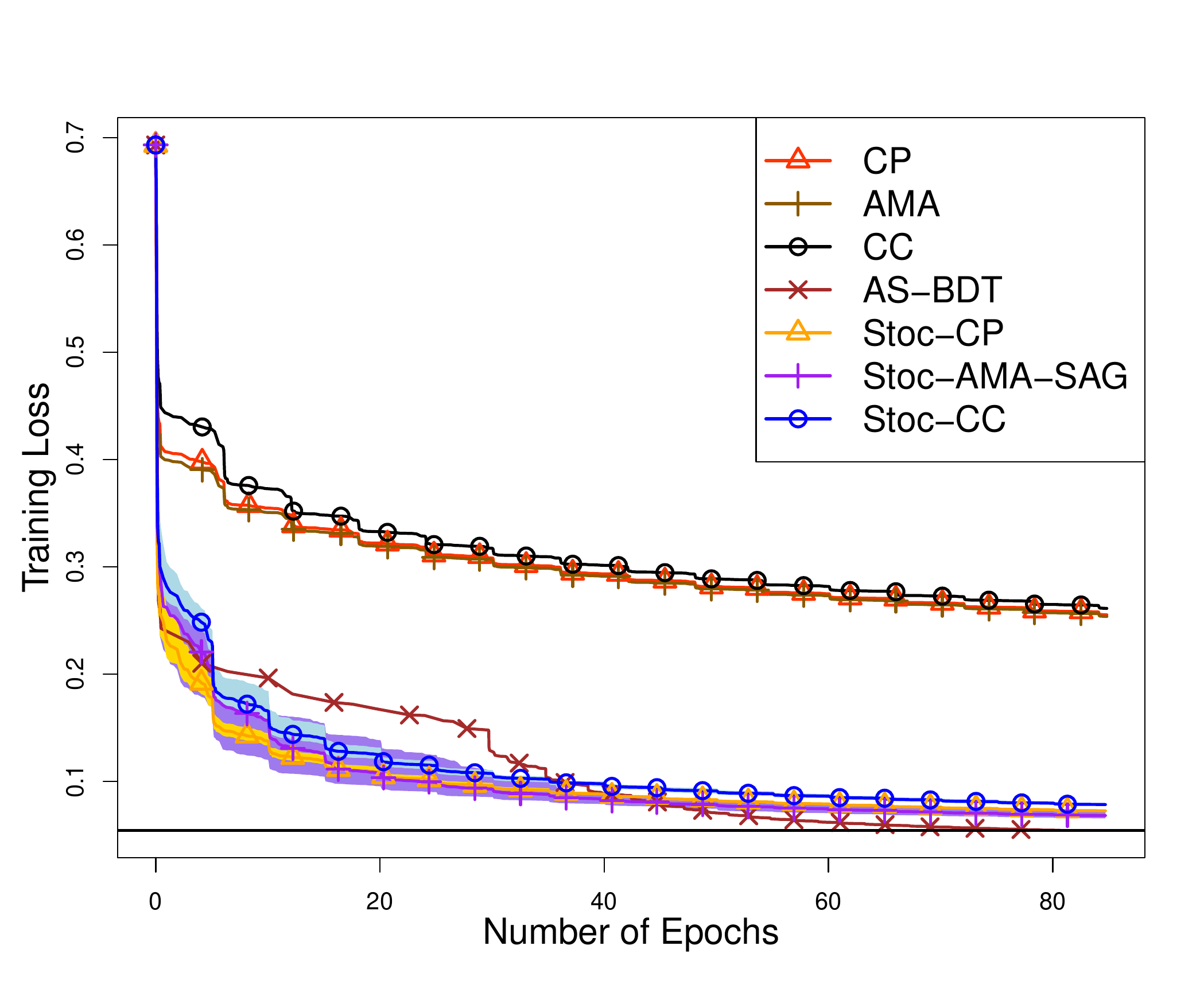}
       \vspace{-.45in}
       \caption{$\rho=2^{-25}$, $\lambda=\|\tilde Y\|_n/2^{13}$}
    \end{subfigure}
    \hfill
    \begin{subfigure}{0.32\textwidth}
       \includegraphics[width=\textwidth]{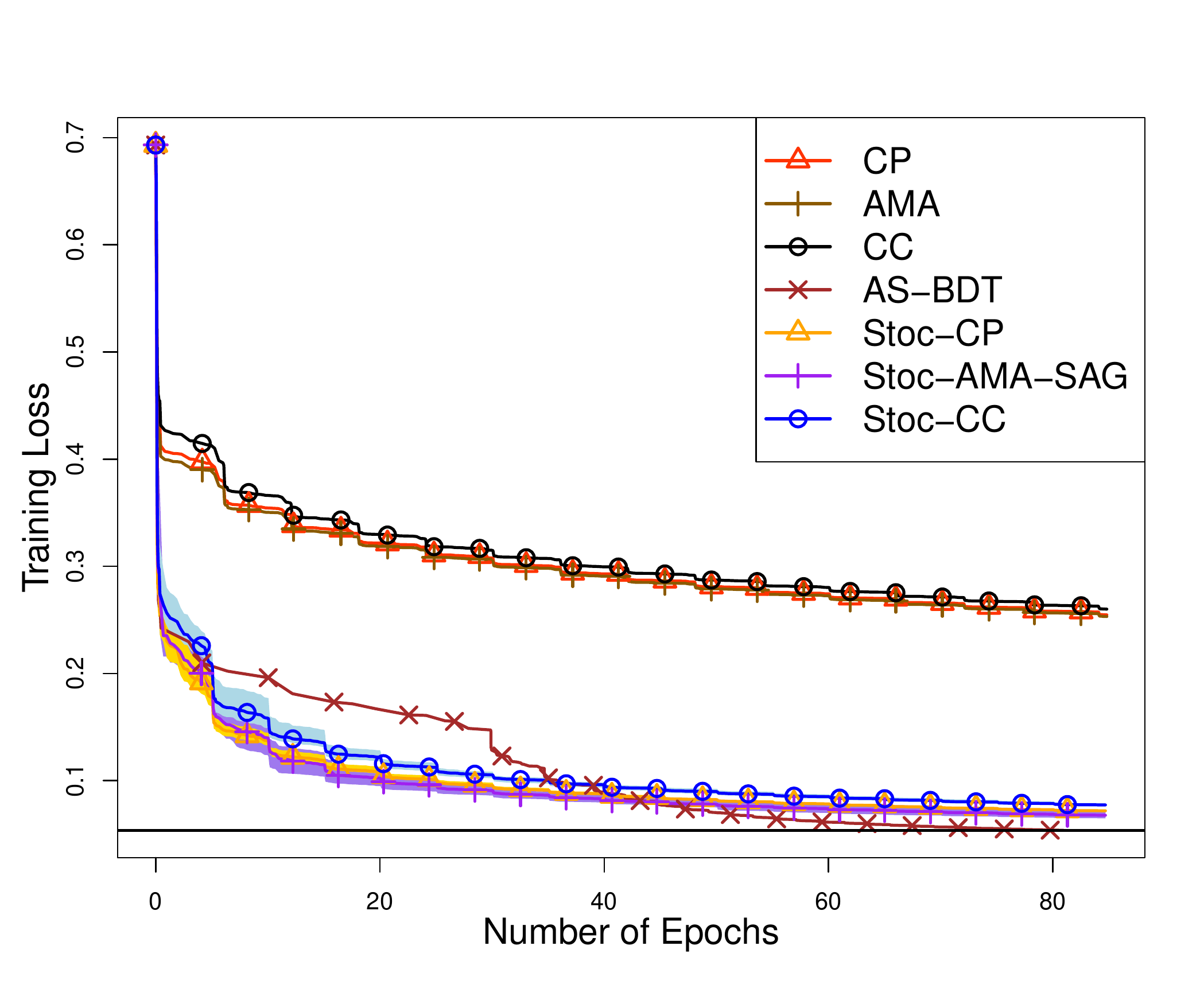}
       \vspace{-.45in}
       \caption{$\rho=2^{-25}$, $\lambda=\|\tilde Y\|_n/2^{15}$}
    \end{subfigure}
    \caption{DPAM training in logistic regression on real data. The minimal training loss achieved from AS-BDT is given by the black solid line. For the stochastic algorithms, we plot the mean as well as the minimum and maximum training losses across 10 repeated runs.}
    \label{fig: logistic on real data}
\end{figure}

For pre-processing,
we standardize the $6$ explanatory variables by subtracting their sample means and then dividing by their sample standard deviations. Then we split the data randomly into a training set of size $n=70870$ (around $80\%$ of the full dataset) and a validation set of size $17718$, to mimic a $5$-fold cross-validation.
We apply the logistic DPAM (\ref{eq:logistic-DPAM}),
with differentiation order $m=2$ (i.e., piecewise cross-linear basis functions), interaction order $K=2$ (main effects and two-way interactions), and
$6$ marginal knots  ($5$ basis functions for each main effect), as in Section \ref{sec:simu-logistic}.
The performances of various algorithms are reported in Figure~\ref{fig: logistic on real data} under different choices of $\rho\in\{2^{-18},2^{-23},2^{-25}\}$ and $\lambda\in\{\|\tilde Y\|_n/2^6,\|\tilde Y\|_n/2^{13},\|\tilde Y\|_n/2^{15}\}$, where $\tilde Y$ is the centered version of $Y$.
For solutions from AS-BDT (after convergence declared), the sparsity levels, the cross-entropy and the misclassification rates under different tuning parameters are summarized in Table~\ref{tab: sparsity, ce and mr of logistic for real data}.
An objective tolerance of $10^{-4}$ is checked in the average log-likelihood after each cycle over all blocks to declare convergence.

\begin{table}[hb]
    \centering
    \caption{Sparsity, cross-entropy and misclassification rates for logistic regression on real data.}
    \begin{tabular}{|l|*{2}{c}{c|}*{2}{c}{c|}*{3}{c}|}
     \hline
     $\log_2(\rho^{-1})$ & \multicolumn{3}{c|}{18} & \multicolumn{3}{c|}{23} & \multicolumn{3}{c|}{25} \\
     \hline $\log_2(\lambda^{-1})$ & 6 & 13 & 15 & 6 & 13 & 15 & 6 & 13 & 15\\
     \hline
     \# nonzero blocks & 17 & 21 & 21 & 19 & 21 & 21 & 19 & 21 & 21 \\
     \# nonzero coefficients & 220 & 240 & 239 & 313 & 332 & 333 & 307 & 336 & 334 \\
     cross-entropy (in $10^{-2}$) & 8.93 & 6.57 & 6.56 & 8.06 & 5.55 & 5.54 & 8.00 & 5.56 & 5.56 \\
     misclassification (\%) & 1.95 & 1.71 & 1.71 & 1.86 & 1.53 & 1.53 & 1.89 & 1.58 & 1.57 \\
     \hline
    \end{tabular}\\[.1in]
    \parbox{1\textwidth}{\small Note: For $K=2$, there are $21$ blocks from which $6$ are main effects and $15$ are two-way interactions. There are a total of $405$ scalar coefficients. The cross-entropy and misclassification rates are evaluated on the validation set.}
    \label{tab: sparsity, ce and mr of logistic for real data}
\end{table}

\subsubsection{Comparisons}

From Figures~\ref{fig: synthetic linear}--\ref{fig: logistic on real data} as well as Figures~\ref{fig: phase shift two-way} and \ref{fig: phase shift three-way} for the phase shift model in the Supplement, we observe the following trends.

\begin{itemize}
    \item \textbf{Batch vs Stochastic:}
     Compared with their batch counterparts, all three stochastic algorithms (Stoc-CP, and Stoc-AMA-SAG, Stoc-CC) achieve considerable improvements. Due to the non-descent nature, batch primal-dual algorithms require more iterations in each block, so that the overall performances are compromised. However, stochastic primal-dual algorithms are less affected by this downside, which can also be seen from the advantages over batch versions in single-block optimization (Figure~\ref{fig: single-block}).

    \item \textbf{Stochastic primal-dual algorithms:} Stochastic primal-dual algorithms perform better than or similarly as Stoc-CC.
        Comparison between stochastic primal-dual algorithms and AS-BDT depends on the penalty parameter $\rho$, which controls the element-wise sparsity of the solution.
        Their performances are similar for relatively large $\rho$.
        As $\rho$ decreases (leading to less sparse solutions), the stochastic primal-dual algorithms achieve a significant advantage over AS-BDT (except in Figure~\ref{fig: logistic on real data} where the acceleration of AS-BDT takes effect as discussed below).

    \item \textbf{Stoc-CC:} The performance of Stoc-CC seems to be sensitive to the penalty parameter $\lambda$, which controls the block-wise sparsity. The performance improves as $\lambda$ decreases. A similar pattern can also be observed on batch CC. A heuristic explanation is that when $\lambda$ is small, the solutions $\beta$ in the nonzero blocks may deviate more from $0$ such that the majorization in CC captures the local curvature more accurately (as illustrated in Figure~\ref{fig:abs}). See Supplement Section \ref{sec:cc} for further discussion.

    \item \textbf{Acceleration of AS-BDT:}
    In Figure~\ref{fig: logistic on real data}, the stochastic algorithms are more efficient at the beginning, but are slightly outperformed by AS-BDT as training continues.
    Such acceleration occurs because AS-BDT exploits the active-set information from the previous cycle to achieve computational savings,
    which can be substantial as many cycles of backfitting are run in logistic regression where quadratic approximations are involved.
    In fact, in Figure~\ref{fig: logistic on real data}, while the three stochastic algorithms go through at most $16$ cycles of backfitting in a $80$-epoch run,
    the number of backfitting cycles completed by AS-BDT are as large as $30$.
    This observation suggests a hybrid strategy combining AS-BDT and stochastic algorithms, as discussed in Section~\ref{subsec:comparison with AS-BDT and hybrid}.

    \item \textbf{Variance:} The three stochastic algorithms have almost invisible variances in Figures~\ref{fig: synthetic linear} and \ref{fig: synthetic logistic}. In Figure~\ref{fig: logistic on real data}, the stochastic Chambolle--Pock algorithm tends to have smaller variances than the other two for small $\lambda$ (leading to block-wise denser solutions).
\end{itemize}

\section{Conclusion}\label{sec:conclusion}
We develop two primal-dual algorithms, including both batch and stochastic versions, for doubly-penalized ANOVA modeling with both HTV and empirical-norm penalties, where existing primal-dual algorithms are not suitable. Our numerical experiments demonstrate considerable gains from the stochastic primal-dual algorithms compared with their batch versions and the previous algorithm AS-BDT in large-scale, especially non-sparse, scenarios. Nevertheless, theoretical convergence remains to be studied. Moreover, a hybrid approach can be explored by combining stochastic primal-dual and AS-BDT algorithms.

\bibliographystyle{apacite}
\bibliography{reference}

\begin{thebibliography}{}

\bibitem [\protect \citeauthoryear {%
Bottou%
, Curtis%
\BCBL {}\ \BBA {} Nocedal%
}{%
Bottou%
\ \protect \BOthers {.}}{%
{\protect \APACyear {2018}}%
}]{%
bottou2018optimization}
\APACinsertmetastar {%
bottou2018optimization}%
\begin{APACrefauthors}%
Bottou, L.%
, Curtis, F\BPBI E.%
\BCBL {}\ \BBA {} Nocedal, J.%
\end{APACrefauthors}%
\unskip\
\newblock
\APACrefYearMonthDay{2018}{}{}.
\newblock
{\BBOQ}\APACrefatitle {Optimization methods for large-scale machine learning}
  {Optimization methods for large-scale machine learning}.{\BBCQ}
\newblock
\APACjournalVolNumPages{SIAM Review}{60}{}{223--311}.
\PrintBackRefs{\CurrentBib}

\bibitem [\protect \citeauthoryear {%
Boyd%
, Parikh%
, Chu%
, Peleato%
\BCBL {}\ \BBA {} Eckstein%
}{%
Boyd%
\ \protect \BOthers {.}}{%
{\protect \APACyear {2011}}%
}]{%
boyd2011distributed}
\APACinsertmetastar {%
boyd2011distributed}%
\begin{APACrefauthors}%
Boyd, S.%
, Parikh, N.%
, Chu, E.%
, Peleato, B.%
\BCBL {}\ \BBA {} Eckstein, J.%
\end{APACrefauthors}%
\unskip\
\newblock
\APACrefYearMonthDay{2011}{}{}.
\newblock
{\BBOQ}\APACrefatitle {Distributed optimization and statistical learning via
  the alternating direction method of multipliers} {Distributed optimization
  and statistical learning via the alternating direction method of
  multipliers}.{\BBCQ}
\newblock
\APACjournalVolNumPages{Foundations and Trends in Machine
  Learning}{3}{}{1--122}.
\PrintBackRefs{\CurrentBib}

\bibitem [\protect \citeauthoryear {%
Chambolle%
\ \BBA {} Pock%
}{%
Chambolle%
\ \BBA {} Pock%
}{%
{\protect \APACyear {2011}}%
}]{%
chambolle2011first}
\APACinsertmetastar {%
chambolle2011first}%
\begin{APACrefauthors}%
Chambolle, A.%
\BCBT {}\ \BBA {} Pock, T.%
\end{APACrefauthors}%
\unskip\
\newblock
\APACrefYearMonthDay{2011}{}{}.
\newblock
{\BBOQ}\APACrefatitle {A first-order primal-dual algorithm for convex problems
  with applications to imaging} {A first-order primal-dual algorithm for convex
  problems with applications to imaging}.{\BBCQ}
\newblock
\APACjournalVolNumPages{Journal of Mathematical Imaging and
  Vision}{40}{}{120--145}.
\PrintBackRefs{\CurrentBib}

\bibitem [\protect \citeauthoryear {%
Chen%
, Huang%
\BCBL {}\ \BBA {} Zhang%
}{%
Chen%
\ \protect \BOthers {.}}{%
{\protect \APACyear {2013}}%
}]{%
chen2013primal}
\APACinsertmetastar {%
chen2013primal}%
\begin{APACrefauthors}%
Chen, P.%
, Huang, J.%
\BCBL {}\ \BBA {} Zhang, X.%
\end{APACrefauthors}%
\unskip\
\newblock
\APACrefYearMonthDay{2013}{}{}.
\newblock
{\BBOQ}\APACrefatitle {A primal--dual fixed point algorithm for convex
  separable minimization with applications to image restoration} {A
  primal--dual fixed point algorithm for convex separable minimization with
  applications to image restoration}.{\BBCQ}
\newblock
\APACjournalVolNumPages{Inverse Problems}{29}{}{025011}.
\PrintBackRefs{\CurrentBib}

\bibitem [\protect \citeauthoryear {%
Condat%
}{%
Condat%
}{%
{\protect \APACyear {2013}}%
}]{%
condat2013primal}
\APACinsertmetastar {%
condat2013primal}%
\begin{APACrefauthors}%
Condat, L.%
\end{APACrefauthors}%
\unskip\
\newblock
\APACrefYearMonthDay{2013}{}{}.
\newblock
{\BBOQ}\APACrefatitle {A primal--dual splitting method for convex optimization
  involving Lipschitzian, proximable and linear composite terms} {A
  primal--dual splitting method for convex optimization involving lipschitzian,
  proximable and linear composite terms}.{\BBCQ}
\newblock
\APACjournalVolNumPages{Journal of Optimization Theory and
  Applications}{158}{}{460--479}.
\PrintBackRefs{\CurrentBib}

\bibitem [\protect \citeauthoryear {%
Defazio%
, Bach%
\BCBL {}\ \BBA {} Lacoste-Julien%
}{%
Defazio%
\ \protect \BOthers {.}}{%
{\protect \APACyear {2014}}%
}]{%
defazio2014saga}
\APACinsertmetastar {%
defazio2014saga}%
\begin{APACrefauthors}%
Defazio, A.%
, Bach, F.%
\BCBL {}\ \BBA {} Lacoste-Julien, S.%
\end{APACrefauthors}%
\unskip\
\newblock
\APACrefYearMonthDay{2014}{}{}.
\newblock
{\BBOQ}\APACrefatitle {{SAGA}: A fast incremental gradient method with support
  for non-strongly convex composite objectives} {{SAGA}: A fast incremental
  gradient method with support for non-strongly convex composite
  objectives}.{\BBCQ}
\newblock
\APACjournalVolNumPages{Advances in Neural Information Processing
  Systems}{27}{}{}.
\PrintBackRefs{\CurrentBib}

\bibitem [\protect \citeauthoryear {%
Drori%
, Sabach%
\BCBL {}\ \BBA {} Teboulle%
}{%
Drori%
\ \protect \BOthers {.}}{%
{\protect \APACyear {2015}}%
}]{%
drori2015simple}
\APACinsertmetastar {%
drori2015simple}%
\begin{APACrefauthors}%
Drori, Y.%
, Sabach, S.%
\BCBL {}\ \BBA {} Teboulle, M.%
\end{APACrefauthors}%
\unskip\
\newblock
\APACrefYearMonthDay{2015}{}{}.
\newblock
{\BBOQ}\APACrefatitle {A simple algorithm for a class of nonsmooth
  convex--concave saddle-point problems} {A simple algorithm for a class of
  nonsmooth convex--concave saddle-point problems}.{\BBCQ}
\newblock
\APACjournalVolNumPages{Operations Research Letters}{43}{}{209--214}.
\PrintBackRefs{\CurrentBib}

\bibitem [\protect \citeauthoryear {%
Esser%
, Zhang%
\BCBL {}\ \BBA {} Chan%
}{%
Esser%
\ \protect \BOthers {.}}{%
{\protect \APACyear {2010}}%
}]{%
esser2010general}
\APACinsertmetastar {%
esser2010general}%
\begin{APACrefauthors}%
Esser, E.%
, Zhang, X.%
\BCBL {}\ \BBA {} Chan, T\BPBI F.%
\end{APACrefauthors}%
\unskip\
\newblock
\APACrefYearMonthDay{2010}{}{}.
\newblock
{\BBOQ}\APACrefatitle {A general framework for a class of first order
  primal-dual algorithms for convex optimization in imaging science} {A general
  framework for a class of first order primal-dual algorithms for convex
  optimization in imaging science}.{\BBCQ}
\newblock
\APACjournalVolNumPages{SIAM Journal on Imaging Sciences}{3}{}{1015--1046}.
\PrintBackRefs{\CurrentBib}

\bibitem [\protect \citeauthoryear {%
Friedman%
}{%
Friedman%
}{%
{\protect \APACyear {1991}}%
}]{%
friedman1991multivariate}
\APACinsertmetastar {%
friedman1991multivariate}%
\begin{APACrefauthors}%
Friedman, J.%
\end{APACrefauthors}%
\unskip\
\newblock
\APACrefYearMonthDay{1991}{}{}.
\newblock
{\BBOQ}\APACrefatitle {Multivariate adaptive regression splines (with
  discussion)} {Multivariate adaptive regression splines (with
  discussion)}.{\BBCQ}
\newblock
\APACjournalVolNumPages{Annals of Statistics}{19}{}{79--141}.
\PrintBackRefs{\CurrentBib}

\bibitem [\protect \citeauthoryear {%
Gu%
}{%
Gu%
}{%
{\protect \APACyear {2013}}%
}]{%
gu2013smoothing}
\APACinsertmetastar {%
gu2013smoothing}%
\begin{APACrefauthors}%
Gu, C.%
\end{APACrefauthors}%
\unskip\
\newblock
\APACrefYear{2013}.
\newblock
\APACrefbtitle {Smoothing {S}pline {ANOVA} {M}odels} {Smoothing {S}pline
  {ANOVA} {M}odels}\ (\PrintOrdinal{Second}\ \BEd).
\newblock
\APACaddressPublisher{}{Springer}.
\PrintBackRefs{\CurrentBib}

\bibitem [\protect \citeauthoryear {%
Hastie%
\ \BBA {} Tibshirani%
}{%
Hastie%
\ \BBA {} Tibshirani%
}{%
{\protect \APACyear {1990}}%
}]{%
hastie1990generalized}
\APACinsertmetastar {%
hastie1990generalized}%
\begin{APACrefauthors}%
Hastie, T.%
\BCBT {}\ \BBA {} Tibshirani, R.%
\end{APACrefauthors}%
\unskip\
\newblock
\APACrefYear{1990}.
\newblock
\APACrefbtitle {Generalized {A}dditive {M}odels} {Generalized {A}dditive
  {M}odels}.
\newblock
\APACaddressPublisher{}{Taylor \& Francis}.
\PrintBackRefs{\CurrentBib}

\bibitem [\protect \citeauthoryear {%
Hunter%
\ \BBA {} Lange%
}{%
Hunter%
\ \BBA {} Lange%
}{%
{\protect \APACyear {2004}}%
}]{%
hunter2004tutorial}
\APACinsertmetastar {%
hunter2004tutorial}%
\begin{APACrefauthors}%
Hunter, D\BPBI R.%
\BCBT {}\ \BBA {} Lange, K.%
\end{APACrefauthors}%
\unskip\
\newblock
\APACrefYearMonthDay{2004}{}{}.
\newblock
{\BBOQ}\APACrefatitle {A tutorial on {MM} algorithms} {A tutorial on {MM}
  algorithms}.{\BBCQ}
\newblock
\APACjournalVolNumPages{American Statistician}{58}{}{30--37}.
\PrintBackRefs{\CurrentBib}

\bibitem [\protect \citeauthoryear {%
Koltchinskii%
\ \BBA {} Yuan%
}{%
Koltchinskii%
\ \BBA {} Yuan%
}{%
{\protect \APACyear {2010}}%
}]{%
koltchinskii2010sparsity}
\APACinsertmetastar {%
koltchinskii2010sparsity}%
\begin{APACrefauthors}%
Koltchinskii, V.%
\BCBT {}\ \BBA {} Yuan, M.%
\end{APACrefauthors}%
\unskip\
\newblock
\APACrefYearMonthDay{2010}{}{}.
\newblock
{\BBOQ}\APACrefatitle {Sparsity in multiple kernel learning} {Sparsity in
  multiple kernel learning}.{\BBCQ}
\newblock
\APACjournalVolNumPages{Annals of Statistics}{38}{}{3660--3695}.
\PrintBackRefs{\CurrentBib}

\bibitem [\protect \citeauthoryear {%
Li%
\ \BBA {} Yan%
}{%
Li%
\ \BBA {} Yan%
}{%
{\protect \APACyear {2021}}%
}]{%
li2021new}
\APACinsertmetastar {%
li2021new}%
\begin{APACrefauthors}%
Li, Z.%
\BCBT {}\ \BBA {} Yan, M.%
\end{APACrefauthors}%
\unskip\
\newblock
\APACrefYearMonthDay{2021}{}{}.
\newblock
{\BBOQ}\APACrefatitle {New convergence analysis of a primal-dual algorithm with
  large stepsizes} {New convergence analysis of a primal-dual algorithm with
  large stepsizes}.{\BBCQ}
\newblock
\APACjournalVolNumPages{Advances in Computational Mathematics}{47}{}{1--20}.
\PrintBackRefs{\CurrentBib}

\bibitem [\protect \citeauthoryear {%
Lin%
\ \BBA {} Zhang%
}{%
Lin%
\ \BBA {} Zhang%
}{%
{\protect \APACyear {2006}}%
}]{%
lin2006component}
\APACinsertmetastar {%
lin2006component}%
\begin{APACrefauthors}%
Lin, Y.%
\BCBT {}\ \BBA {} Zhang, H\BPBI H.%
\end{APACrefauthors}%
\unskip\
\newblock
\APACrefYearMonthDay{2006}{}{}.
\newblock
{\BBOQ}\APACrefatitle {Component selection and smoothing in multivariate
  nonparametric regression} {Component selection and smoothing in multivariate
  nonparametric regression}.{\BBCQ}
\newblock
\APACjournalVolNumPages{Annals of Statistics}{34}{}{2272--2297}.
\PrintBackRefs{\CurrentBib}

\bibitem [\protect \citeauthoryear {%
Mammen%
\ \BBA {} Van De~Geer%
}{%
Mammen%
\ \BBA {} Van De~Geer%
}{%
{\protect \APACyear {1997}}%
}]{%
mammen1997locally}
\APACinsertmetastar {%
mammen1997locally}%
\begin{APACrefauthors}%
Mammen, E.%
\BCBT {}\ \BBA {} Van De~Geer, S.%
\end{APACrefauthors}%
\unskip\
\newblock
\APACrefYearMonthDay{1997}{}{}.
\newblock
{\BBOQ}\APACrefatitle {Locally adaptive regression splines} {Locally adaptive
  regression splines}.{\BBCQ}
\newblock
\APACjournalVolNumPages{Annals of Statistics}{25}{}{387--413}.
\PrintBackRefs{\CurrentBib}

\bibitem [\protect \citeauthoryear {%
Meier%
, Van~de Geer%
\BCBL {}\ \BBA {} B{\"u}hlmann%
}{%
Meier%
\ \protect \BOthers {.}}{%
{\protect \APACyear {2009}}%
}]{%
meier2009high}
\APACinsertmetastar {%
meier2009high}%
\begin{APACrefauthors}%
Meier, L.%
, Van~de Geer, S.%
\BCBL {}\ \BBA {} B{\"u}hlmann, P.%
\end{APACrefauthors}%
\unskip\
\newblock
\APACrefYearMonthDay{2009}{}{}.
\newblock
{\BBOQ}\APACrefatitle {High-dimensional additive modeling} {High-dimensional
  additive modeling}.{\BBCQ}
\newblock
\APACjournalVolNumPages{Annals of Statistics}{37}{}{3779--3821}.
\PrintBackRefs{\CurrentBib}

\bibitem [\protect \citeauthoryear {%
Osborne%
, Presnell%
\BCBL {}\ \BBA {} Turlach%
}{%
Osborne%
\ \protect \BOthers {.}}{%
{\protect \APACyear {2000}}%
}]{%
osborne2000new}
\APACinsertmetastar {%
osborne2000new}%
\begin{APACrefauthors}%
Osborne, M\BPBI R.%
, Presnell, B.%
\BCBL {}\ \BBA {} Turlach, B\BPBI A.%
\end{APACrefauthors}%
\unskip\
\newblock
\APACrefYearMonthDay{2000}{}{}.
\newblock
{\BBOQ}\APACrefatitle {A new approach to variable selection in least squares
  problems} {A new approach to variable selection in least squares
  problems}.{\BBCQ}
\newblock
\APACjournalVolNumPages{IMA Journal of Numerical Analysis}{20}{}{389--403}.
\PrintBackRefs{\CurrentBib}

\bibitem [\protect \citeauthoryear {%
Parikh%
\ \BBA {} Boyd%
}{%
Parikh%
\ \BBA {} Boyd%
}{%
{\protect \APACyear {2014}}%
}]{%
parikh2014proximal}
\APACinsertmetastar {%
parikh2014proximal}%
\begin{APACrefauthors}%
Parikh, N.%
\BCBT {}\ \BBA {} Boyd, S.%
\end{APACrefauthors}%
\unskip\
\newblock
\APACrefYearMonthDay{2014}{}{}.
\newblock
{\BBOQ}\APACrefatitle {Proximal algorithms} {Proximal algorithms}.{\BBCQ}
\newblock
\APACjournalVolNumPages{Foundations and Trends in Optimization}{1}{}{127--239}.
\PrintBackRefs{\CurrentBib}

\bibitem [\protect \citeauthoryear {%
Petersen%
, Witten%
\BCBL {}\ \BBA {} Simon%
}{%
Petersen%
\ \protect \BOthers {.}}{%
{\protect \APACyear {2016}}%
}]{%
petersen2016fused}
\APACinsertmetastar {%
petersen2016fused}%
\begin{APACrefauthors}%
Petersen, A.%
, Witten, D.%
\BCBL {}\ \BBA {} Simon, N.%
\end{APACrefauthors}%
\unskip\
\newblock
\APACrefYearMonthDay{2016}{}{}.
\newblock
{\BBOQ}\APACrefatitle {Fused lasso additive model} {Fused lasso additive
  model}.{\BBCQ}
\newblock
\APACjournalVolNumPages{Journal of Computational and Graphical
  Statistics}{25}{}{1005--1025}.
\PrintBackRefs{\CurrentBib}

\bibitem [\protect \citeauthoryear {%
Radchenko%
\ \BBA {} James%
}{%
Radchenko%
\ \BBA {} James%
}{%
{\protect \APACyear {2010}}%
}]{%
radchenko2010variable}
\APACinsertmetastar {%
radchenko2010variable}%
\begin{APACrefauthors}%
Radchenko, P.%
\BCBT {}\ \BBA {} James, G\BPBI M.%
\end{APACrefauthors}%
\unskip\
\newblock
\APACrefYearMonthDay{2010}{}{}.
\newblock
{\BBOQ}\APACrefatitle {Variable selection using adaptive nonlinear interaction
  structures in high dimensions} {Variable selection using adaptive nonlinear
  interaction structures in high dimensions}.{\BBCQ}
\newblock
\APACjournalVolNumPages{Journal of the American Statistical
  Association}{105}{}{1541--1553}.
\PrintBackRefs{\CurrentBib}

\bibitem [\protect \citeauthoryear {%
Raskutti%
, Wainwright%
\BCBL {}\ \BBA {} Yu%
}{%
Raskutti%
\ \protect \BOthers {.}}{%
{\protect \APACyear {2012}}%
}]{%
raskutti2012minimax}
\APACinsertmetastar {%
raskutti2012minimax}%
\begin{APACrefauthors}%
Raskutti, G.%
, Wainwright, M\BPBI J.%
\BCBL {}\ \BBA {} Yu, B.%
\end{APACrefauthors}%
\unskip\
\newblock
\APACrefYearMonthDay{2012}{}{}.
\newblock
{\BBOQ}\APACrefatitle {Minimax-Optimal Rates For Sparse Additive Models Over
  Kernel Classes Via Convex Programming} {Minimax-optimal rates for sparse
  additive models over kernel classes via convex programming}.{\BBCQ}
\newblock
\APACjournalVolNumPages{Journal of Machine Learning Research}{13}{}{}.
\PrintBackRefs{\CurrentBib}

\bibitem [\protect \citeauthoryear {%
Ravikumar%
, Lafferty%
, Liu%
\BCBL {}\ \BBA {} Wasserman%
}{%
Ravikumar%
\ \protect \BOthers {.}}{%
{\protect \APACyear {2009}}%
}]{%
ravikumar2009sparse}
\APACinsertmetastar {%
ravikumar2009sparse}%
\begin{APACrefauthors}%
Ravikumar, P.%
, Lafferty, J.%
, Liu, H.%
\BCBL {}\ \BBA {} Wasserman, L.%
\end{APACrefauthors}%
\unskip\
\newblock
\APACrefYearMonthDay{2009}{}{}.
\newblock
{\BBOQ}\APACrefatitle {Sparse additive models} {Sparse additive models}.{\BBCQ}
\newblock
\APACjournalVolNumPages{Journal of the Royal Statistical Society, Series
  B}{71}{}{1009--1030}.
\PrintBackRefs{\CurrentBib}

\bibitem [\protect \citeauthoryear {%
Roux%
, Schmidt%
\BCBL {}\ \BBA {} Bach%
}{%
Roux%
\ \protect \BOthers {.}}{%
{\protect \APACyear {2012}}%
}]{%
roux2012stochastic}
\APACinsertmetastar {%
roux2012stochastic}%
\begin{APACrefauthors}%
Roux, N.%
, Schmidt, M.%
\BCBL {}\ \BBA {} Bach, F.%
\end{APACrefauthors}%
\unskip\
\newblock
\APACrefYearMonthDay{2012}{}{}.
\newblock
{\BBOQ}\APACrefatitle {A stochastic gradient method with an exponential
  convergence rate for finite training sets} {A stochastic gradient method with
  an exponential convergence rate for finite training sets}.{\BBCQ}
\newblock
\APACjournalVolNumPages{Advances in Neural Information Processing
  Systems}{25}{}{}.
\PrintBackRefs{\CurrentBib}

\bibitem [\protect \citeauthoryear {%
Ryu%
\ \BBA {} Yin%
}{%
Ryu%
\ \BBA {} Yin%
}{%
{\protect \APACyear {2022}}%
}]{%
ryu2022large}
\APACinsertmetastar {%
ryu2022large}%
\begin{APACrefauthors}%
Ryu, E\BPBI K.%
\BCBT {}\ \BBA {} Yin, W.%
\end{APACrefauthors}%
\unskip\
\newblock
\APACrefYearMonthDay{2022}{}{}.
\newblock
\APACrefbtitle {Large-{S}cale {C}onvex {O}ptimization via {M}onotone
  {O}perators.} {Large-{S}cale {C}onvex {O}ptimization via {M}onotone
  {O}perators.}
\newblock
\APACaddressPublisher{}{Cambridge University Press}.
\PrintBackRefs{\CurrentBib}

\bibitem [\protect \citeauthoryear {%
Stone%
}{%
Stone%
}{%
{\protect \APACyear {1986}}%
}]{%
stone1986dimensionality}
\APACinsertmetastar {%
stone1986dimensionality}%
\begin{APACrefauthors}%
Stone, C\BPBI J.%
\end{APACrefauthors}%
\unskip\
\newblock
\APACrefYearMonthDay{1986}{}{}.
\newblock
{\BBOQ}\APACrefatitle {The dimensionality reduction principle for generalized
  additive models} {The dimensionality reduction principle for generalized
  additive models}.{\BBCQ}
\newblock
\APACjournalVolNumPages{Annals of Statistics}{}{}{590--606}.
\PrintBackRefs{\CurrentBib}

\bibitem [\protect \citeauthoryear {%
Tan%
\ \BBA {} Zhang%
}{%
Tan%
\ \BBA {} Zhang%
}{%
{\protect \APACyear {2019}}%
}]{%
tan2019doubly}
\APACinsertmetastar {%
tan2019doubly}%
\begin{APACrefauthors}%
Tan, Z.%
\BCBT {}\ \BBA {} Zhang, C\BHBI H.%
\end{APACrefauthors}%
\unskip\
\newblock
\APACrefYearMonthDay{2019}{}{}.
\newblock
{\BBOQ}\APACrefatitle {Doubly penalized estimation in additive regression with
  high-dimensional data} {Doubly penalized estimation in additive regression
  with high-dimensional data}.{\BBCQ}
\newblock
\APACjournalVolNumPages{Annals of Statistics}{47}{}{2567--2600}.
\PrintBackRefs{\CurrentBib}

\bibitem [\protect \citeauthoryear {%
Tseng%
}{%
Tseng%
}{%
{\protect \APACyear {1988}}%
}]{%
tseng1988coordinate}
\APACinsertmetastar {%
tseng1988coordinate}%
\begin{APACrefauthors}%
Tseng, P.%
\end{APACrefauthors}%
\unskip\
\newblock
\APACrefYearMonthDay{1988}{}{}.
\newblock
{\BBOQ}\APACrefatitle {Coordinate ascent for maximizing nondifferentiable
  concave functions} {Coordinate ascent for maximizing nondifferentiable
  concave functions}.{\BBCQ}
\newblock
\APACjournalVolNumPages{Technical Report LIDS-P-1840, MIT}{}{}{}.
\PrintBackRefs{\CurrentBib}

\bibitem [\protect \citeauthoryear {%
Tseng%
}{%
Tseng%
}{%
{\protect \APACyear {1991}}%
}]{%
tseng1991applications}
\APACinsertmetastar {%
tseng1991applications}%
\begin{APACrefauthors}%
Tseng, P.%
\end{APACrefauthors}%
\unskip\
\newblock
\APACrefYearMonthDay{1991}{}{}.
\newblock
{\BBOQ}\APACrefatitle {Applications of a splitting algorithm to decomposition
  in convex programming and variational inequalities} {Applications of a
  splitting algorithm to decomposition in convex programming and variational
  inequalities}.{\BBCQ}
\newblock
\APACjournalVolNumPages{SIAM Journal on Control and
  Optimization}{29}{}{119--138}.
\PrintBackRefs{\CurrentBib}

\bibitem [\protect \citeauthoryear {%
V{\~u}%
}{%
V{\~u}%
}{%
{\protect \APACyear {2013}}%
}]{%
vu2013splitting}
\APACinsertmetastar {%
vu2013splitting}%
\begin{APACrefauthors}%
V{\~u}, B\BPBI C.%
\end{APACrefauthors}%
\unskip\
\newblock
\APACrefYearMonthDay{2013}{}{}.
\newblock
{\BBOQ}\APACrefatitle {A splitting algorithm for dual monotone inclusions
  involving cocoercive operators} {A splitting algorithm for dual monotone
  inclusions involving cocoercive operators}.{\BBCQ}
\newblock
\APACjournalVolNumPages{Advances in Computational Mathematics}{38}{}{667--681}.
\PrintBackRefs{\CurrentBib}

\bibitem [\protect \citeauthoryear {%
Wahba%
, Wang%
, Gu%
, Klein%
\BCBL {}\ \BBA {} Klein%
}{%
Wahba%
\ \protect \BOthers {.}}{%
{\protect \APACyear {1995}}%
}]{%
wahba1995smoothing}
\APACinsertmetastar {%
wahba1995smoothing}%
\begin{APACrefauthors}%
Wahba, G.%
, Wang, Y.%
, Gu, C.%
, Klein, R.%
\BCBL {}\ \BBA {} Klein, B.%
\end{APACrefauthors}%
\unskip\
\newblock
\APACrefYearMonthDay{1995}{}{}.
\newblock
{\BBOQ}\APACrefatitle {Smoothing spline {ANOVA} for exponential families, with
  application to the {W}isconsin {E}pidemiological {S}tudy of {D}iabetic
  {R}etinopathy} {Smoothing spline {ANOVA} for exponential families, with
  application to the {W}isconsin {E}pidemiological {S}tudy of {D}iabetic
  {R}etinopathy}.{\BBCQ}
\newblock
\APACjournalVolNumPages{Annals of Statistics}{23}{}{1865--1895}.
\PrintBackRefs{\CurrentBib}

\bibitem [\protect \citeauthoryear {%
Yang%
\ \BBA {} Tan%
}{%
Yang%
\ \BBA {} Tan%
}{%
{\protect \APACyear {2018}}%
}]{%
yang2018backfitting}
\APACinsertmetastar {%
yang2018backfitting}%
\begin{APACrefauthors}%
Yang, T.%
\BCBT {}\ \BBA {} Tan, Z.%
\end{APACrefauthors}%
\unskip\
\newblock
\APACrefYearMonthDay{2018}{}{}.
\newblock
{\BBOQ}\APACrefatitle {Backfitting algorithms for total-variation and
  empirical-norm penalized additive modelling with high-dimensional data}
  {Backfitting algorithms for total-variation and empirical-norm penalized
  additive modelling with high-dimensional data}.{\BBCQ}
\newblock
\APACjournalVolNumPages{Stat}{7}{}{e198}.
\PrintBackRefs{\CurrentBib}

\bibitem [\protect \citeauthoryear {%
Yang%
\ \BBA {} Tan%
}{%
Yang%
\ \BBA {} Tan%
}{%
{\protect \APACyear {2021}}%
}]{%
yang2021hierarchical}
\APACinsertmetastar {%
yang2021hierarchical}%
\begin{APACrefauthors}%
Yang, T.%
\BCBT {}\ \BBA {} Tan, Z.%
\end{APACrefauthors}%
\unskip\
\newblock
\APACrefYearMonthDay{2021}{}{}.
\newblock
{\BBOQ}\APACrefatitle {Hierarchical total variations and doubly penalized
  {ANOVA} modeling for multivariate nonparametric regression} {Hierarchical
  total variations and doubly penalized {ANOVA} modeling for multivariate
  nonparametric regression}.{\BBCQ}
\newblock
\APACjournalVolNumPages{Journal of Computational and Graphical
  Statistics}{30}{}{848--862}.
\PrintBackRefs{\CurrentBib}

\bibitem [\protect \citeauthoryear {%
Zhang%
\ \BBA {} Xiao%
}{%
Zhang%
\ \BBA {} Xiao%
}{%
{\protect \APACyear {2017}}%
}]{%
zhang2017stochastic}
\APACinsertmetastar {%
zhang2017stochastic}%
\begin{APACrefauthors}%
Zhang, Y.%
\BCBT {}\ \BBA {} Xiao, L.%
\end{APACrefauthors}%
\unskip\
\newblock
\APACrefYearMonthDay{2017}{}{}.
\newblock
{\BBOQ}\APACrefatitle {Stochastic Primal-Dual Coordinate Method for Regularized
  Empirical Risk Minimization} {Stochastic primal-dual coordinate method for
  regularized empirical risk minimization}.{\BBCQ}
\newblock
\APACjournalVolNumPages{Journal of Machine Learning Research}{18}{}{1--42}.
\PrintBackRefs{\CurrentBib}

\end{thebibliography}

\clearpage

\setcounter{page}{1}

\setcounter{section}{0}
\setcounter{equation}{0}

\setcounter{figure}{0}
\setcounter{table}{0}
\setcounter{lemma}{0}

\setcounter{algorithm}{0}
\setcounter{pro}{0}

\renewcommand{\theequation}{S\arabic{equation}}
\renewcommand{\thesection}{\Roman{section}}

\renewcommand\thefigure{S\arabic{figure}}
\renewcommand\thetable{S\arabic{table}}

\renewcommand\thealgorithm{S\arabic{algorithm}}
\renewcommand\thepro{S\arabic{pro}}

\begin{center}
{\Large Supplementary Material for

``Block-wise Primal-dual Algorithms for Large-scale Doubly Penalized ANOVA Modeling"}

\vspace{.1in} {\large Penghui Fu and Zhiqiang Tan}
\end{center}

\section{Three-operator splitting method}\label{sec:additional discussions}

Condat--V\~u (Condat, 2013; V\~u, 2013) 
is a three-operator splitting method which can be viewed as a generalization of the Chambolle--Pock and proximal gradient method. Consider an optimization problem in the form
\begin{equation}\label{eq:generic 3-operator}
    \min_\beta \, \tilde h(\beta) + \tilde g(\beta)  + \tilde f(X\beta),
\end{equation}
where $\tilde h$, $\tilde g$, and $\tilde f$ are CCP functions, and $\tilde h$ is $L$-smooth. The primal-dual Lagrangian associated with
(\ref{eq:generic 3-operator}) is
\begin{equation}\label{eq:primal-dual lagrangian of 3-operator}
   L(\beta; u) = \tilde{h}(\beta) + \tilde g(\beta) + \langle u, X\beta \rangle - \tilde{f}^*(u).
\end{equation}
Then the dual problem of (\ref{eq:generic 3-operator}) is $\max_{u} \, -(\tilde{h} + \tilde{g})^*(-X^\T u) - \tilde{f}^*(u)$. For solving (\ref{eq:generic 3-operator}), Condat--V\~u iterations are defined as follows: for $k\geq 0$,
\begin{equation}\label{eq:Condat-Vu}
    \begin{split}
        \beta^{k+1} &= \prox_{\tau\tilde{g}/n}\left(\beta^k-\frac{\tau}{n}(X^\T u^k+\nabla \tilde{h}(\beta^k))\right), \\
        u^{k+1} &= \prox_{\alpha \tilde{f}^*}\left(u^k+\alpha X(2\beta^{k+1}-\beta^k)\right).
    \end{split}
\end{equation}
If $\tilde h=0$, then (\ref{eq:Condat-Vu}) becomes the Chambolle--Pock method (\ref{eq:CP}). If $\tilde f=0$, then (\ref{eq:Condat-Vu}) becomes the proximal gradient method (\ref{eq:proximal gradient method}), because $\tilde f^*(u)=\delta_{\{0\}}(u)$ and consequently $u^{k+1}=0$. Therefore, Condat--V\~u generalizes the CP and proximal gradient method. Assuming that total duality holds, $\alpha>0$,  $\tau>0$, and $(\alpha+1/2)\tau L<n$, then the iterates $(\beta^{k+1},u^{k+1})$ in (\ref{eq:Condat-Vu}) converge to a saddle point of (\ref{eq:primal-dual lagrangian of 3-operator}) (Ryu \& Yin, 2022). 

To apply Condat--V\~u, we reformulate the single-block problem (\ref{eq:rescaled single-block}) as
\begin{equation}\label{eq:single-block for 3-operator splitting}
    \min_{\beta} \, \underbrace{\frac{1}{2}\|r-X\beta\|_2^2}_{\tilde{h}(\beta)} + \underbrace{n\|\Gamma\beta\|_1}_{\tilde g(\beta)} + \underbrace{\lambda\sqrt{n} \|X\beta\|_2}_{\tilde{f}(X\beta)},
\end{equation}
where $\tilde{h}(\beta)=\|r-X\beta\|_2^2/2$ is $\|X\|_2^2$-smooth, $\tilde g(\beta)=n\|\Gamma\beta\|_1$, and $\tilde{f}(z)=\lambda\sqrt{n} \|z\|_2$ for $z\in\bbR^n$.
The conjugate $\tilde{f}^*(u)$ can be calculated in a closed form as $\tilde{f}^*(u)=\delta(\|u\|_2\leq \lambda\sqrt{n})$. Applying Condat--V\~u (\ref{eq:Condat-Vu}) to problem (\ref{eq:single-block for 3-operator splitting}), we obtain
\begin{subequations}\label{eq:batch Condat-Vu}
\begin{align}
    \beta^{k+1} &= \prox_{\tau\tilde{g}/n}\left(\beta^k-\frac{\tau}{n}(X^\T u^k+\nabla \tilde{h}(\beta^k))\right) \\
    &= \mathcal{S}\left(\beta^k-\frac{\tau}{n}X^\T(u^k+X\beta^k-r), \tau\diag(\Gamma)\right), \\
    u^{k+1} &= \prox_{\alpha \tilde{f}^*}\left(u^k+\alpha X(2\beta^{k+1}-\beta^k)\right)\label{subeq:dual update in Condat-Vu} \\
    &= \prod\nolimits_{\{z\colon\|z\|_2\leq \lambda\sqrt{n}\}}(u^k+\alpha X(2\beta^{k+1}-\beta^k)),\label{subeq:dual formula in Condat-Vu}
\end{align}
\end{subequations}
where $\prod_{\{z\colon\|z\|_2\leq \lambda\sqrt{n}\}}$ is the orthogonal projection onto the set $\{z\in\bbR^n\colon \|z\|_2\leq \lambda\sqrt{n}\}$. By the preceding discussion, the iterates  $(\beta^{k+1},u^{k+1})$ in (\ref{eq:batch Condat-Vu}) converge to a saddle point of (\ref{eq:primal-dual lagrangian of 3-operator}) provided that $\alpha>0$, $\tau>0$, and $(\alpha+1/2)\tau\|X\|_2^2<n$.

While the batch algorithm (\ref{eq:batch Condat-Vu}) is tractable, a proper randomization of the algorithm seems to be difficult. The randomization techniques used in Sections~\ref{subsec:stoc-CP} and \ref{subsec:stoc-linear-AMA} are not applicable for deriving a randomized version of the dual update (\ref{subeq:dual formula in Condat-Vu}), which will be denoted as $u^{k+1}_{\text{B}}$. On one hand, $\tilde{f}^*$ above is not separable, so that evaluating one coordinate of $u^{k+1}_{\text{B}}$ has a cost of $O(nd)$, the same as that of evaluating the full update. The randomized coordinate update which is unbiased for $u^{k+1}_{\text{B}}$ as in stochastic linearized AMA is infeasible. On the other hand, $\tilde{f}^*$ is not differentiable either. Hence a randomized coordinate minimization update as used in stochastic CP may not be valid (Tseng, 1988). 

\section{Concave conjugate method}\label{sec:cc}
In Sections~\ref{subsec:PD-single block} and \ref{subsec:stoc-PD},
the single-block optimization problem (\ref{eq:rescaled single-block}) is solved through a saddle-point problem by using primal-dual methods.
In fact, we replace $f(X\beta)=(1/2)\|r-X\beta\|^2_2+\lambda\sqrt{n}\|X\beta\|_2$ by its Fenchel conjugate
$$
f(X\beta) = \max_{u}\,\langle u,X\beta \rangle - f^*(u),
$$
and keep $g(\beta)=n\|\Gamma\beta\|_1$ unchanged. Then the original problem (\ref{eq:rescaled single-block})
is transformed to the min-max (i.e., saddle-point) problem of the Lagrangian
$$
\min_{\beta}\max_{u}\,L(\beta,u)=g(\beta)+ \langle u,X\beta \rangle - f^*(u).
$$
In this section, we present a different approach by considering the conjugate of the square root, which is a concave function.
This transforms (\ref{eq:rescaled single-block}) to a min-inf problem, in contrast with the min-max problem using primal-dual methods.

\subsection{Concave conjugate and MM}
By Fenchel duality on the square root function, we have
\begin{equation}\label{eq:conjugate of square root}
    \sqrt{x} = \inf_{u<0}\, (\psi^*(u) - ux), \quad x\geq 0,
\end{equation}
where $\psi^*(u)=-1/(4u)$ if $u<0$ and $\infty$ otherwise. The optimal $u(x)$ achieving the infimum is $-1/(2\sqrt{x})$ if $x>0$ and $-\infty$ if $x=0$. By (\ref{eq:conjugate of square root}), problem (\ref{eq:single-block}) is equivalent to
\begin{equation}\label{eq: cc reformulation}
    \min_{\beta}\inf_{u<0} \, \frac{1}{2n}\|r-X\beta\|_2^2 + \|\Gamma\beta\|_1 + \lambda (\psi^*(u)-u\|X\beta\|_n^2).
\end{equation}
Naturally, we consider alternately minimizing or decreasing the objective in (\ref{eq: cc reformulation}) with respect to
$u$ and $\beta$. For fixed $\beta\neq 0$, the optimal $u$ is given by
$\hat u(\beta) = -1/(2\|X\beta\|_n)$. For fixed $u$, minimization over $\beta$ is equivalent to, after rescaling by $1/(1-2\lambda u)>0$,
\begin{equation}\label{eq:cc-lasso}
   \min_{\beta} \, \frac{1}{2n}\|X\beta - \frac{1}{1-2\lambda u}r\|_2^2 + \frac{1}{1-2\lambda u} \|\Gamma\beta\|_1.
\end{equation}
Note that (\ref{eq:cc-lasso}) is a Lasso problem similar to (\ref{eq:single-block lasso}), but with $r$ and $\Gamma$ rescaled.
By the min-inf structure of (\ref{eq: cc reformulation}), any update of $\beta$ decreasing the objective in (\ref{eq:cc-lasso})
with $u = \hat u(\beta^k )$ given the current $\beta^k$ can be shown to also reduce the objective in the original problem (\ref{eq:single-block}). For example, consider a proximal gradient update as in (\ref{eq:proximal gradient method})
\begin{equation}\label{eq:batch-cc}
    \beta^{k+1}=\mathcal{S}\left(\beta^k-\frac{\tau}{n}X^\T(X\beta^k-\frac{1}{1-2\lambda u}r), \frac{\tau}{1-2\lambda u}\cdot\diag(\Gamma)\right).
\end{equation}
For $0<\tau\leq n/\|X\|_2^2 $, this update is guaranteed to decrease the objective in (\ref{eq:cc-lasso}) and hence also the objective in problem (\ref{eq:single-block}) (Parikh \& Boyd, 2014). 

The alternate updating procedure above will be called concave conjugate (CC), to reflect our original motivation.
Interestingly, the procedure can also be interpreted as an MM algorithm (Hunter \& Lange, 2004). 
The basic idea of MM is to replace an objective function $h(\beta)$ by a suitable surrogate $\tilde{h}(\beta, \beta^k)$ at the current $\beta^k$. The surrogate is called a majorization if
the surrogate contacts the original objective at $\beta^k$, $\tilde{h}(\beta^k, \beta^k)=h(\beta^k)$, and is an upper bound, $\tilde{h}(\beta, \beta^k)\geq h(\beta)$ for all $\beta$. If the surrogate is a majorization, then any $\beta^{k+1}$ that decreases $\tilde{h}(\beta, \beta^k)$ will also decrease $h$, i.e., $h(\beta^{k+1})\leq h(\beta^k)$.

\begin{figure}[ht]
   \centering
   \vspace{-\baselineskip}
   \begin{subfigure}{0.49\textwidth}
      \includegraphics[width=\textwidth]{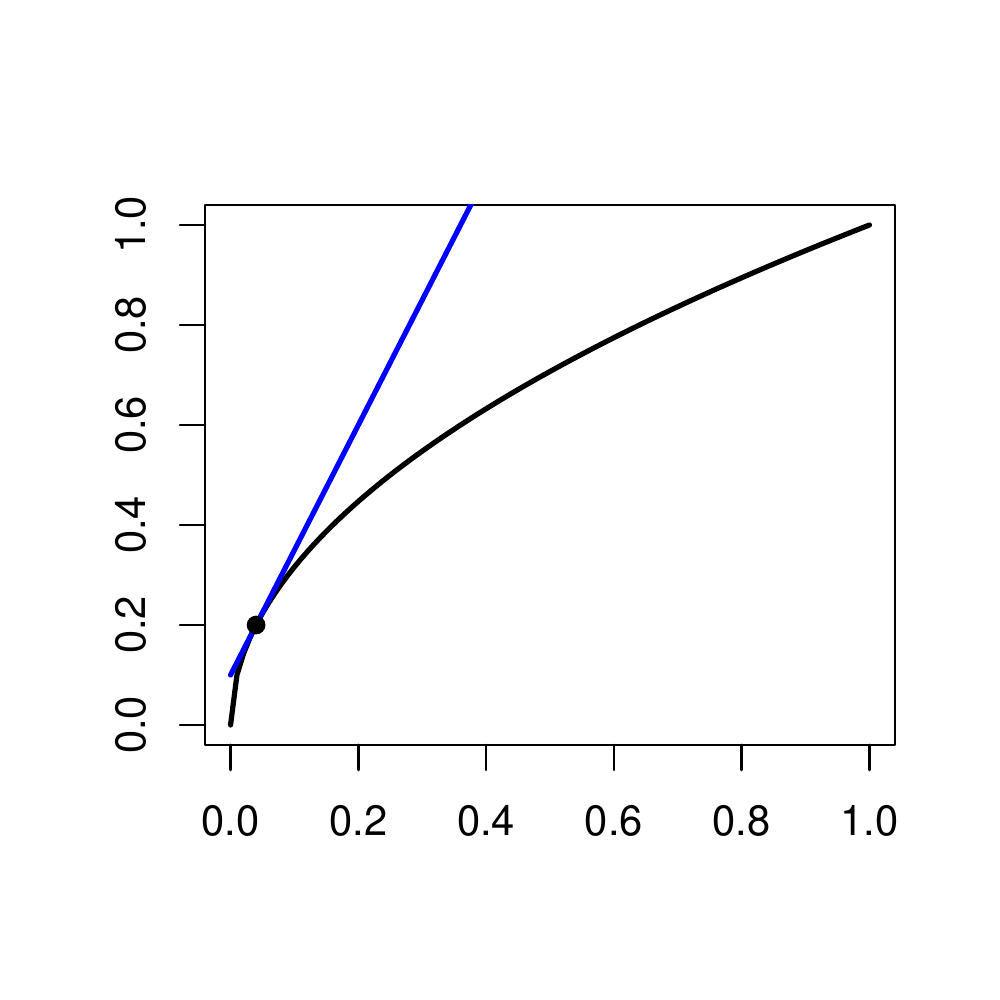}
   \end{subfigure}
   \hfill
   \begin{subfigure}{0.49\textwidth}
      \includegraphics[width=\textwidth]{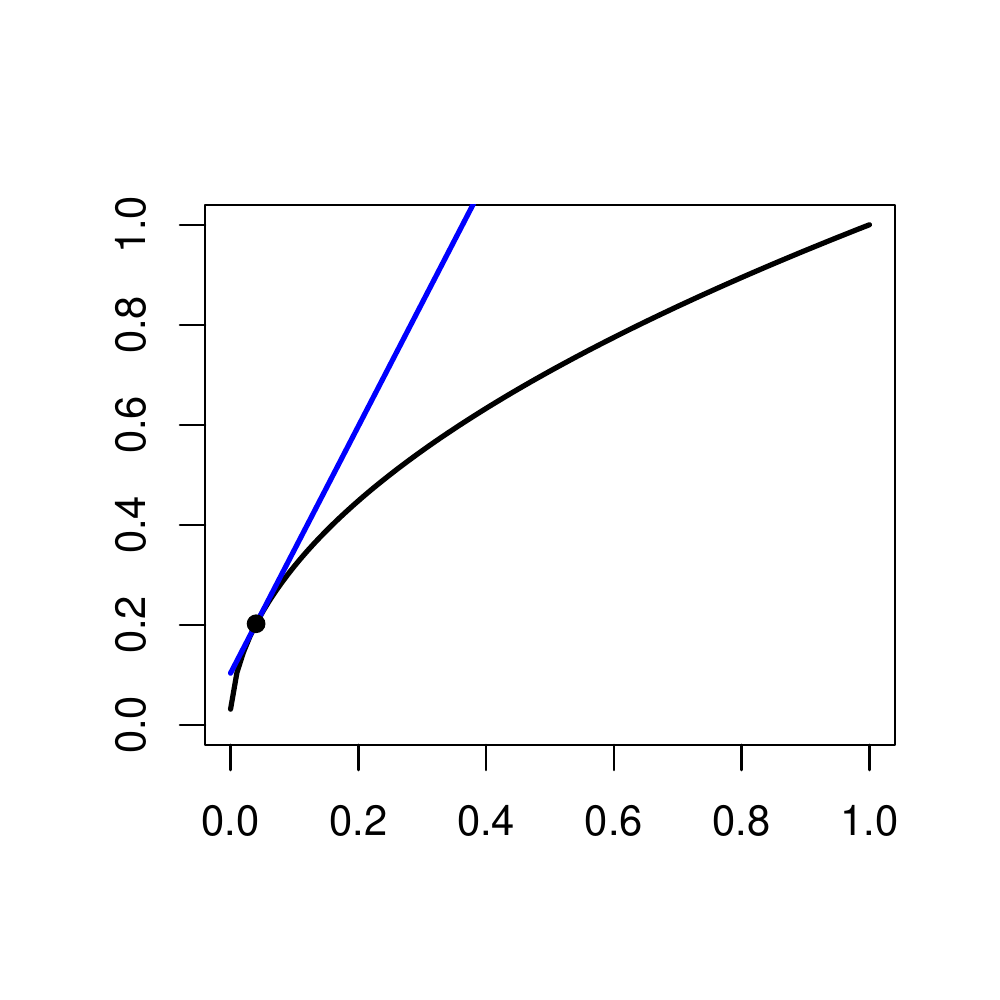}
   \end{subfigure}
   \vspace{-.5in}
   \caption{Linear majorizations of $\sqrt{x}$ and $\sqrt{x+0.001}$ at $x=0.04$.}
   \label{fig:sqrt}
\end{figure}

\begin{figure}[hb]
    \centering
    \vspace{-\baselineskip}
    \begin{subfigure}{0.49\textwidth}
       \includegraphics[width=\textwidth]{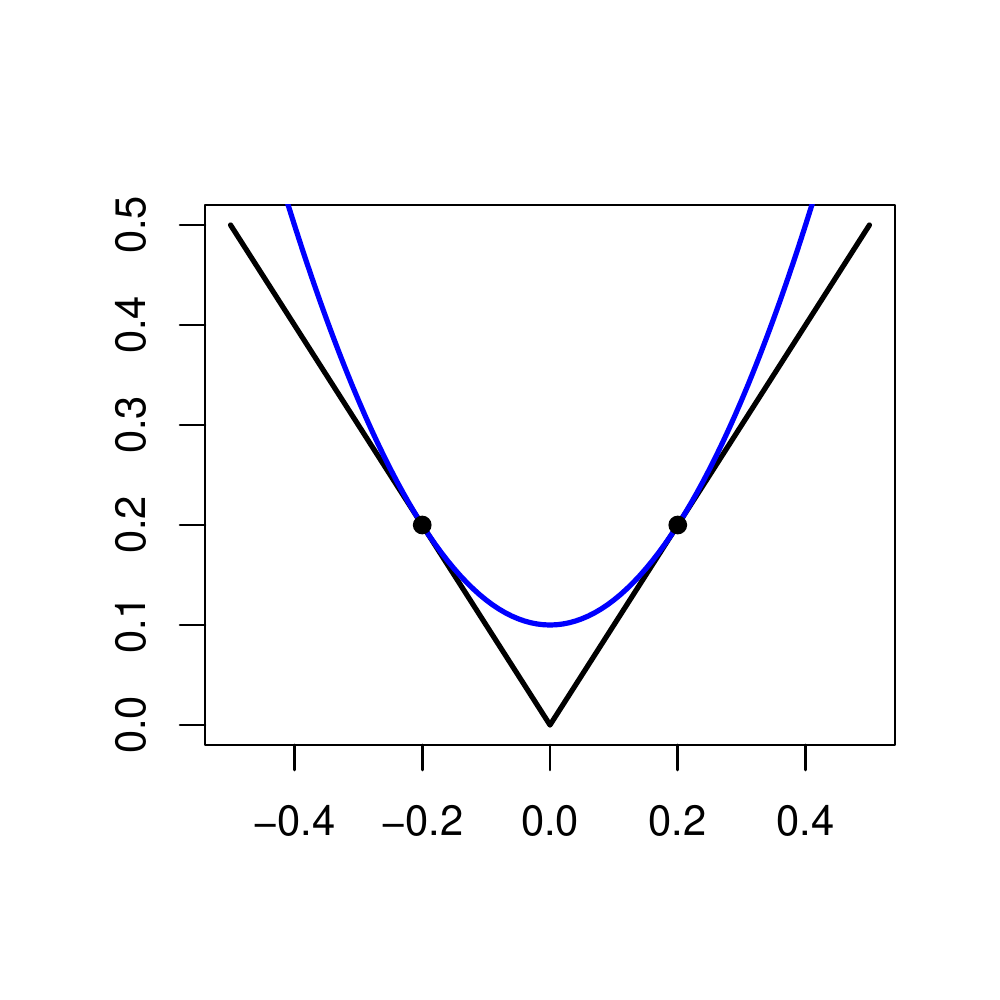}
    \end{subfigure}
    \hfill
    \begin{subfigure}{0.49\textwidth}
       \includegraphics[width=\textwidth]{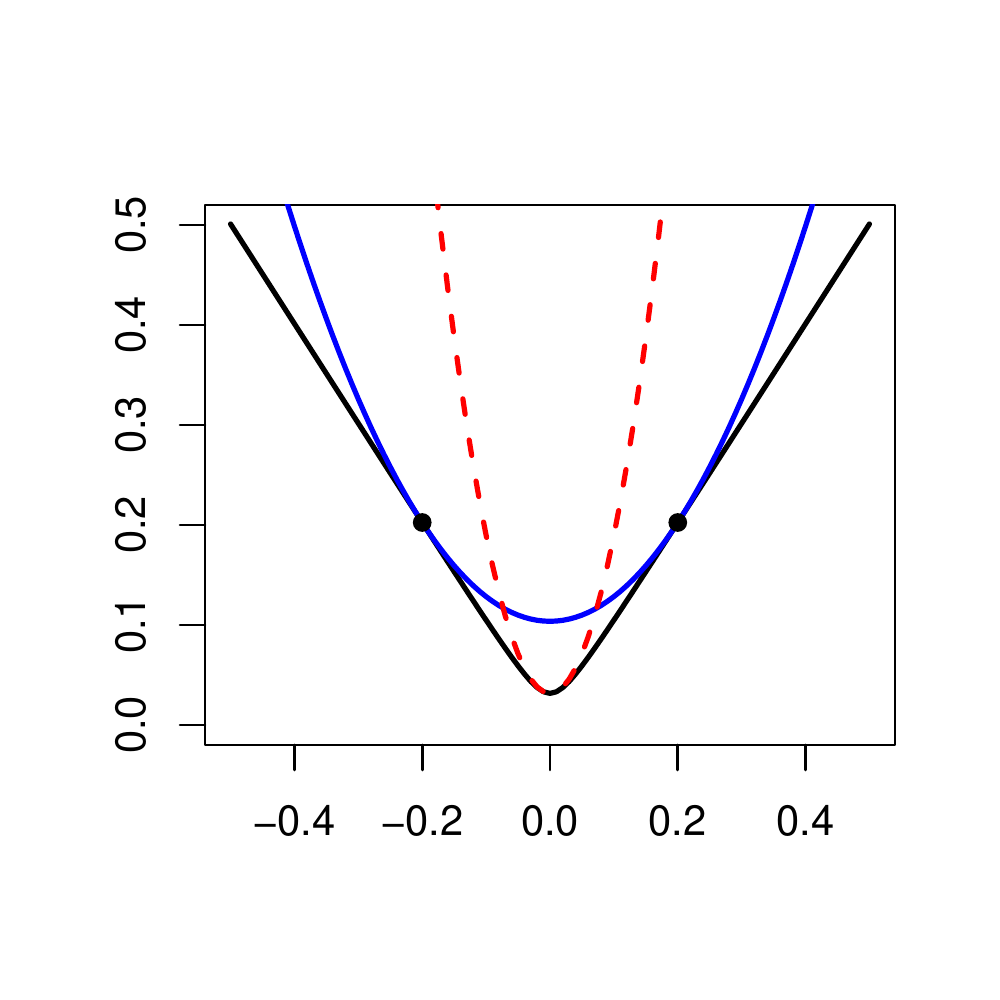}
    \end{subfigure}
    \vspace{-.5in}
    \caption{Quadratic majorizations of $|t|$ and $\sqrt{t^2+0.001}$ at $t=0.2$. The red dotted curve is the quadratic majorization of $\sqrt{t^2+0.001}$ at $t=0$.}
    \label{fig:abs}
\end{figure}

The objective in (\ref{eq: cc reformulation}) with $u=\hat u(\beta^k)$ can be shown to provide
a majorization function at $\beta^k$ for the objective in the original problem (\ref{eq:single-block}).
In fact, a majorization function for $\|X\beta\|_n$ is constructed as follows:
\begin{align*}
    \|X\beta\|_n & \leq \Big\{ \psi^*(u)-u\|X\beta\|_n^2 \Big\} \Big|_{ u=\hat u(\beta^k) } \\
    & =\frac{1}{2\sqrt{\|X\beta^k\|_n^2}}(\|X\beta\|_n^2 - \|X\beta^k\|_n^2) + \sqrt{\|X\beta^k\|_n^2}.
\end{align*}
For $x = \|X\beta\|_n^2 $,
the preceding display can be stated directly as a linear majorization of the square root function $\sqrt{x}$ at $x^k=\|X\beta^k\|_n^2$:
\begin{align}
 \sqrt{x} \le \frac{1}{2 \sqrt{x^k}} (x-x^k) + \sqrt{x^k}, \label{eq:sqrt-major}
\end{align}
See Figure~\ref{fig:sqrt} for an illustration. Equivalently, (\ref{eq:sqrt-major}) can be expressed as a quadratic majorization of the absolute function $|t|$
at some $t^k$:
\begin{align}
 |t| \le \frac{1}{2 |t^k| } (t^2 -(t^k)^2) + |t^k|. \label{eq:abs-major}
\end{align}
See Figure~\ref{fig:abs} for an illustration.
Quadratic majorizations related to (\ref{eq:abs-major}) were used in Hunter \& Li (2005) 
to derive MM algorithms for handling non-concave penalties.

A major limitation of CC or MM discussed above is that the procedure would break down when $\beta^k=0$.
From the CC perspective, if $\beta^k=0$, the associated $\hat u(\beta^k)$ becomes  $-\infty$, so that
$\beta^{k+1}$ would stay at $0$, even when the solution for $\beta$ is nonzero.
From the MM perspective, the majorization in (\ref{eq:sqrt-major}) or (\ref{eq:abs-major}) becomes improper if
$x^k = 0$ or $t^k=0$.

To address this issue, a possible approach is to add a small perturbation to the empirical norm, that is,
modifying the objective in (\ref{eq:single-block}) to
\begin{align}
\frac{1}{2n}\|r-X\beta\|_2^2 + \|\Gamma\beta\|_1 + \lambda \sqrt{\|X\beta\|_n^2+\delta},  \label{eq:perturbed-obj}
\end{align}
for some small $\delta>0$.
For the alternate updating procedure, the $u$ step leads to $\hat u_\delta (\beta) =-1/(2\sqrt{\|X\beta\|_n^2+\delta})>-\infty$.
The proximal gradient update for $\beta$ remains the same as (\ref{eq:batch-cc}), but with $u$ set to $\hat u_\delta (\beta)$.
As illustrated in Figures \ref{fig:abs} and \ref{fig:sqrt}, the corresponding majorization can be stated as
\begin{align*}
 \sqrt{x + \delta} \le \frac{1}{2 \sqrt{x^k + \delta}} (x-x^k) + \sqrt{x^k+\delta},
\end{align*}
for $x = \|X\beta\|_n^2 $, or equivalently
\begin{align*}
 \sqrt{t^2 + \delta} \le \frac{1}{2 \sqrt{(t^k)^2 + \delta}} (t^2 - (t^k)^2)  + \sqrt{ (t^k)^2+\delta}.
\end{align*}
for $t = \pm \|X\beta\|_n $.
The resulting CC algorithm is presented in Algorithm~\ref{alg:batch-CC}. For easy comparison with other algorithms,the updates are also expressed as (\ref{eq:batch cc-tau}) and (\ref{eq:batch cc-beta}).
The step size $\tau^k$ in (\ref{eq:batch cc-tau})
corresponds to the choice $\tau=n/\|X\|_2^2$ in (\ref{eq:batch-cc}) to guarantee descent of the objective, and the
$\beta$-update in (\ref{eq:batch cc-beta}) is re-expressed from (\ref{eq:batch-cc}) with $u=\hat u_\delta (\beta^k)$ absorbed.

A stochastic version of batch CC is presented in Algorithm~\ref{alg:stoc-CC}, with outer and inner loops.
Given the current $\beta^k$ in the outer loop,
$u$ is updated as $\hat u _\delta (\beta^k)$ and then
SAGA is applied in an inner loop to the proximal gradient update (\ref{eq:batch-cc}) for $\beta$,
with $u=\hat u _\delta (\beta^k)$ fixed.
Each inner iteration costs $O(d)$. Therefore, one outer iteration
with an inner loop of $n$ iterations costs $O(nd)$, the same as one batch step in Algorithm~\ref{alg:batch-CC}.

\newpage
\begin{algorithm} [htb]
\caption{Single-block Batch CC}\label{alg:batch-CC}
\begin{algorithmic}
\STATE {\bf Input} Initial $\beta^0$, number of batch steps $B$, perturbation $\delta$.
\FOR{$k=0,\ldots, B-1$}
\STATE Set $\tau = n/\|X\|_2^2$. Perform the following updates:
\begin{align*}
    u^{k+1} &= \hat u_\delta (\beta^k) =-1/(2\sqrt{\|X\beta^k\|_n^2+\delta}), \\
    \beta^{k+1} &= \mathcal{S}\left(\beta^k-\frac{\tau}{n}X^\T(X\beta^k-\frac{1}{1-2\lambda u^{k+1}}r), \frac{\tau}{1-2\lambda u^{k+1}}\cdot\diag(\Gamma)\right).
\end{align*}
Equivalently, the updates can be expressed as
\begin{align}
    \tau^k &= \frac{n/\|X\|_2^2}{(1+\lambda / \sqrt{\|X\beta^{k}\|_n^2+\delta})}, \label{eq:batch cc-tau}\\
    \beta^{k+1} &= \mathcal{S}\left(\beta^{k} - \frac{\tau^k}{n}X^\T\left( X\beta^k-r+\frac{\lambda X\beta^k}{\sqrt{\|X\beta^{k}\|_n^2+\delta}} \right)  , \tau^k\cdot\diag(\Gamma)\right). \label{eq:batch cc-beta}
\end{align}
\ENDFOR
\end{algorithmic}
\end{algorithm}

\begin{algorithm}[h]
\caption{Single-block Stochastic CC-SAGA}\label{alg:stoc-CC}
\begin{algorithmic}
\STATE {\bf Input}  Initial $\beta^0$, number of batch steps $B$, step size $\tau$,
perturbation $\delta$.
\FOR{$k=0,\ldots, B-1$}
\STATE Update $$u^{k+1} = \hat u_\delta (\beta^k) = -\frac{1}{2\sqrt{\|X\beta^k\|_n^2+\delta}}.$$ \\
Set $v^0 = X\beta^k-r/(1-2\lambda u^k)$, $w^0=X^\T v^0/n$, and $\tilde \beta^0=\beta^k$.
\FOR{$s=0,\ldots, n-1$}
\STATE Pick $i$ uniformly from $\{1,\ldots,n\}$, and perform the following updates:
\begin{align*}
    v_i^{s+1}&=X_{i,\cdot}^\T\beta^{s}-r_i/(1-2\lambda u^k) \quad \text{or}\quad  v_j^{s+1}=v_j^s \text{ for } j\neq i, \\
    \tilde{\beta}^{s+1} &= \mathcal{S}\left(\tilde{\beta}^{s}-\tau \left(w^s + X_{i,\cdot}(v_i^{s+1}-v_i^s)\right),  \frac{\tau}{1-2\lambda u^k }\cdot\diag(\Gamma)\right), \\
    w^{s+1} &= w^s + X_{i,\cdot}(v_i^{s+1}-v_i^s)/n .
\end{align*}
\ENDFOR
\STATE Update $\beta^{k+1}=\tilde{\beta}^{n}$.
\ENDFOR
\end{algorithmic}
\end{algorithm}

\subsection{Comparison with proximal gradient method}

After the empirical norm is modified, the perturbed objective (\ref{eq:perturbed-obj}) becomes amenable to a broader range of methods in convex optimization, for example, the proximal gradient method, the proximal Newton method, and so on. We compare CC with
the proximal gradient method applied directly to (\ref{eq:perturbed-obj}) with a constant step size,
and show that the majorization allows the proximal gradient step in CC to adapt to the local curvature.

The perturbed objective (\ref{eq:perturbed-obj}) can be split into two terms:
the differential term $h_1(\beta) =  \|r-X\beta\|_n^2/2+\lambda\sqrt{\|X\beta\|_n^2+\delta}$
and the non-differentiable term $h_2 (\beta) = \| \Gamma\beta \|_1$.
With this split, the proximal gradient method can be applied directly to minimize (\ref{eq:perturbed-obj}).
The $\beta$-update can be shown to be of the same form as (\ref{eq:batch cc-beta}) in Algorithm~\ref{alg:batch-CC}, but with the step size $\tau^k$ in (\ref{eq:batch cc-tau}) replaced by a constant $\tau$.
To guarantee convergence,
the choice of the step size $\tau$ depends on the global curvature/smoothness of the differentiable term $h_1(\beta)$.

\begin{pro}\label{pro:smoothness of perturbed l2 norm}
Let $h_0(z) = \lambda\sqrt{\|z\|_2^2+\delta}$. Then $h_0$ is $\lambda/\sqrt{\delta}$-smooth:
$$
\|\nabla h_0(x)-\nabla h_0(y)\|_2 \leq (\lambda/\sqrt{\delta})\|x-y\|_2, \quad \forall x,y.
$$
As a result, $h_1(\beta) = \|r-X\beta\|_n^2/2+\lambda\sqrt{\|X\beta\|_n^2+\delta}$ is $(1+\lambda/\sqrt{\delta})\|X\|_2^2/n$-smooth.
\end{pro}

From Proposition \ref{pro:smoothness of perturbed l2 norm},
the step size of the proximal gradient method should be no greater than $n/\{(1+\lambda/\sqrt{\delta})\|X\|_2^2\}$, to ensure descent of the objective (Parikh \& Boyd, 2014). 
By comparison, the varying step size $\tau^k$ in (\ref{eq:batch cc-tau}) is no smaller than $n/\{(1+\lambda/\sqrt{\delta})\|X\|_2^2\}$
and can be considerably larger if $\|X\beta^k\|_n^2\gg\delta$. 

This phenomenon can be explained geometrically. As seen from Figure~\ref{fig:abs}, the quadratic majorization captures the local curvature of $\sqrt{t^2+\delta}$,
depending on the current point $t^k$:
it becomes flatter as $|t^k|$ is larger, and is sharpest at the origin. Hence the implied step size $\tau^k$ for the proximal gradient in Algorithm~\ref{alg:batch-CC} can be adaptive to  the local curvature.
In contrast, the direct proximal gradient uses a constant global curvature, which corresponds to the largest curvature (at $0$).
This leads to a smaller (conservative) step size and hence a slower convergence if the solution for $\beta$ is far away from $0$.

\section{Technical details}

\subsection{Relation between linearized ADMM and CP}

We derive the Chambolle--Pock (CP) algorithm (\ref{eq:CP}) from linearized ADMM (\ref{eq:linear-ADMM}). By (\ref{subeq:linear-ADMM-z}) and Moreau's identity (\ref{eq:Moreau's identity}), we have
\begin{equation}\label{eq:l-admm-z-moreau}
    z^{k+1} = Ax^k + u^k/\alpha - \prox_{\alpha f^*}(\alpha Ax^k+u^k)/\alpha.
\end{equation}
Let $v^{k+1}=\prox_{\alpha f^*}(\alpha Ax^k+u^k)$, which is (\ref{subeq:v-from-ADMM}). Then (\ref{eq:l-admm-z-moreau}) becomes
\begin{equation}\label{eq:l-admm-z-u-v}
    z^{k+1} = Ax^k + u^k/\alpha -v^{k+1}/\alpha.
\end{equation}
Substituting (\ref{eq:l-admm-z-u-v}) into (\ref{subeq:linear-ADMM-x}), we obtain (\ref{subeq:CP-x}). Combining (\ref{eq:l-admm-z-u-v}) with $u^{k+1}=u^k+\alpha(Ax^{k+1}-z^{k+1})$ in (\ref{eq:linear-ADMM}) we have
\begin{equation}\label{eq:u-v-relation}
    u^{k+1} = \alpha A(x^{k+1}-x^k)+v^{k+1}.
\end{equation}
If $u^0=\alpha A(x^0-x^{-1})+v^0$, then (\ref{eq:u-v-relation}) holds for $k\geq -1$. Hence we obtain (\ref{subeq:u-from-CP}). Substituting (\ref{eq:u-v-relation}) into the definition of $v^{k+1}$, we obtain (\ref{subeq:CP-v}). Collecting (\ref{subeq:CP-v}) and (\ref{subeq:CP-x}) yields the CP algorithm.

For matching between linear ADMM and CP, based on the reasoning above, $(v^{k+1},x^{k+1})$ in CP (\ref{eq:CP}) can be recovered from linearized ADMM (\ref{eq:linear-ADMM}) as $(\prox_{\alpha f^*}(\alpha Ax^k+u^k), x^{k+1})$ or $(u^k+\alpha(Ax^k-z^{k+1}),x^{k+1})$ for all $k\geq 0$. Reversely, $(z^{k+1},x^{k+1},u^{k+1})$ in linearized ADMM (\ref{eq:linear-ADMM}) can be recovered from  CP (\ref{eq:CP}) as $(A(2x^{k}-x^{k-1})+(v^k-v^{k+1})/\alpha,x^{k+1},\alpha A(x^{k+1}-x^k)+v^{k+1})$, for all $k\geq 0$.

For completeness, we briefly discuss the relation between re-ordered linearized ADMM and a dual version of CP. After exchanging the order of (\ref{subeq:linear-ADMM-z}) and (\ref{subeq:linear-ADMM-x}), through a similar argument as above, the re-ordered linearized ADMM can be simplified as
\begin{equation}\label{eq:dual CP}
 \begin{split}
    x^{k+1} &= \prox_{\tau g}(x^k-\tau A^\T(2u^k-u^{k-1})), \\
    u^{k+1} &=\prox_{\alpha f^*}(u^k+\alpha Ax^{k+1}),
 \end{split}
\end{equation}
with $z^{k+1}$ absorbed. It can be shown that (\ref{eq:dual CP}) is equivalent to CP (\ref{eq:CP}) applied to the dual problem (\ref{eq:generic dual problem for review}). Compared with the original CP (\ref{eq:CP}), both the update order of primal and dual variables and the extrapolation are flipped in (\ref{eq:dual CP}).

\subsection{Proof of Proposition~\ref{pro:linear-AMA}}

Applying PAPC/$\mathrm{PDFP^2O}$ to the dual problem (\ref{eq:generic dual problem for review}), we obtain
\begin{align*}
    x^{k+1} &=\prox_{\tau g}(x^k-\tau A^\T (u^k-\alpha \nabla f^*(u^k)+\alpha Ax^k)) , \\
    u^{k+1} &= u^k-\alpha \nabla f^*(u^k) + \alpha Ax^{k+1}.
\end{align*}
Letting $z^{k+1}=\nabla f^*(u^k)$, we immediately recover (\ref{eq:linear-AMA}).

For the convergence, by Theorem~2 in Li \& Yan (2021), 
if $0<\alpha<2\mu$ and $0<\alpha\tau\|A\|_2^2\leq 4/3$, then $u^{k+1}$ converges to $u^*$ which is optimal for (\ref{eq:generic dual problem for review}), and $x^{k+1}$ converges to $x^*\in\partial g^*(-A^\T u^*)$ such that $\nabla f^*(u^*)-Ax^*=0$.
Let $z^*=\nabla f^*(u^*)$. Then $(z^*,x^*,u^*)$ is a saddle point for (\ref{eq:generic lagrangian}). By the continuity of $\nabla f^*$, $z^{k+1}$ converges to $\nabla f^*(u^*)=z^*$. Hence $(z^{k+1},x^{k+1},u^{k+1})$ converges to $(z^*,x^*,u^*)$, as desired.

\subsection{Proof of Proposition~\ref{pro:cd of proximal}}
The coordinate descent problem is to solve
$$
\argmin_{v_i}\, \frac{1}{2}(v_i-b_i)^2 + \alpha f^*(v_i,v^k_{-i}),
$$
which is the proximal mapping of $f^*$ with respect to a single coordinate $v_i$. Since $f^*$ is closed, convex and proper, the proximal mapping uniquely exists, and we denote it as $v_i^*$. Because $f^*$ is smooth, it suffices to solve the first-order condition
\begin{equation}\label{eq:first-order condition}
    v_i-b_i + \alpha\left(1-\frac{\lambda\sqrt{n}}{\sqrt{(v_i+r_i)^2+\|v^k+r\|_{-i}^2}} \right)_+ (v_i+r_i)=0.
\end{equation}
If $(v_i+r_i)^2+\|v^k+r\|_{-i}^2\leq n\lambda^2$, (\ref{eq:first-order condition}) reduces to $v_i-b_i=0$, so that $v_i^*=b_i$. Next, assume that $(b_i+r_i)^2+\|v^k+r\|_{-i}^2> n\lambda^2$. Then the term inside $(\cdot)_+$ in (\ref{eq:first-order condition}) must be positive. Let $\tilde v_i=v_i+r_i$. Then (\ref{eq:first-order condition}) becomes
$$
b_i + r_i = \left( 1+\alpha-\frac{\alpha\lambda\sqrt{n}}{\tilde v_i^2+\|v^k+r\|_{-i}^2}\right)\tilde v_i.
$$
If $b_i+r_i=0$, then $\tilde v_i=0$ by the fact that $\|v^k+r\|_{-i}^2>n\lambda^2$. If $b_i+r_i\neq 0$, then $\tilde v_i\neq 0$ and $\tilde v_i$ must take the form of $c(b_i+r_i)$, where $c\neq 0$ is the root of equation
\begin{equation}\label{eq:c equation}
    \left(1+\alpha - \frac{\alpha\lambda\sqrt{n}}{\sqrt{c^2(b_i+r_i)^2+\|\mymu^k+r\|^2_{-i}}} \right)c=1.
\end{equation}
The left-hand side (LHS) of (\ref{eq:c equation}) is strictly increasing on $[0,\infty)$. When $c=0$ the LHS of (\ref{eq:c equation}) is $0$. When $c=1$, the LHS of (\ref{eq:c equation}) is greater than $1$ by the assumption $(b_i+r_i)^2+\|v^k+r\|_{-i}^2> n\lambda^2$. Hence there is a unique $c^*$ in $(0,1)$ which solves (\ref{eq:c equation}). In the case of $\|v^k+r\|_{-i}^2=0$ and $(b_i+r_i)^2> n\lambda^2$, $c^*$ has a close-form expression
$$
c^*=\left(1+\frac{\alpha\lambda\sqrt{n}}{|b_i+r_i|}\right)/(1+\alpha).
$$

\subsection{Proof of Proposition~\ref{pro:smoothness of perturbed l2 norm}}
By the chain rule, it suffices to show that $h_0(z)=\lambda \sqrt{\|z\|_2^2+\delta}$ is $(\lambda/\sqrt{\delta})$-smooth. Without loss of generality, let $\lambda=1$. Then
$$
\nabla h_0(z)=z/ \sqrt{\|z\|_2^2+\delta},
$$ and
$$
\nabla^2 h_0(z)=c\cdot I-c^3\cdot zz',
$$ where $c=1/\sqrt{\|z\|_2^2+\delta}\leq 1/\sqrt{\delta}$. When $z=0$, $\nabla^2 h_0(0)$ is a rescaled identity matrix $(1/\sqrt{\delta})\cdot I$. Next assume that $z\neq 0$. Then for any $y$ we have $\nabla^2 h_0(z)y=c\cdot y-c^3\cdot zz'y$. Let $y=\hat{y}+y^{\bot}$ where $\hat{y}=zz'y/(z'z)$ is the projection of $y$ on $z$. Then $zz'y=(z'z)\cdot\hat{y}=\|z\|_2^2\cdot\hat{y}$, and
\begin{align*}
    \nabla^2 h_0(z)y&=c\cdot y - c^3\cdot zz'y=c\cdot y - (c^3\|z\|_2^2)\cdot \hat{y} \\
    &=(c-c^3\|z\|_2^2)\cdot \hat{y}+c\cdot y^{\bot}=(\delta c^3)\cdot \hat{y}+c\cdot y^\bot.
\end{align*}
Therefore, for any $y$,
\begin{align*}
    \|\nabla^2 h_0(z)y\|_2^2&=\|(\delta c^3)\cdot \hat{y}+c\cdot y^\bot\|_2^2=(\delta^2c^6)\|\hat{y}\|_2^2+c^2\|y^\bot\|_2^2 \\
    &\leq (\delta^2/\delta^3)\|\hat{y}\|_2^2+(1/\delta)\|y^\bot\|_2^2=(1/\delta)\|y\|_2^2.
\end{align*}
This shows that $\|\nabla^2 h_0(z)\|_2\leq 1/\sqrt{\delta}$ for all $z$, which is equivalent to $\nabla h_0$ being $(1/\sqrt{\delta})$-Lipschitz.

\section{Experiments with phase shift model}\label{sec:phase shift}

\begin{figure}[htp]
    \centering
    \begin{subfigure}{0.32\textwidth}
       \includegraphics[width=\textwidth]{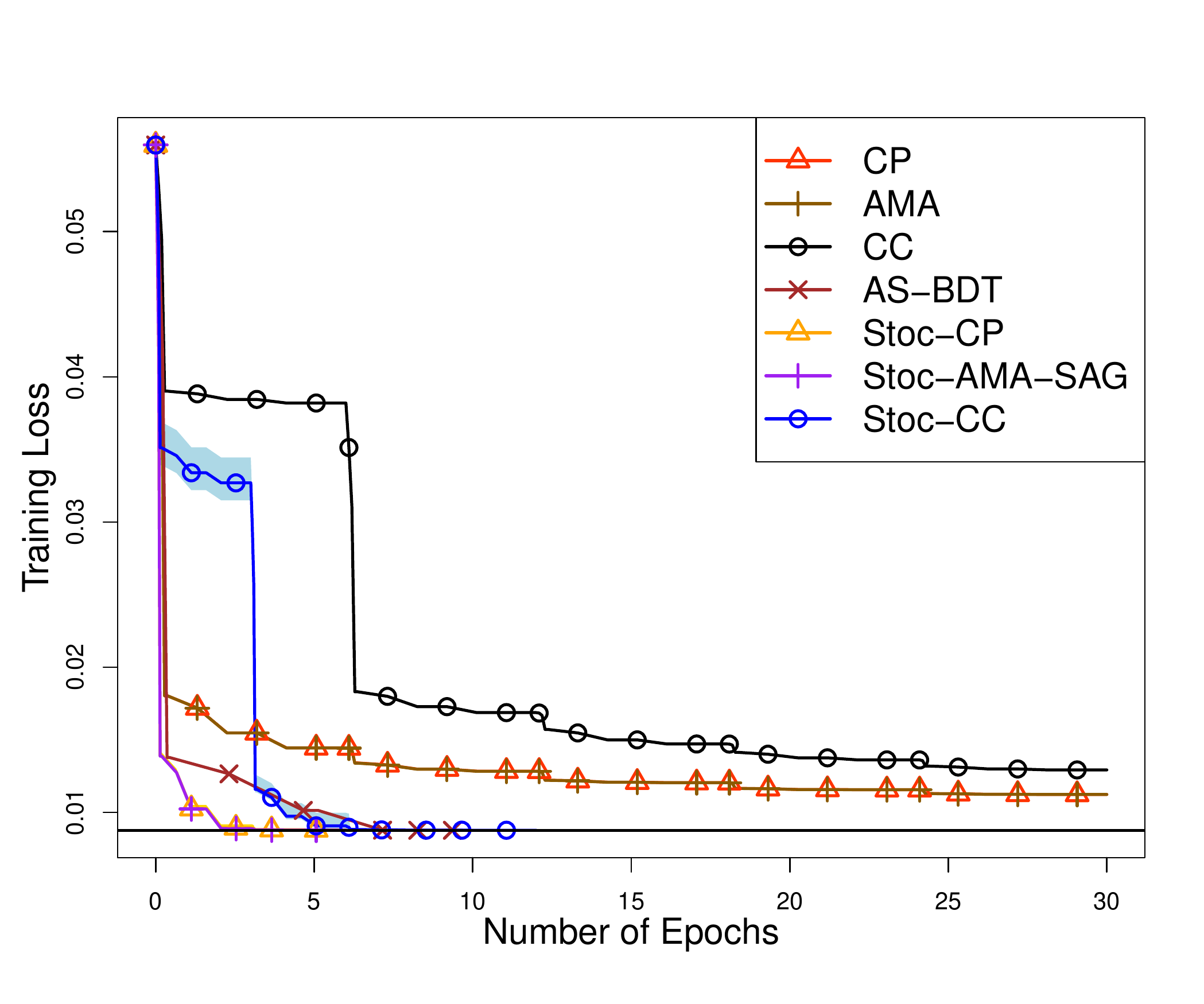}
       \vspace{-.45in}
       \caption{$\rho=2^{-17}$, $\lambda=\|\tilde \phi\|_n/2^{7}$}
    \end{subfigure}
    \hfill
    \begin{subfigure}{0.32\textwidth}
       \includegraphics[width=\textwidth]{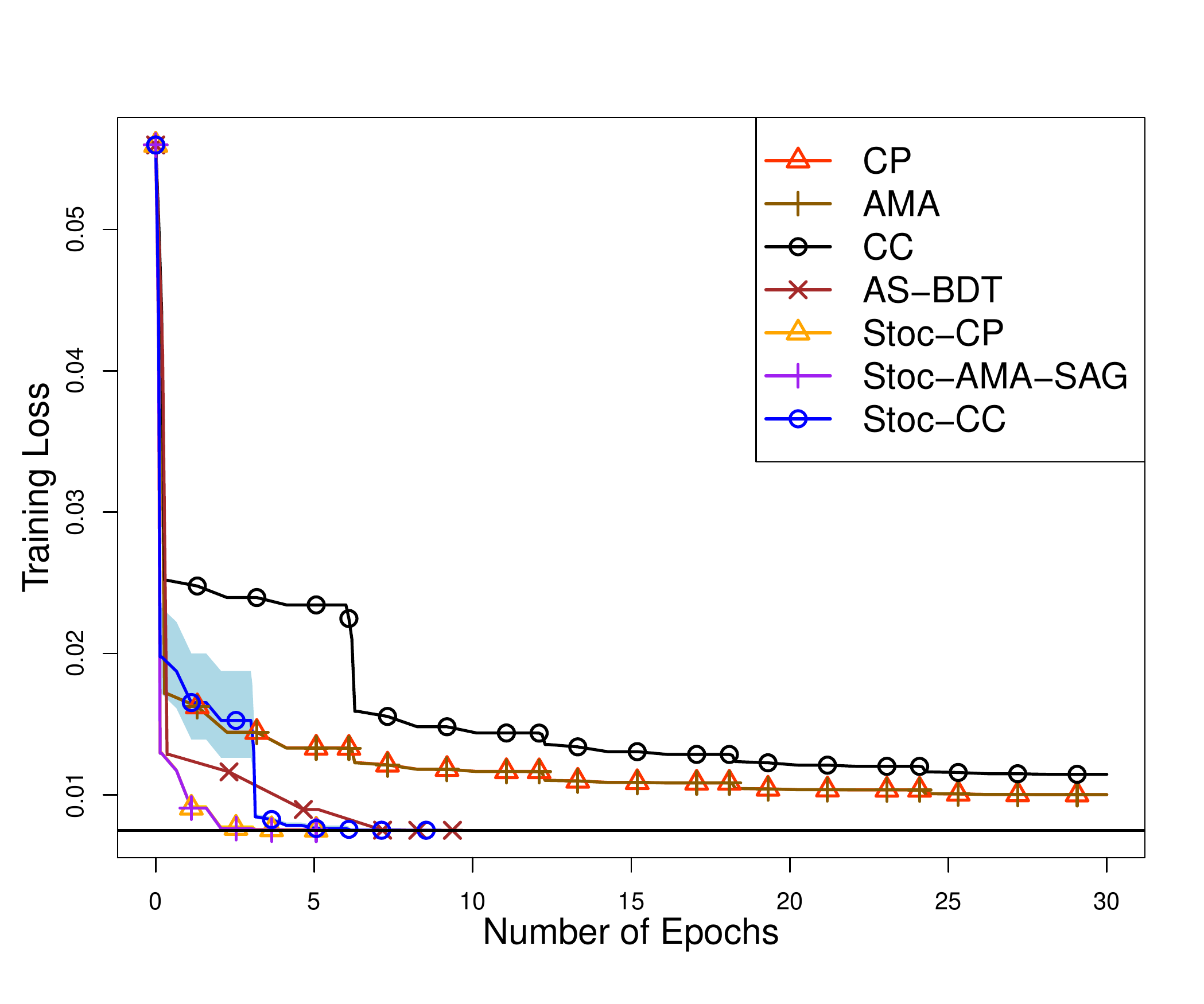}
       \vspace{-.45in}
       \caption{$\rho=2^{-17}$, $\lambda=\|\tilde \phi\|_n/2^{9}$}
    \end{subfigure}
    \hfill
    \begin{subfigure}{0.32\textwidth}
       \includegraphics[width=\textwidth]{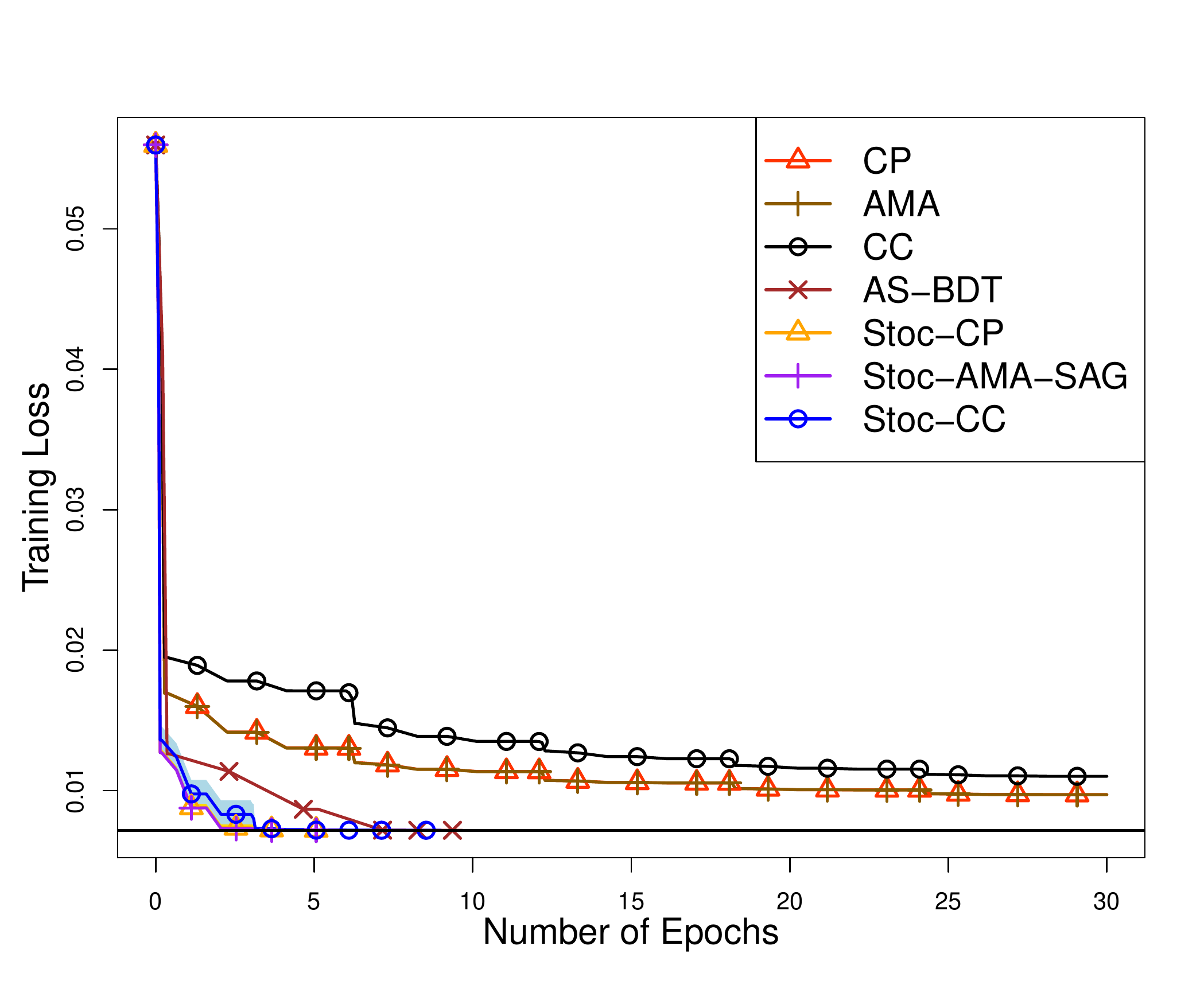}
       \vspace{-.45in}
       \caption{$\rho=2^{-17}$, $\lambda=\|\tilde \phi\|_n/2^{11}$}
    \end{subfigure}

    \begin{subfigure}{0.32\textwidth}
       \includegraphics[width=\textwidth]{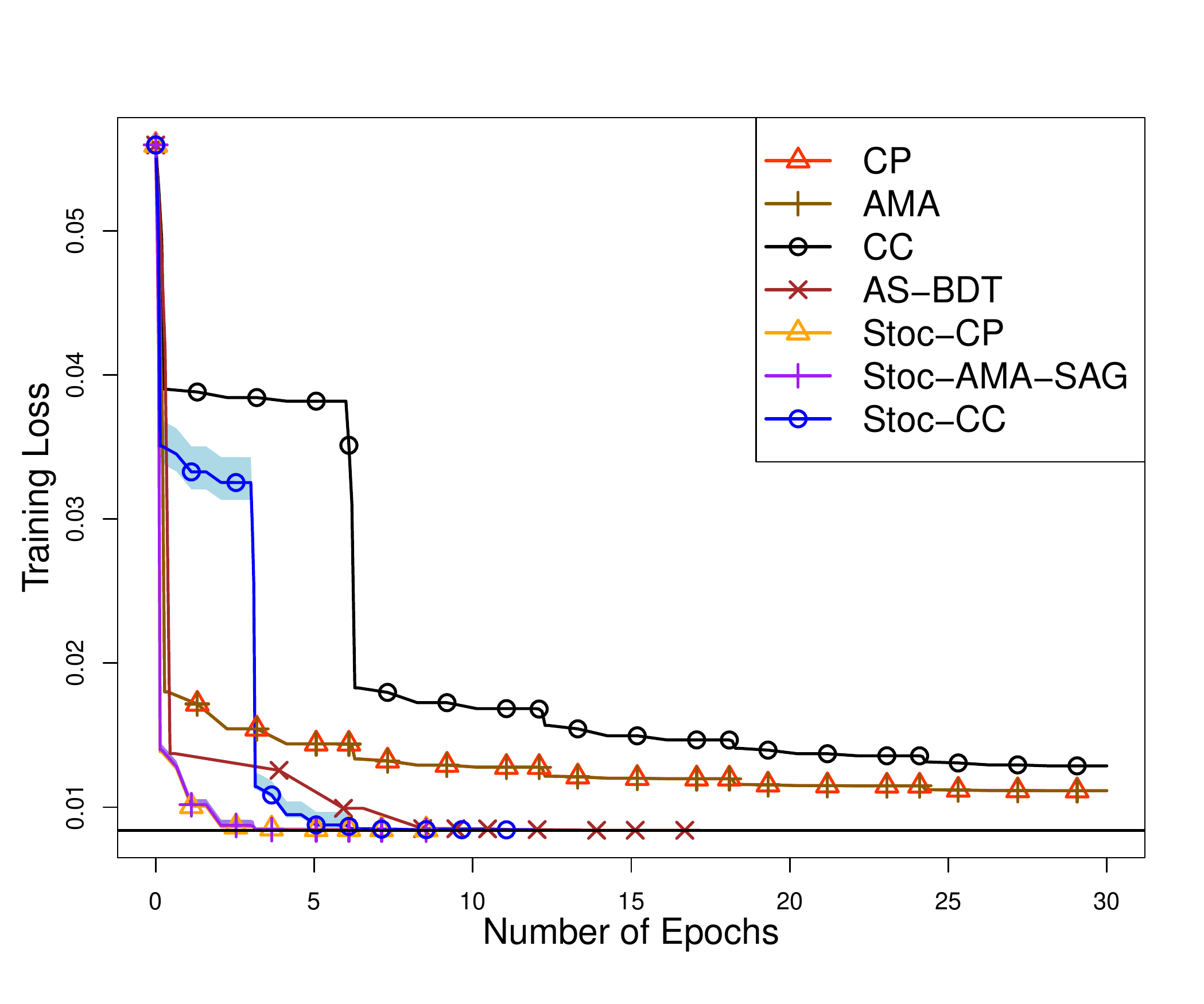}
       \vspace{-.45in}
       \caption{$\rho=2^{-21}$, $\lambda=\|\tilde \phi\|_n/2^{7}$}
    \end{subfigure}
    \hfill
    \begin{subfigure}{0.32\textwidth}
       \includegraphics[width=\textwidth]{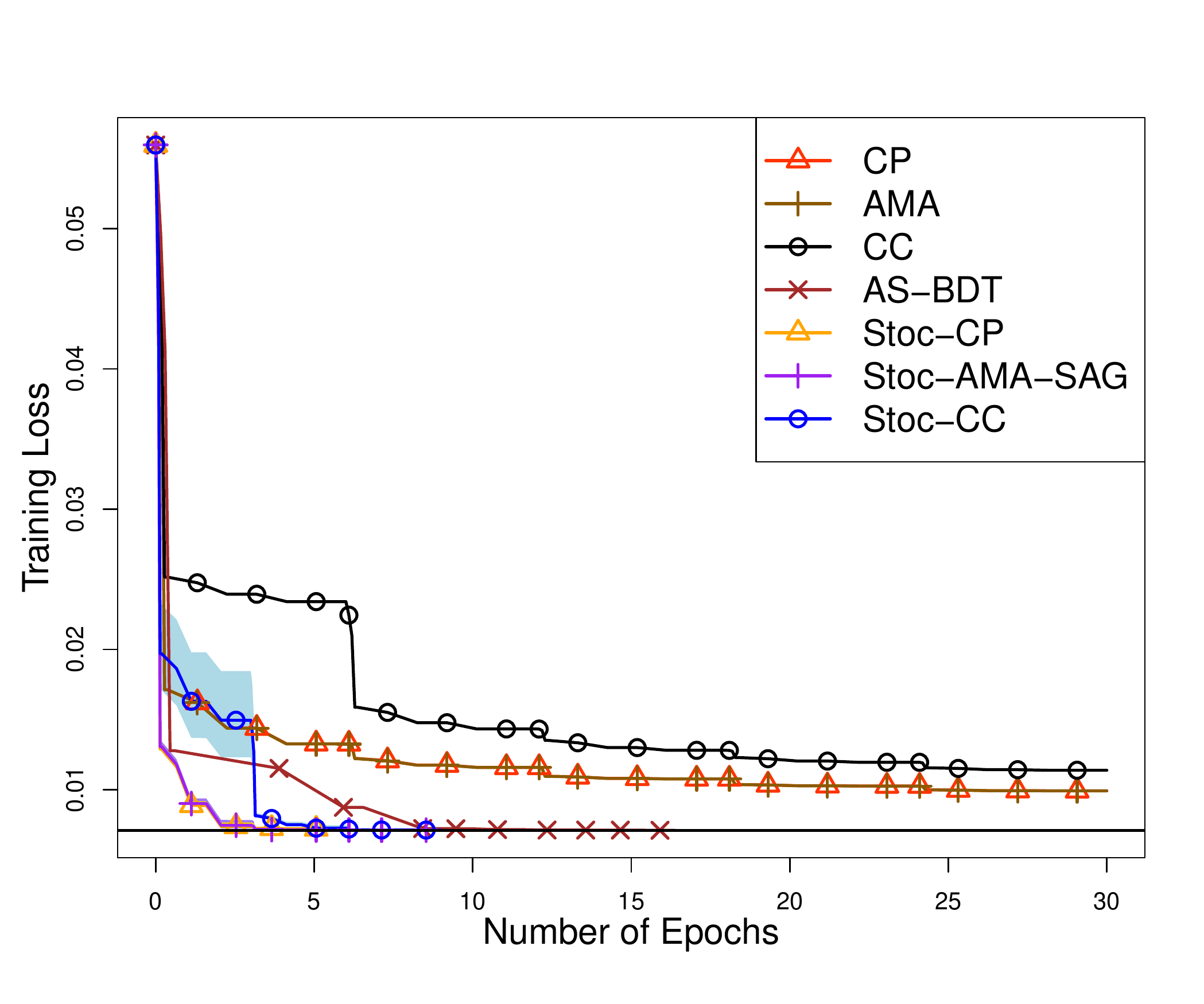}
       \vspace{-.45in}
       \caption{$\rho=2^{-21}$, $\lambda=\|\tilde \phi\|_n/2^{9}$}
    \end{subfigure}
    \hfill
    \begin{subfigure}{0.32\textwidth}
       \includegraphics[width=\textwidth]{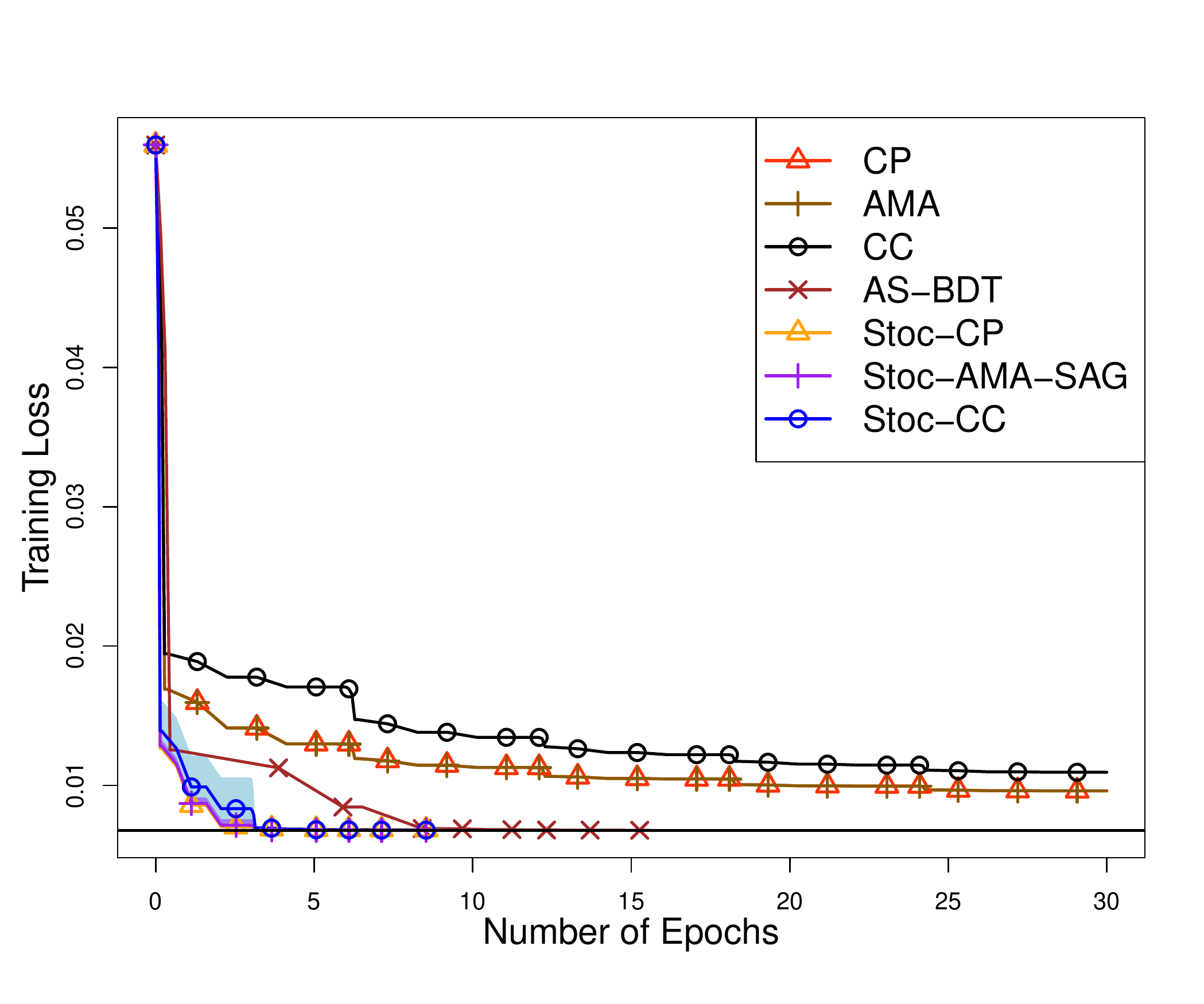}
       \vspace{-.45in}
       \caption{$\rho=2^{-21}$, $\lambda=\|\tilde \phi\|_n/2^{11}$}
    \end{subfigure}

    \begin{subfigure}{0.32\textwidth}
       \includegraphics[width=\textwidth]{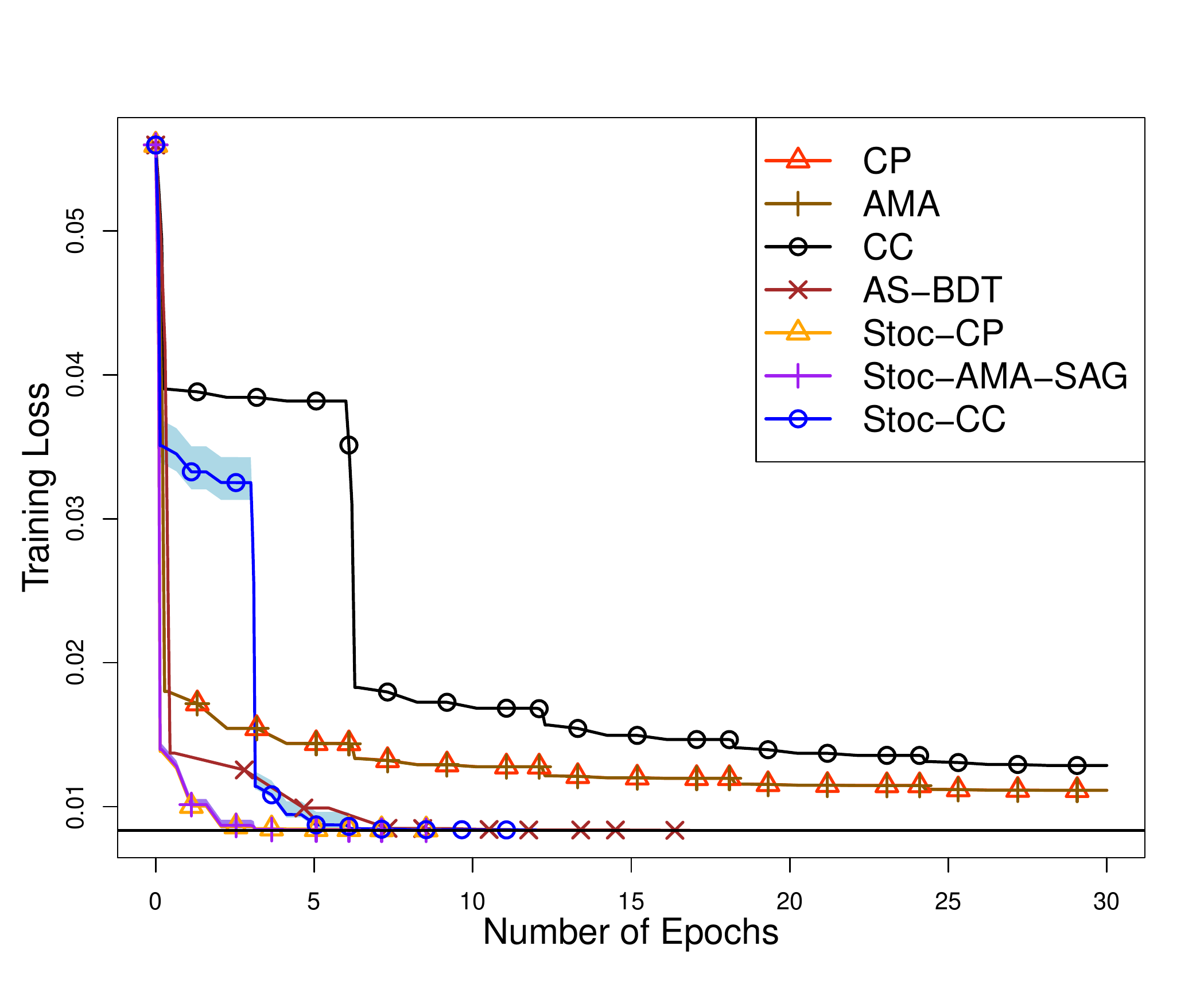}
       \vspace{-.45in}
       \caption{$\rho=2^{-25}$, $\lambda=\|\tilde \phi\|_n/2^{7}$}
    \end{subfigure}
    \hfill
    \begin{subfigure}{0.32\textwidth}
       \includegraphics[width=\textwidth]{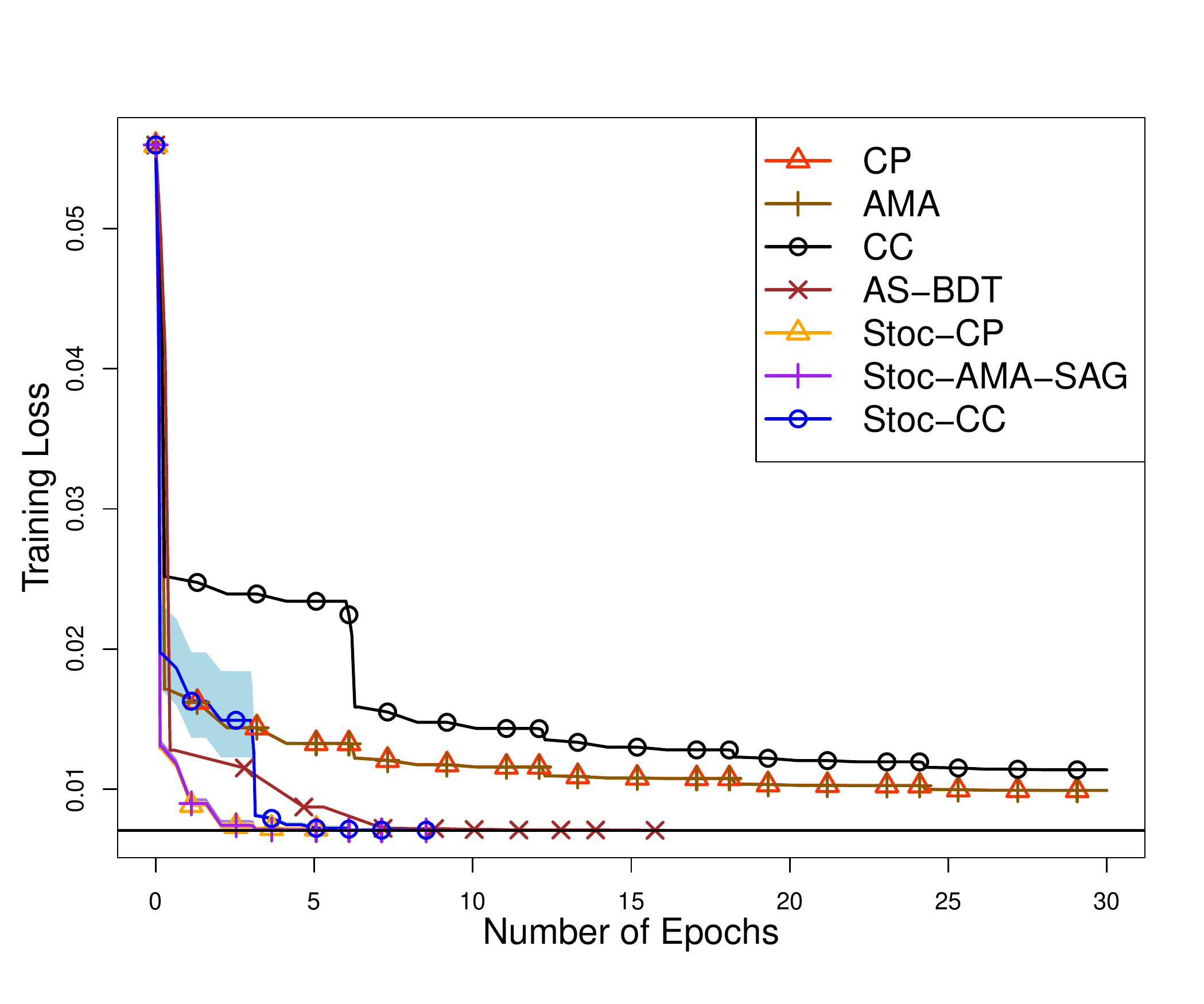}
       \vspace{-.45in}
       \caption{$\rho=2^{-25}$, $\lambda=\|\tilde \phi\|_n/2^{9}$}
    \end{subfigure}
    \hfill
    \begin{subfigure}{0.32\textwidth}
       \includegraphics[width=\textwidth]{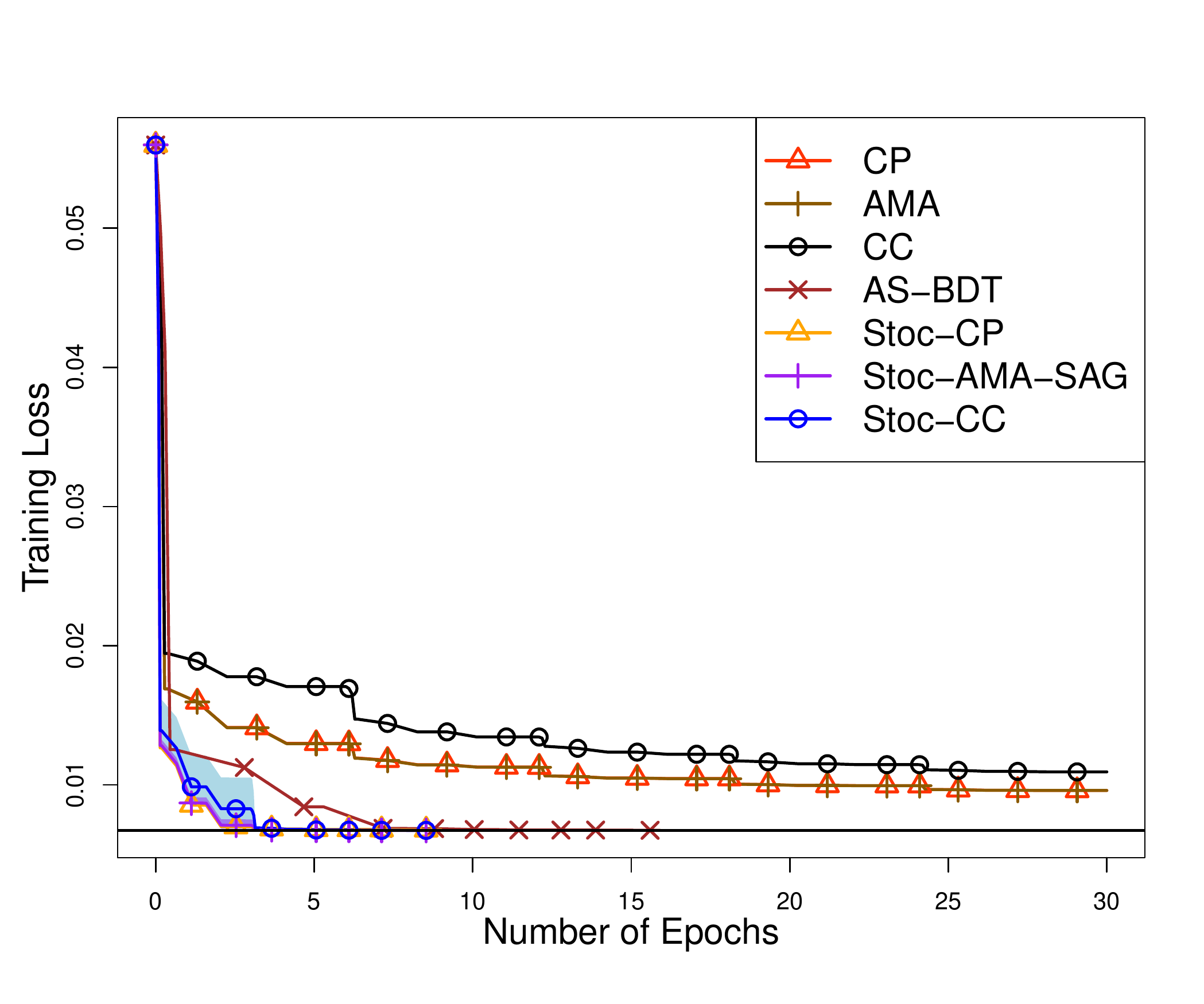}
       \vspace{-.45in}
       \caption{$\rho=2^{-25}$, $\lambda=\|\tilde \phi\|_n/2^{11}$}
    \end{subfigure}
    \caption{DPAM training in phase shift model with $K=2$ (up to two-way interactions). The minimal training loss achieved from AS-BDT is given by the black solid line. For the stochastic algorithms, we plot the mean as well as the minimum and maximum training losses across 10 repeated runs.}
    \label{fig: phase shift two-way}
\end{figure}

\begin{figure}[htp]
    \centering
    \begin{subfigure}{0.32\textwidth}
       \includegraphics[width=\textwidth]{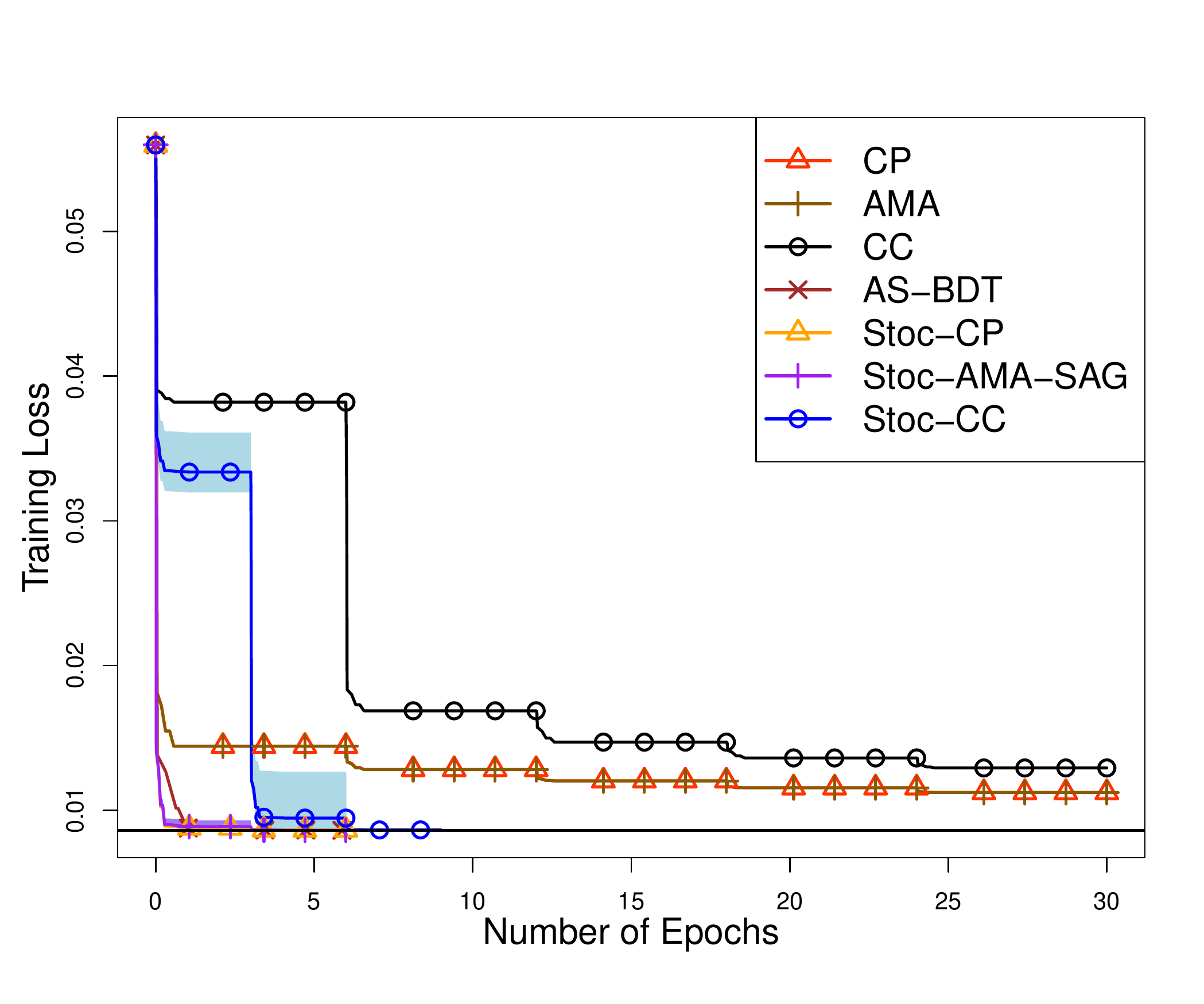}
       \vspace{-.45in}
       \caption{$\rho=2^{-17}$, $\lambda=\|\tilde \phi\|_n/2^{7}$}
    \end{subfigure}
    \hfill
    \begin{subfigure}{0.32\textwidth}
       \includegraphics[width=\textwidth]{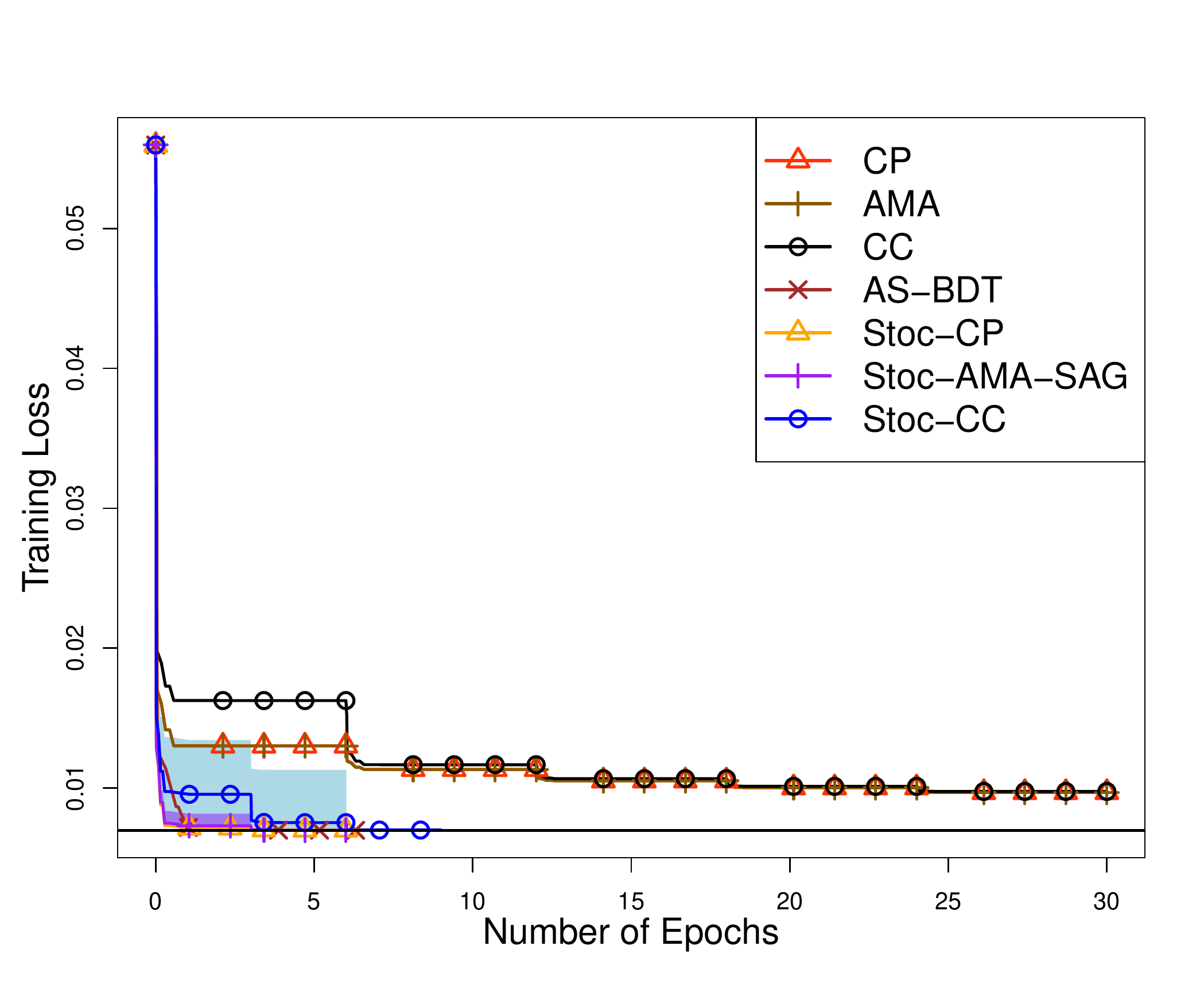}
       \vspace{-.45in}
       \caption{$\rho=2^{-17}$, $\lambda=\|\tilde \phi\|_n/2^{11}$}
    \end{subfigure}
    \hfill
    \begin{subfigure}{0.32\textwidth}
       \includegraphics[width=\textwidth]{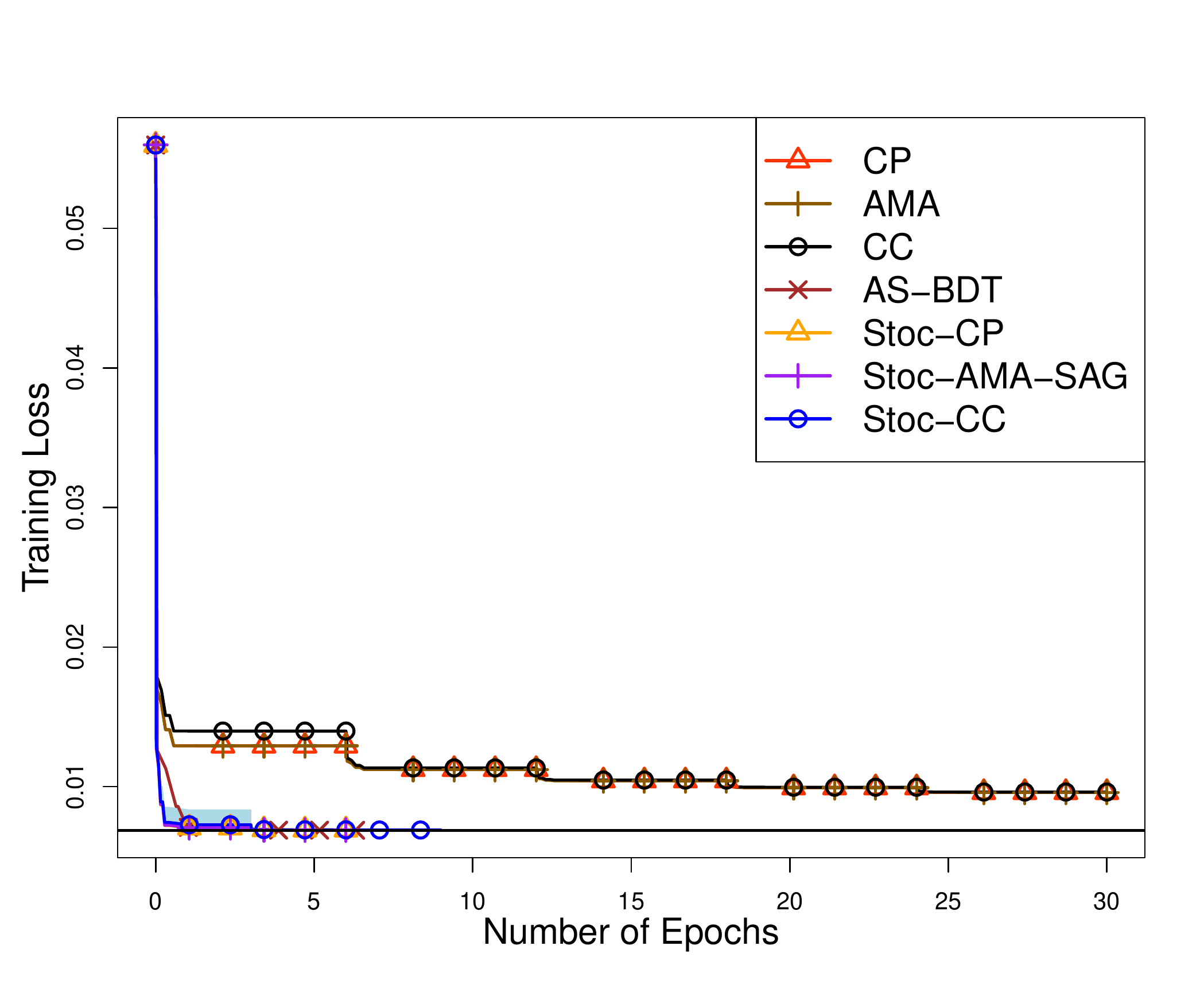}
       \vspace{-.45in}
       \caption{$\rho=2^{-17}$, $\lambda=\|\tilde \phi\|_n/2^{15}$}
    \end{subfigure}

    \centering
    \begin{subfigure}{0.32\textwidth}
       \includegraphics[width=\textwidth]{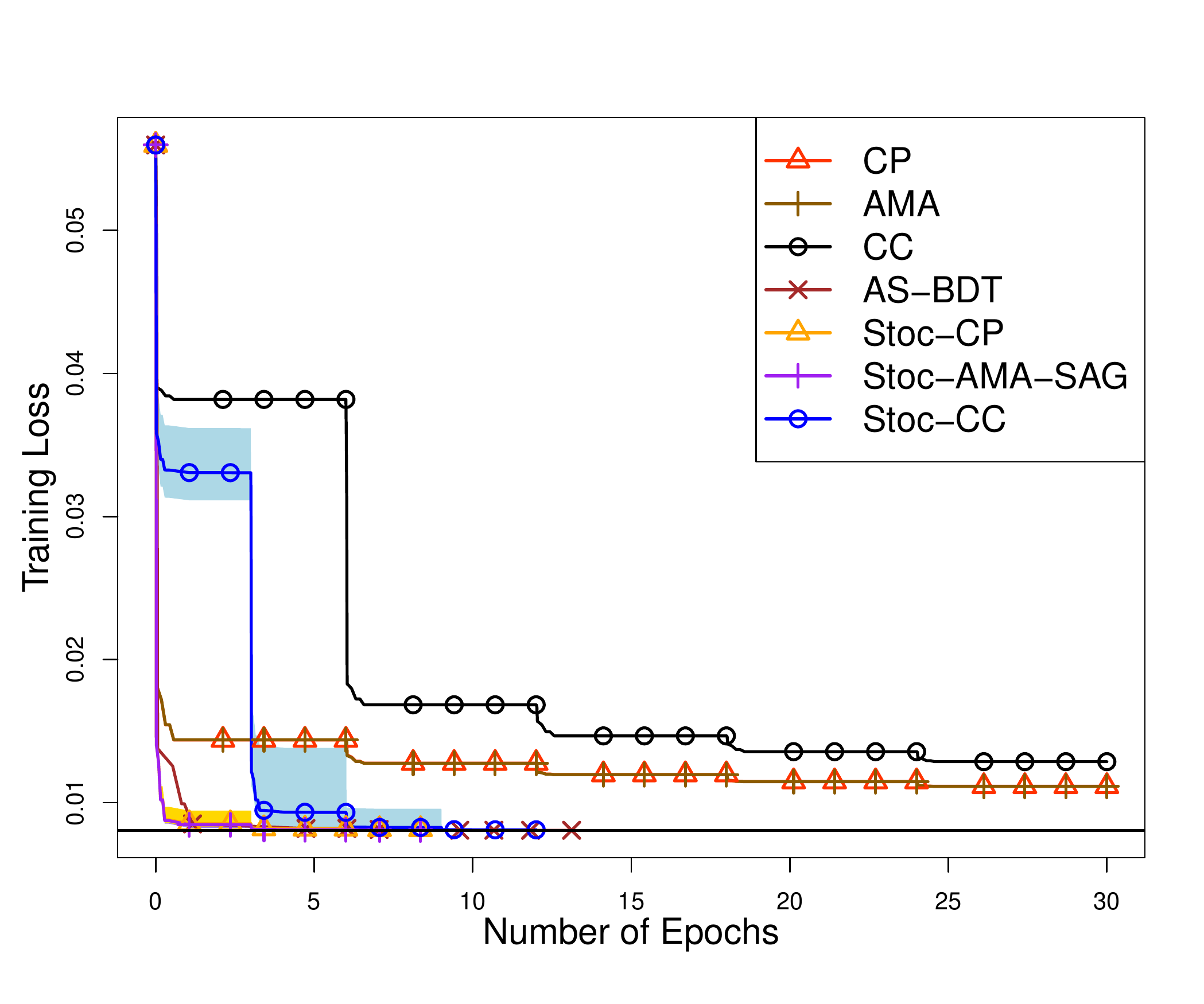}
       \vspace{-.45in}
       \caption{$\rho=2^{-21}$, $\lambda=\|\tilde \phi\|_n/2^{7}$}
    \end{subfigure}
    \hfill
    \begin{subfigure}{0.32\textwidth}
       \includegraphics[width=\textwidth]{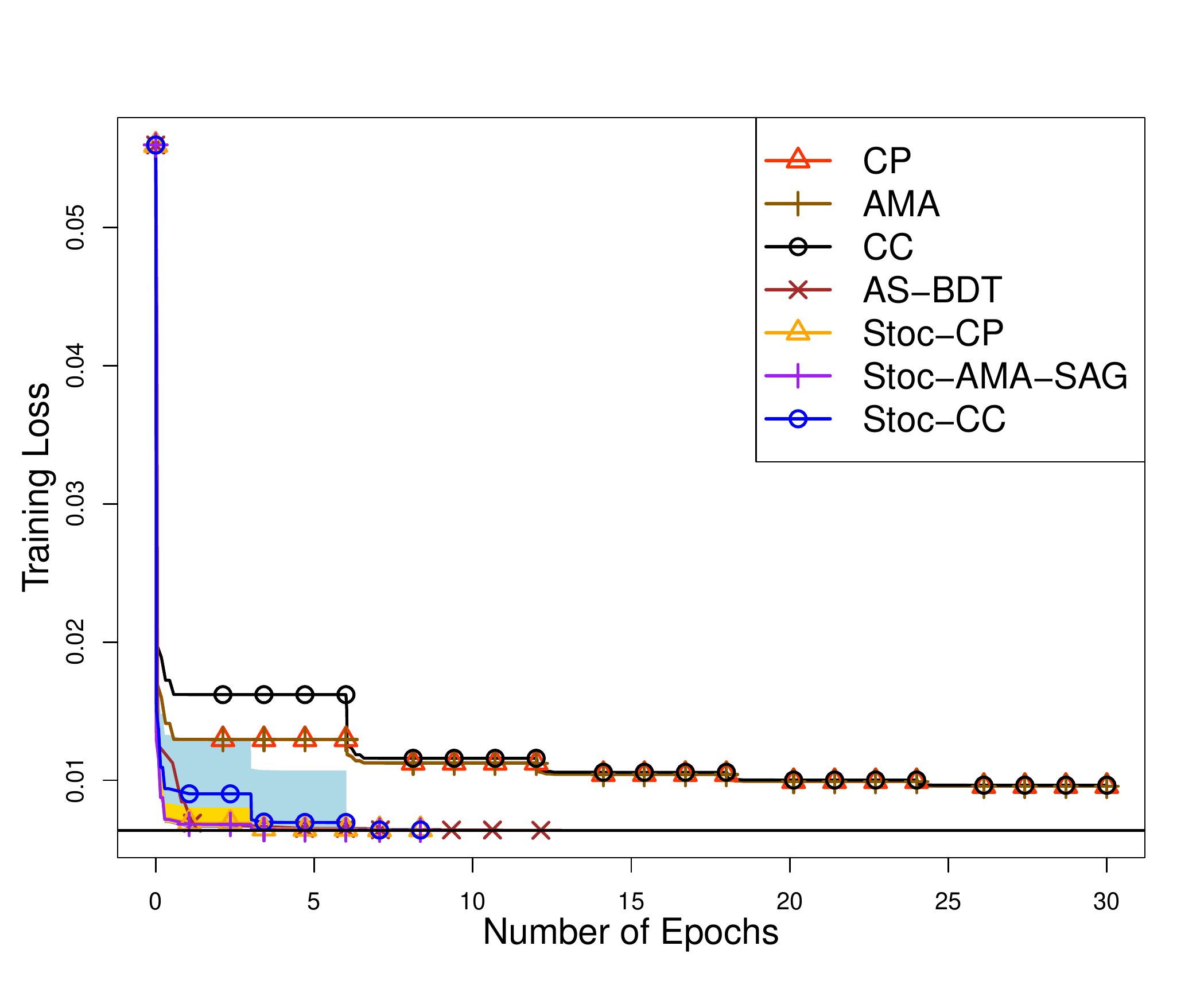}
       \vspace{-.45in}
       \caption{$\rho=2^{-21}$, $\lambda=\|\tilde \phi\|_n/2^{11}$}
    \end{subfigure}
    \hfill
    \begin{subfigure}{0.32\textwidth}
       \includegraphics[width=\textwidth]{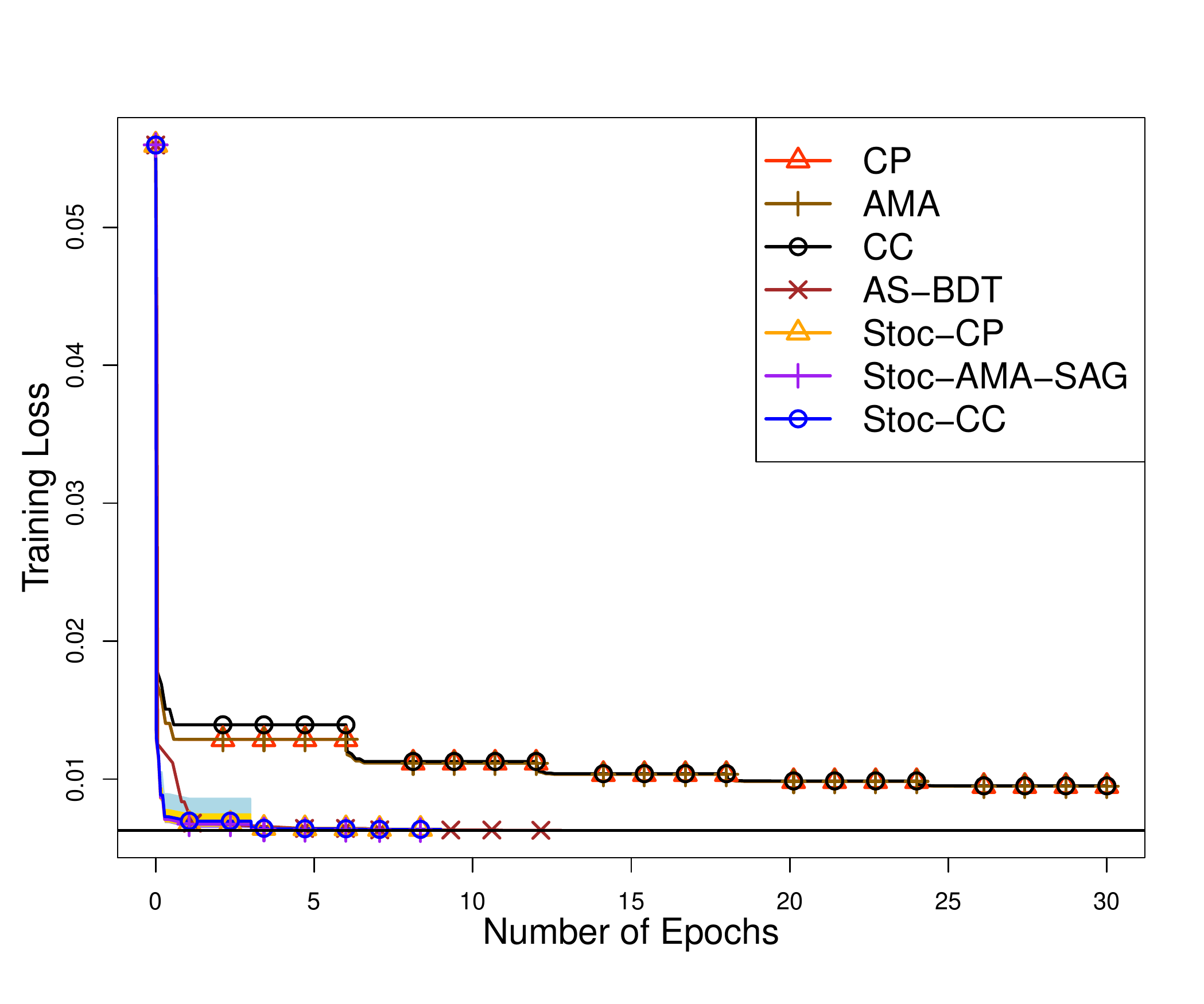}
       \vspace{-.45in}
       \caption{$\rho=2^{-21}$, $\lambda=\|\tilde \phi\|_n/2^{15}$}
    \end{subfigure}

    \centering
    \begin{subfigure}{0.32\textwidth}
       \includegraphics[width=\textwidth]{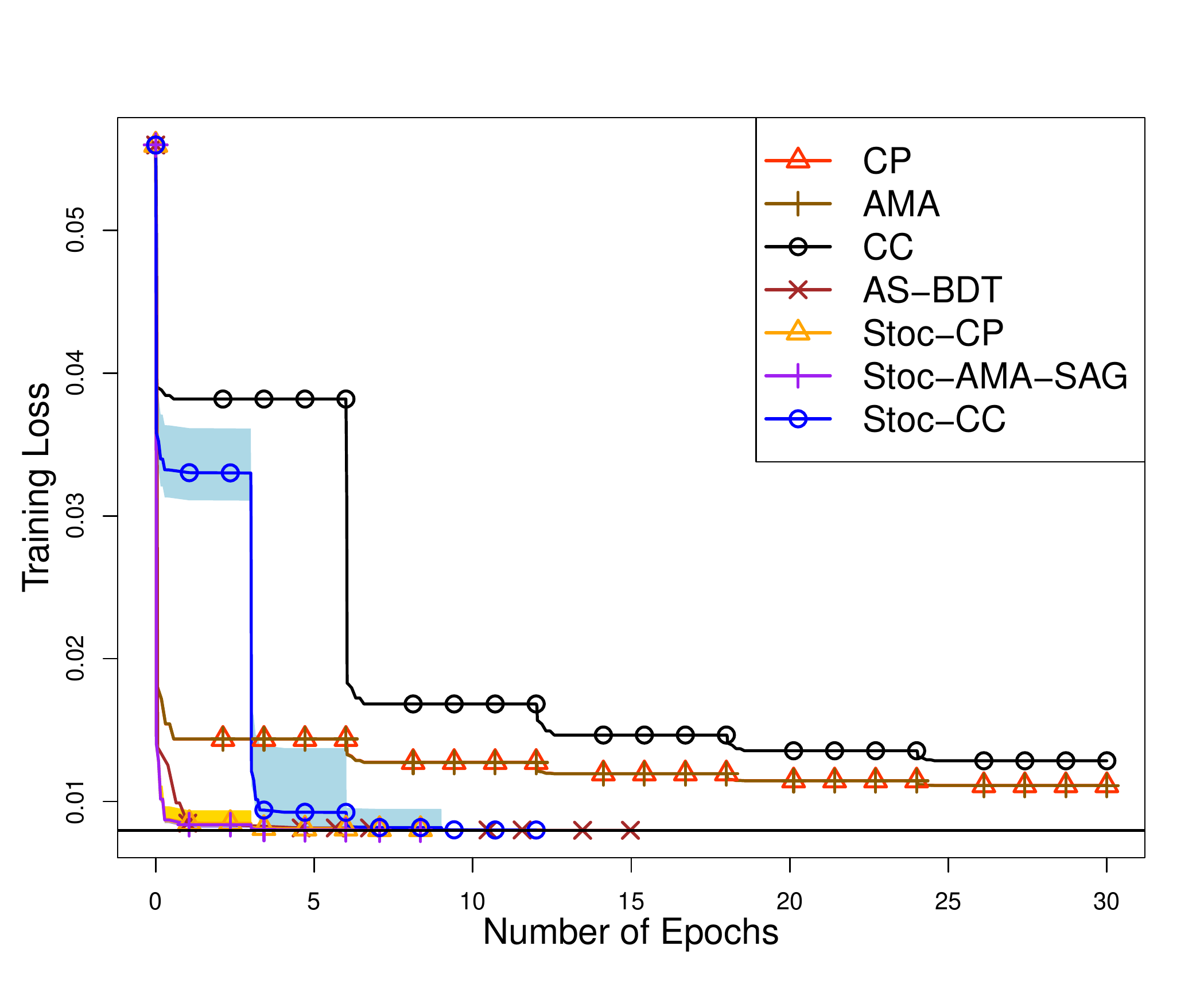}
       \vspace{-.45in}
       \caption{$\rho=2^{-25}$, $\lambda=\|\tilde \phi\|_n/2^{7}$}
    \end{subfigure}
    \hfill
    \begin{subfigure}{0.32\textwidth}
       \includegraphics[width=\textwidth]{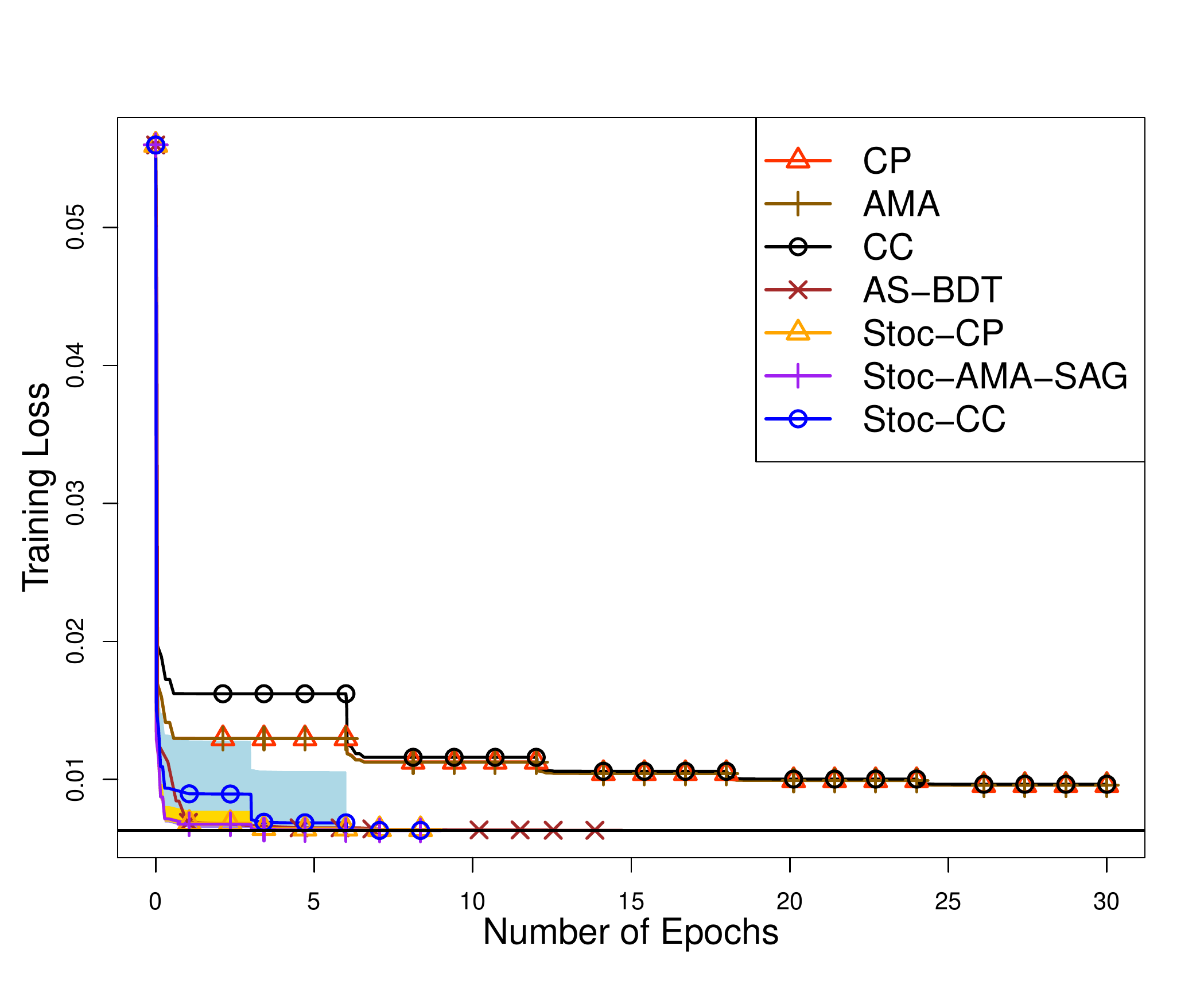}
       \vspace{-.45in}
       \caption{$\rho=2^{-25}$, $\lambda=\|\tilde \phi\|_n/2^{11}$}
    \end{subfigure}
    \hfill
    \begin{subfigure}{0.32\textwidth}
       \includegraphics[width=\textwidth]{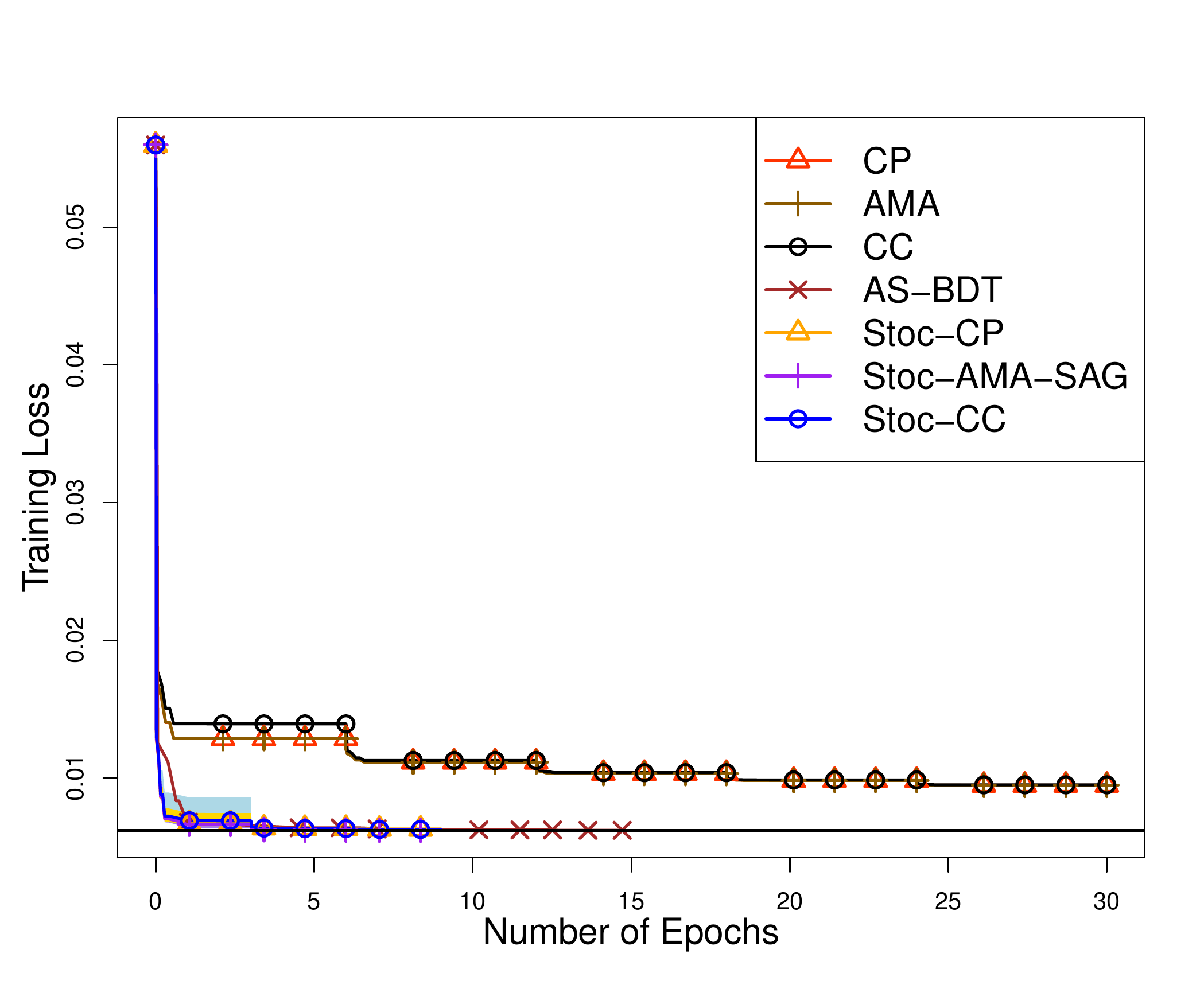}
       \vspace{-.45in}
       \caption{$\rho=2^{-25}$, $\lambda=\|\tilde \phi\|_n/2^{15}$}
    \end{subfigure}
    \caption{DPAM training in phase shift model with $K=3$ (up to three-way interactions). The minimal training loss achieved from AS-BDT is given by the black solid line. For the stochastic algorithms, we plot the mean as well as the minimum and maximum training losses across 10 repeated runs.}
    \label{fig: phase shift three-way}
\end{figure}

Consider the following example in Friedman (1991),
which models the dependence of the phase shift $\phi$ on the components in a circuit:
\begin{equation}\label{eq:phase shift}
    \phi=\arctan\left(\frac{\omega L-1/(\omega C)}{R}\right).
\end{equation}
The input variables are independently uniform in the  range $0\leq R\leq 100$, $40\pi\leq \omega\leq 560\pi$, $0\leq L\leq 1$ and $1\leq C\leq 11$. We add a normal noise with the standard deviation determined to give a signal-to-noise level 3:1.  We first transform the input region to $[0,1]^4$ and then apply the linear DPAM (\ref{eq:linear-DPAM}). We take $n=50000$ and use $m=2$ (i.e., piecewise cross-linear basis functions) and $11$ knots for each raw input. Since all orders of interactions are present in the relationship (\ref{eq:phase shift}), we consider both $K=2$ (up to two-way interactions) and $K=3$ (up to three-way interactions). The performances of various algorithms are reported in Figures~\ref{fig: phase shift two-way} and \ref{fig: phase shift three-way} under different choices of $\rho\in\{2^{-17},2^{-21},2^{-25}\}$, $\lambda\in\{\|\tilde \phi\|_n/2^{7},\|\tilde \phi\|_n/2^{9},\|\tilde \phi\|_n/2^{11}\}$ for $K=2$ and $\lambda\in\{\|\tilde \phi\|_n/2^{7},\|\tilde \phi\|_n/2^{11},\|\tilde \phi\|_n/2^{15}\}$ for $K=3$, where $\tilde\phi$ is the centered version of $\phi$. For solutions from AS-BDT (after convergence declared), the sparsity levels and MSEs under different tuning parameters are summarized in Table~\ref{tab: sparsity and MSE of phase shift}. The sparsity levels from stochastic primal-dual algorithms are reported in Tables~\ref{tab:sparsity for phase shift k=2} and \ref{tab:sparsity for phase shift k=3}. The number of batch steps fixed within each block is $6$ for three batch algorithms and $3$ for three stochastic algorithms as in Section \ref{sec:simu-linear}. A tolerance of $10^{-4}$ in the objective value is checked at the end of each cycle over all blocks to declare convergence.

\begin{table}[h]
    \centering
    \caption{Sparsity and MSEs (in units of $10^{-3}$) for the phase shift model.}
    \begin{tabular}{|l|*{2}{c}{c|}*{2}{c}{c|}*{3}{c}|}
    \hline
    \multicolumn{10}{|c|}{$K=2$ (up to two-way interactions)} \\
    \hline
    $\log_2(\rho^{-1})$ & \multicolumn{3}{c|}{17} & \multicolumn{3}{c|}{21} & \multicolumn{3}{c|}{25} \\
    \hline
    $\log_2(\lambda^{-1})$ & 7 & 9 & 11 & 7 & 9 & 11 & 7 & 9 & 11 \\
    \hline
    \# nonzero blocks & 6 & 8 & 10 & 6 & 9 & 10 & 6 & 8 & 10  \\
    \# nonzero coefficients & 50 & 56 & 58 & 79 & 97 & 96 & 75 & 93 & 96 \\
    MSEs ($10^{-3}$) & 13.72 & 13.66 & 13.66 & 13.43 & 13.4 & 13.40 & 13.43 & 13.39 & 13.40 \\
    \hline
    \multicolumn{10}{|c|}{$K=3$ (up to three-way interactions)} \\
    \hline
     $\log_2(\rho^{-1})$ & \multicolumn{3}{c|}{17} & \multicolumn{3}{c|}{21} & \multicolumn{3}{c|}{25} \\
    \hline
    $\log_2(\lambda^{-1})$ & 7 & 11 & 15 & 7 & 11 & 15 & 7 & 11 & 15 \\
    \hline
    \# nonzero blocks & 7 & 13 & 14 & 7 & 14 & 14 & 7 & 13 & 14 \\
    \# nonzero coefficients & 56 & 68 & 69 & 92 & 122 & 122 & 95 & 132 & 132 \\
    MSEs ($10^{-3}$) & 13.13 & 13.03 & 13.03 & 12.65 & 12.61 & 12.61 & 12.63 & 12.59 & 12.59 \\
    \hline
    \end{tabular}\\[.1in]
    \parbox{1\textwidth}{\small Note: For $K=2$, there are $10$ blocks from which $4$ are main effects and $6$ are two-way interactions. There are a total of $640$ scalar coefficients. For $K=3$, there are $14$ blocks with $4$ additional three-way interactions. There are a total of $4640$ scalar coefficients. The MSEs are evaluated on a validation set of $n=50000$ data points.}
    \label{tab: sparsity and MSE of phase shift}
\end{table}

\newpage
\section{Sparsity levels in numerical experiments}

We report the sparsity levels of solutions from stochastic primal-dual algorithms (after the convergences are declared) for the synthetic linear and logistic regression in Section~\ref{sec:experiments} and the phase shift model in Section~\ref{sec:phase shift}. The sparsity levels for logistic regression on the run-or-walk
data in Section \ref{sec:simu-logistic-real} are omitted, because the stochastic primal-dual algorithms may not be declared convergent by the maximum number of epochs attempted.
Means and standard deviations (in parenthesis) of the numbers of nonzero blocks and nonzero coefficients across $10$ repeated runs are summarized in Tables~\ref{tab:sparsity for synthetic linear}--\ref{tab:sparsity for phase shift k=3}.
For easy comparison, we replicate the sparsity levels of solutions from AS-BDT as shown in the main paper.

\begin{table}[p]
    \caption{Sparsity for synthetic linear regression}
    \centering
    \begin{tabular}{*5{c}}
    \hline
    $\log_2(\rho^{-1})$ & $\log_2(\lambda^{-1})$ & AS-BDT & Stoc-CP & Stoc-AMA-SAG \\
    \multicolumn{5}{c}{\# Nonzero blocks} \\
    16 & 6 & 14 & 14 (0) & 14 (0) \\
    16 & 8 & 29 & 35.2 (1.6) & 32 (2.3) \\
    16 & 10 & 54 & 54 (0) & 54.1 (0.3) \\
    19 & 6 & 14 & 14 (0) & 14 (0) \\
    19 & 8 & 54 & 53.8 (0.4) & 52.4 (0.8) \\
    19 & 10 & 54 & 55 (0) & 55 (0) \\
    22 & 6 & 14 & 14 (0) & 14 (0) \\
    22 & 8 & 54 & 53.8 (0.4) & 52.5 (0.7) \\
    22 & 10 & 55 & 55 (0) & 55 (0) \\
    \multicolumn{5}{c}{\# Nonzero coefficients} \\
    16 & 6 & 119 & 131.4 (2.3) & 127.60 (1.2) \\
    16 & 8 & 195 & 251.2 (10.6) & 226.80 (12.6) \\
    16 & 10 & 335 & 357.2 (1.9) & 352.20 (1.2) \\
    19 & 6 & 169 & 183.3 (2.0) & 183.00 (2.8) \\
    19 & 8 & 670 & 832.8 (8.5) & 842.4 (14.3) \\
    19 & 10 & 656 & 850.5 (5.1) & 832.8 (3.9) \\
    22 & 6 & 185 & 209.2 (0.8) & 209.50 (0.5) \\
    22 & 8 & 841 & 1157.6 (3.7) & 1149.7 (6.7) \\
    22 & 10 & 885 & 1167.5 (2.3) & 1163.3 (3.0) \\
    \hline
    \end{tabular}
    \label{tab:sparsity for synthetic linear}\\[.1in]
    \parbox{1\textwidth}{\small Note: For $K=2$, there are $55$ blocks from which $10$ are main effects and $45$ are two-way interactions. There are a total of $1175$ scalar coefficients. For the underlying regression function (\ref{eq:f in single-block simulation}), $14$ of the $55$ components are true. }
\end{table}

\begin{table}[p]
    \centering
    \caption{Sparsity for synthetic logistic regression}
    \begin{tabular}{*5{c}}
    \hline
    $\log_2(\rho^{-1})$ & $\log_2(\lambda^{-1})$ & AS-BDT & Stoc-CP & Stoc-AMA-SAG \\
    \multicolumn{5}{c}{\# Nonzero blocks} \\
    16 & 6 & 13 & 13.9 (0.3) & 13.9 (0.3) \\
    16 & 8 & 46 & 48 (0.7) & 47.7 (0.7) \\
    16 & 10 & 55 & 55 (0) & 55 (0) \\
    19 & 6 & 18 & 16.2 (0.4) & 16 (0) \\
    19 & 8 & 55 & 55 (0) & 55 (0) \\
    19 & 10 & 55 & 55 (0) & 55 (0) \\
    22 & 6 & 39 & 28.7 (0.5) & 27.1 (0.7) \\
    22 & 8 & 55 & 55 (0) & 55 (0) \\
    22 & 10 & 55 & 55 (0) & 55 (0) \\
    \multicolumn{5}{c}{\# Nonzero coefficients} \\
    16 & 6 & 91 & 92.8 (0.9) & 92.6 (1.0) \\
    16 & 8 & 226 & 231.1 (2.1) & 230.6 (2.2) \\
    16 & 10 & 258 & 261.3 (1.1) & 258.5 (0.7) \\
    19 & 6 & 205 & 182.9 (6.0) & 177.9 (2.9) \\
    19 & 8 & 658 & 670.9 (3.6) & 665.5 (1.6) \\
    19 & 10 & 655 & 667.8 (1.8) & 661.5 (2.5) \\
    22 & 6 & 721 & 542.8 (12.6) & 506.4 (16.3) \\
    22 & 8 & 1011 & 1086.5 (3.1) & 1073.0 (2.9) \\
    22 & 10 & 1020 & 1098.8 (2.9) & 1076.4 (1.3) \\
    \hline
    \end{tabular}\\[.1in]
    \parbox{1\textwidth}{\small Note: For $K=2$, there are $55$ blocks from which $10$ are main effects and $45$ are two-way interactions. There are a total of $1175$ scalar coefficients. For the underlying regression function (\ref{eq:f in single-block simulation}), $14$ of the $55$ components are true.}
    \label{tab:sparsity for synthetic logistic}
\end{table}

\begin{table}[p]
    \centering
    \caption{Sparsity for phase shift model with $K=2$}
    \begin{tabular}{*5{c}}
    \hline
    $\log_2(\rho^{-1})$ & $\log_2(\lambda^{-1})$ & AS-BDT & Stoc-CP & Stoc-AMA-SAG \\
    \multicolumn{5}{c}{\# Nonzero blocks} \\
    17 & 7 & 6 & 6 (0) & 6 (0) \\
    17 & 9 & 8 & 10 (0) & 10 (0) \\
    17 & 11 & 10 & 10 (0) & 10 (0) \\
    21 & 7 & 6 & 6.3 (0.5) & 6 (0) \\
    21 & 9 & 9 & 10 (0) & 10 (0) \\
    21 & 11 & 10 & 10 (0) & 10 (0) \\
    25 & 7 & 6 & 9 (0) & 9 (0) \\
    25 & 9 & 8 & 10 (0) & 10 (0) \\
    25 & 11 & 10 & 10 (0) & 10 (0) \\
    \multicolumn{5}{c}{\# Nonzero coefficients} \\
    17 & 7 & 50 & 124.2 (8.0) & 99.8 (6.1) \\
    17 & 9 & 56 & 147.3 (8.3) & 114.9 (6.3) \\
    17 & 11 & 58 & 143.8 (9.7) & 113.9 (5.5) \\
    21 & 7 & 79 & 301.2 (29.5) & 208.8 (8.9) \\
    21 & 9 & 97 & 484.9 (35.1) & 326.0 (13.3) \\
    21 & 11 & 96 & 431.2 (15.7) & 325.7 (11.0) \\
    25 & 7 & 75 & 628.3 (1.2) & 629.8 (0.4) \\
    25 & 9 & 93 & 639.4 (0.5) & 637.9 (0.6) \\
    25 & 11 & 96 & 638.9 (0.6) & 637.1 (1.2) \\
    \hline
    \end{tabular}\\[.1in]
    \parbox{1\textwidth}{\small Note: For $K=2$, there are $10$ blocks from which $4$ are main effects and $6$ are two-way interactions. There are a total of $640$ scalar coefficients. For $K=3$, there are $14$ blocks with $4$ additional three-way interactions. There are a total of $4640$ scalar coefficients.}
    \label{tab:sparsity for phase shift k=2}
\end{table}

\begin{table}[p]
    \centering
    \caption{Sparsity for phase shift model with $K=3$}
    \begin{tabular}{*5{c}}
    \hline
    $\log_2(\rho^{-1})$ & $\log_2(\lambda^{-1})$ & AS-BDT & Stoc-CP & Stoc-AMA-SAG \\
    \multicolumn{5}{c}{\# Nonzero blocks} \\
    17 & 7 & 7 & 7 (0) & 7 (0) \\
    17 & 11 & 13 & 12 (0) & 12.4 (0.7) \\
    17 & 15 & 14 & 14 (0) & 14 (0) \\
    21 & 7 & 7 & 8.6 (0.5) & 7 (0) \\
    21 & 11 & 14 & 14 (0) & 14 (0) \\
    21 & 15 & 14 & 14 (0) & 14 (0) \\
    25 & 7 & 7 & 13 (0) & 11.6 (0.5) \\
    25 & 11 & 13 & 14 (0) & 14 (0) \\
    25 & 15 & 14 & 14 (0) & 14 (0) \\
    \multicolumn{5}{c}{\# Nonzero coefficients} \\
    17 & 7 & 56 & 148.7 (7.4) & 113 (4) \\
    17 & 11 & 68 & 169.8 (7.8) & 129.7 (3.4) \\
    17 & 15 & 69 & 176.1 (7.7) & 139.3 (2.6) \\
    21 & 7 & 92 & 567.8 (58.2) & 360.8 (9.4) \\
    21 & 11 & 122 & 764.3 (69.9) & 550.7 (10.8) \\
    21 & 15 & 122 & 825.3 (85.9) & 553.5 (12.7) \\
    25 & 7 & 95 & 3238.2 (138.0) & 2715.3 (376.7) \\
    25 & 11 & 132 & 3535.3 (94.5) & 3178.9 (58.9) \\
    25 & 15 & 132 & 3596.4 (86.3) & 3182.7 (53.7) \\
    \hline
    \end{tabular}\\[.1in]
    \parbox{1\textwidth}{\small Note: For $K=2$, there are $10$ blocks from which $4$ are main effects and $6$ are two-way interactions. There are a total of $640$ scalar coefficients. For $K=3$, there are $14$ blocks with $4$ additional three-way interactions. There are a total of $4640$ scalar coefficients.}
    \label{tab:sparsity for phase shift k=3}
\end{table}

\section{Tuning information}\label{sec:tuning}

We tune three batch algorithms (Algorithms~\ref{alg:batch-CP}, \ref{alg:batch-linearized-AMA}, and \ref{alg:batch-CC}) and their stochastic versions (Algorithms~\ref{alg:stoc-CP}, \ref{alg:stoc-linear-AMA}, and \ref{alg:stoc-CC}) in all numerical experiments.
In repeated runs for stochastic algorithms, tuning is performed only in the first run. The selected step sizes are then used in all the subsequent runs.

For each algorithm, the step sizes attempted are taken from the set $\{a\times 10^{-k}\colon a\in\{1,2,\ldots,9\},k\in\mathbb{Z}\}$ unless otherwise noted
in Table \ref{tab:step sizes of synthetic linear and logistic}.
We start from a candidate set $S=\{0.1,0.2,\ldots,1\}$. The best step size in $S$ which gives the minimal training loss after a given number of epochs is selected. If the best step size is the largest element in $S$ (say $1$), modify the candidate set to be $S\leftarrow 10\cdot S$ (i.e., all elements in $S$ are multiplies by 10). If the best step size is the smallest element (say $0.1$), modify the candidate set as $S\leftarrow 0.1\cdot S$  (i.e., all elements in $S$ are multiplies by $0.1$). The procedure is repeated until a non-monotone change of training loss is detected. For example when $0.5$ is the best in $S=\{0.1,0.2,\ldots,1\}$ or $0.1$ is the best in both $\{0.1,0.2,\ldots,1\}$ and $\{0.01,0.02,\ldots,0.1\}$. To limit the amount of search we restrict the range of step sizes between $0.001$ and $100$.

For Algorithms~\ref{alg:batch-CC} and \ref{alg:stoc-CC}, only a single step size $\tau$ needs tuning. For primal-dual algorithms (Algorithm~\ref{alg:batch-CP}, \ref{alg:batch-linearized-AMA}, \ref{alg:stoc-CP}, and \ref{alg:stoc-linear-AMA}), there are two step sizes $\tau$ and $\alpha$. For simplicity, we tune one step size while fixing the other at a preliminary value, and then tune the second step size while fixing the first at the selected value.
From our numerical experience, Algorithm~\ref{alg:stoc-linear-AMA} (Stoc-AMA) seems to be more sensitive to $\alpha$ than to $\tau$. Hence for Algorithm~\ref{alg:stoc-linear-AMA} we first tune $\alpha$ then $\tau$. For the other primal-dual algorithms we first tune $\tau$ then $\alpha$.

In the simulated, multi-block experiments (Sections \ref{sec:simu-linear}, \ref{sec:simu-logistic}, and \ref{sec:phase shift}),
all inputs are sampled uniformly from $[0,1]$. For simplicity, tuning is performed with the same step sizes used in all blocks. For the real-data experiments (Section \ref{sec:simu-logistic-real}), tuning is performed separately within each block while removing the other blocks (as in single-block experiments).
If ignoring any difference in how many times the candidate sets are adjusted, the two procedures roughly have similar costs, because performing one epoch in multi-block experiments has a similar cost as
performing single-block epochs across all blocks.

\subsection{Single-block experiments}\label{subsec:tuning single-block}
For the single-block experiments (Figure~\ref{fig: single-block}), the first $5$ epochs are for tuning, and the results are summarized in Table~\ref{tab:step sizes for single-block}.
\begin{table}[H]
    \centering
    \caption{Step sizes for the single-block experiments.}
    \begin{tabular}{|l|*{2}{c}{c|}*{3}{c}|}
    \hline
    $\lambda$ & \multicolumn{3}{c|}{$\lambda_0(\rho)/4$} & \multicolumn{3}{c|}{$2\lambda_0(\rho)$} \\
    \hline
    $\log_2(\rho^{-1})$ & 15 & 18 & 21 & 15 & 18 & 21 \\
    \hline
    CP & 100,0.8 & 100,0.8 & 100,0.8 & 0.001,100 & 0.001,100 & 0.001,100 \\
    AMA & 100,0.5 & 100,0.5 & 100,0.5 & 0.001,100 & 0.001,100 & 0.001,100 \\
    CC & 100 & 100 & 100 & 0.001 & 0.001 & 0.001 \\
    Stoc-CP & 20,1 & 20,1 & 30,0.8 & 6,6 & 30,1 & 5,8 \\
    Stoc-AMA-SAG & 30,1 & 30,1 & 40,1 & 1,50 & 0.5,30 & 0.9,60 \\
    Stoc-AMA-SAGA & 9,0.7 & 20,0.6 & 20,0.6 & 0.8,30 & 0.6,30 & 0.5,30 \\
    Stoc-CC & 20 & 20 & 20 & 0.001 & 0.001 & 0.001 \\
    \hline
    \end{tabular}\\[.1in]
    \parbox{1\textwidth}{\small Note: There is one step size $\tau$ for CC and Stoc-CC. For primal-dual methods there are two step sizes $\tau,\alpha$, listed from left to right.}
    \label{tab:step sizes for single-block}
\end{table}

\subsection{Multi-block experiments}\label{subsec:tuning multi-block}
For the phase shift model (Figure~\ref{fig: phase shift two-way} and Figure~\ref{fig: phase shift three-way}), the first $2$ epochs are for tuning. The results are summarized in Table~\ref{tab:step sizes for phase shift}.

\begin{table}
    \centering
    \caption{Step sizes for phase shift model.}
    \begin{tabular}{|l|*{2}{c}{c|}*{2}{c}{c|}*{3}{c}|}
    \hline
    \multicolumn{10}{|c|}{$K=2$ (up to two-way interactions)} \\
    \hline
    $\log_2(\rho^{-1})$ & \multicolumn{3}{c|}{17} & \multicolumn{3}{c|}{21} & \multicolumn{3}{c|}{25} \\
    \hline
    $\log_2(\lambda^{-1})$ & 7 & 9 & 11 & 7 & 9 & 11 & 7 & 9 & 11 \\
    \hline
    CP & 20,0.7 & 20,0.7 & 20,0.7 & 20,0.7 & 20,0.7 & 20,0.7 & 20,0.7 & 20,0.7 & 20,0.7 \\
    AMA & 20,0.5 & 20,0.5 & 20,0.5 & 20,0.5 & 20,0.5 & 20,0.5 & 20,0.5 & 20,0.5 & 20,0.5 \\
    CC & 10 & 10 & 10 & 10 & 10 & 10 & 10 & 10 & 10 \\
    Stoc-CP & 0.4,4	& 0.3,5 & 0.4,4 & 0.6,4 & 0.4,6 & 0.6,4 & 0.6,4	& 0.4,6 & 0.6,4 \\
    Stoc-AMA-SAG &1,0.8 & 1,0.8 & 1,0.8	& 2,0.8 & 2,0.8 & 2,0.8 &	2,0.8 &2,0.8 & 2,0.8 \\
    Stoc-CC & 1 & 2 & 1 & 1 & 2 & 2 & 1 & 2 & 2  \\
    \hline
    \multicolumn{10}{|c|}{$K=3$ (up to three-way interactions)} \\
    \hline
     $\log_2(\rho^{-1})$ & \multicolumn{3}{c|}{17} & \multicolumn{3}{c|}{21} & \multicolumn{3}{c|}{25} \\
    \hline
    $\log_2(\lambda^{-1})$ & 7 & 11 & 15 & 7 & 11 & 15 & 7 & 11 & 15 \\
    \hline
    CP & 20,0.6 & 20,0.6 & 20,0.6 & 20,0.6 & 20,0.6 & 20,0.6 & 20,0.6 & 20,0.6 & 20,0.6 \\
    AMA & 20,0.4 & 20,0.4 & 20,0.4 & 20,0.4 & 20,0.4 & 20,0.4 & 20,0.4 & 20,0.4 & 20,0.4 \\
    CC & 10 & 20 & 20 & 10 & 20 & 20 & 10 & 20 & 20\\
    Stoc-CP & 0.6,4	& 0.6,4 & 0.5,8 & 0.8,4 & 0.9,4 & 0.6,9 & 0.8,4	& 0.8,4 & 0.6,9 \\
    Stoc-AMA-SAG & 2,0.8 & 2,0.8 & 1,0.8 & 2,0.8 & 2,0.8 & 2,0.8 & 2,0.8 & 2,0.8 & 2,0.8 \\
    Stoc-CC & 1 & 3 & 0.6 & 2 & 3 & 0.8 & 2 & 3 & 0.8 \\
    \hline
    \end{tabular}\\[.1in]
    \parbox{1\textwidth}{\small Note: There is one step size $\tau$ for CC and Stoc-CC. For primal-dual methods there are two step sizes $\tau,\alpha$, listed from left to right.}
    \label{tab:step sizes for phase shift}
\end{table}

For the synthetic linear regression model (Figure~\ref{fig: synthetic linear}), the first $6$ epochs are for tuning. For the synthetic logistic model (Figure~\ref{fig: synthetic logistic}), the first $10$ epochs are for tuning. The corresponding step sizes are summarized in Table~\ref{tab:step sizes of synthetic linear and logistic}.

\begin{table}
    \centering
    \caption{Step sizes for synthetic linear and logistic regressions.}
    \begin{tabular}{|l|*{2}{c}{c|}*{2}{c}{c|}*{3}{c}|}
    \hline
    \multicolumn{10}{|c|}{Linear regression} \\
    \hline
    $\log_2(\rho^{-1})$ & \multicolumn{3}{c|}{16} & \multicolumn{3}{c|}{19} & \multicolumn{3}{c|}{22} \\
    \hline
    $\log_2(\lambda^{-1})$ & 6 & 8 & 10 & 6 & 8 & 10 & 6 & 8 & 10 \\
    \hline
    CP & 20,0.2 & 20,0.2 & 20,0.2 & 20,0.2 & 20,0.2 & 20,0.2 & 20,0.2 & 20,0.2 & 20,0.2 \\
    AMA & 20,0.4 & 20,0.4 & 20,0.4 & 20,0.4 & 20,0.4 & 20,0.4 & 20,0.4 & 20,0.4 & 20,0.4 \\
    CC & 10 & 20 & 20 & 10 & 20 & 20 & 10 & 20 & 20 \\
    Stoc-CP & 2,2.5 & 4,1.5 & 2,3.5 & 2,2.5 & 4,1.5 & 2,3.5 & 2,2.5	& 4,1.5 & 2,3.5 \\
    Stoc-AMA-SAG & 3,0.8 & 10,0.9 & 3,0.7 & 4,0.8 & 20,0.9 & 4,0.7	& 4,0.8 & 20,0.9 & 4,0,7 \\
    Stoc-CC & 2 & 2 & 2 & 2 & 2 & 2 & 2 & 2 & 2  \\
    \hline
    \multicolumn{10}{|c|}{Logistic regression} \\
    \hline
     $\log_2(\rho^{-1})$ & \multicolumn{3}{c|}{16} & \multicolumn{3}{c|}{19} & \multicolumn{3}{c|}{22} \\
    \hline
    $\log_2(\lambda^{-1})$ & 6 & 8 & 10 & 6 & 8 & 10 & 6 & 8 & 10 \\
    \hline
    CP & 20,0.4 & 20,0.2 & 20,0.2 & 20,0.4 & 20,0.2 & 20,0.2 & 20,0.4 & 20,0.2 & 20,0.2 \\
    AMA & 10,2 & 20,0.2 & 20,0.2 & 10,2 & 20,0.2 & 20,0.2 & 10,2 & 20,0.2 & 20,0.2 \\
    CC & 10 & 20 & 20 & 10 & 20 & 20 & 10 & 20 & 20 \\
    Stoc-CP & 2,1 & 2,1 & 2,1 & 5,1 & 5,1 & 5,1 & 5,1 & 6,1 & 6,1\\
    Stoc-AMA-SAG & 5,0.9 & 3,1 & 6,0.9 & 5,0.9 & 8,0.4 & 8,0.4 &	10,0.9 & 10,0.4 & 10,0.4 \\
    Stoc-CC & 4 & 3 & 3 & 4 & 3 & 3 & 4 & 3 & 4 \\
    \hline
    \end{tabular}\\[.1in]
    \parbox{1\textwidth}{\small Note: There is one step size $\tau$ for CC and Stoc-CC. For primal-dual methods there are two step sizes $\tau,\alpha$, listed from left to right. The step size $\alpha$ for Stoc-CP is picked from $\{1,1.5,2,2.5,3,3.5,4\}$.}
    \label{tab:step sizes of synthetic linear and logistic}
\end{table}

For logistic regression on the run-or-walk data (Figure~\ref{fig: logistic on real data}), step sizes are tuned separately within each block as mentioned earlier. The first $10$ epochs are for tuning. For conciseness, we only report the step sizes under $\rho=2^{-23}$ and $\lambda=\|\tilde Y\|_n/2^{13}$ in Table~\ref{tab:step sizes of logistic model on real data}, which are the tuning parameters with the smallest misclassification rate on the validation set. From Table~\ref{tab:step sizes of logistic model on real data} we observe that the magnitude of step sizes varies widely among the blocks associated with main and two-way effects.

\begin{table}
    \centering
    \caption{Step sizes for logistic regression on real data with $\rho=2^{-23}$ and $\lambda=\|\tilde Y\|_n/2^{13}$.}
    \begin{tabular}{|c|cc|cc|c|cc|cc|c|}
    \hline
    Algorithms & \multicolumn{2}{c|}{CP} & \multicolumn{2}{c|}{AMA} & CC & \multicolumn{2}{c|}{Stoc-CP} & \multicolumn{2}{c|}{Stoc-AMA-SAG} & Stoc-CC   \\
    \hline
    Block index & $\tau$ & $\alpha$ & $\tau$ & $\alpha$ & $\tau$ & $\tau$ & $\alpha$ & $\tau$ & $\alpha$ & $\tau$  \\
    \hline
    1 & 1 & 0.2 & 1 & 0.2 & 1 & 0.07 & 6& 0.4 & 0.2 & 0.08 \\
    2 & 1 & 0.3 & 1 & 0.2 & 1 & 0.2 & 1 & 0.5 & 0.5 & 0.07  \\
    3 & 0.6 & 0.2 & 0.6 & 0.1 & 0.6 & 0.1 & 0.9 & 0.05 & 1 & 0.05 \\
    4 & 1 & 0.5 & 1 & 0.4 & 1 & 0.04 & 0.6 & 0.03 & 0.4 & 0.1 \\
    5 & 1 & 0.8 & 1 & 0.6 & 1 & 0.1 & 0.4 & 0.3 & 0.3 & 0.04  \\
    6 & 2 & 0.5 & 2 & 0.4 & 2 & 0.5 & 0.9 & 0.4 & 0.3 & 0.2  \\
    1,2 & 0.7 & 0.2 & 0.7 & 0.2 & 0.7 & 0.009 & 1 & 0.003 & 0.04 & 0.001  \\
    1,3 & 0.1 & 0.1 & 0.1 & 0.1 & 0.1 & 0.007 & 0.01 & 0.2 & 0.02 & 0.003  \\
    1,4 & 0.3 & 0.07 & 0.3 & 0.06 & 0.3 & 0.01 & 0.2 & 0.03 & 0.3 & 0.005  \\
    1,5 & 0.6 & 0.3 & 0.6 & 0.3 & 0.6 & 0.01 & 1 & 0.4 & 0.02 & 0.005 \\
    1,6 & 0.7 & 0.2 & 0.7 & 0.2 & 0.7 & 0.03 & 0.2 & 0.08 & 0.3 & 0.01 \\
    2,3 & 0.3 & 0.2 & 0.3 & 0.1 & 0.3 & 0.01 & 0.09 & 0.05 & 0.2 & 0.004 \\
    2,4 & 1 & 0.5 & 1 & 0.4 & 1 & 0.01 & 2 & 0.03 & 0.2 & 0.006 \\
    2,5 & 1 & 0.3 & 1 & 0.3 & 1 & 0.01 & 0.5 & 0.02 & 0.5 & 0.006 \\
    2,6 & 1 & 0.2 & 1 & 0.2 & 1 & 0.06 & 0.3 & 0.08 & 0.3 & 0.02 \\
    3,4 & 1 & 0.3 & 1 & 0.3 & 1 & 0.01 & 1 & 0.02 & 0.7 & 0.005 \\
    3,5 & 0.3 & 0.4 & 0.3 & 0.3 & 0.3 & 0.005 & 0.3 & 0.004 & 0.1 & 0.002 \\
    3,6 & 0.4 & 0.3 & 0.4 & 0.2 & 0.4 & 0.01 & 0.2 & 0.06 & 0.2 & 0.006 \\
    4,5 & 1 & 0.3 & 1 & 0.2 & 1 & 0.03 & 0.4 & 0.06 & 0.4 & 0.01 \\
    4,6 & 1 & 0.2 & 1 & 0.2 & 1 & 0.06 & 0.3 & 0.1 & 0.3 & 0.03 \\
    5,6 & 1 & 0.2 & 1 & 0.2 & 1 & 0.04 & 0.3 & 0.1 & 0.4 & 0.02 \\
    \hline
    \end{tabular}\\[.1in]
    \parbox{1\textwidth}{\small Note: Blocks with indices $1$ to $6$ refer to main effects of $6$ explanatory variables $acc_x$, $acc_y$, $acc_z$, $gyro_x$, $gyro_y$, and $gyro_z$ respectively. Blocks with double indices refer to the corresponding two-way interactions.}
    \label{tab:step sizes of logistic model on real data}
\end{table}

\subsection{Recoveries}

We perform a check-and-recovery after the update of each block as follows. If the training loss increases rather than decreases, we skip the update for the current block and directly move to the next block. The cost of checking is negligible, and this operation
leads to more stable performances (especially for stochastic algorithms).
In particular, we find the check-and-recovery useful in the real-data experiments where step sizes are tuned block-wise.
Table~\ref{tab: recovery for logistic regression on real data} reports the recovery information of stochastic algorithms in the real-data experiments across $10$ repeated runs. From Table~\ref{tab: recovery for logistic regression on real data} we observe that Stoc-CP is more robust against bad step sizes than the other two stochastic algorithms.

\begin{table}[h]
    \centering
    \caption{Recovery for logistic regression on real data.}
    \begin{tabular}{|l|*{2}{c}{c|}*{2}{c}{c|}*{3}{c}|}
     \hline
     \multicolumn{10}{|c|}{Total numbers of recoveries} \\
     \hline
     $\log_2(\rho^{-1})$ & \multicolumn{3}{c|}{18} & \multicolumn{3}{c|}{23} & \multicolumn{3}{c|}{25} \\
     \hline $\log_2(\lambda^{-1})$ & 6 & 13 & 15 & 6 & 13 & 15 & 6 & 13 & 15\\
     \hline
     Stoc-CC & 398 & 177 & 169 & 504 & 163 & 167 & 503 & 162 & 167 \\
     Stoc-CP & 342 & 42 & 41 & 287 & 42 & 43 & 284 & 42 & 43 \\
     Stoc-AMA-SAG & 532 & 159 & 161 & 572 & 166 & 132 & 571 & 165 & 133 \\
     \hline
     \multicolumn{10}{|c|}{Numbers of blocks with recoveries} \\
     \hline
     $\log_2(\rho^{-1})$ & \multicolumn{3}{c|}{18} & \multicolumn{3}{c|}{23} & \multicolumn{3}{c|}{25} \\
     \hline $\log_2(\lambda^{-1})$ & 6 & 13 & 15 & 6 & 13 & 15 & 6 & 13 & 15\\
     \hline
     Stoc-CC & 11 & 11 & 10 & 14 & 11 & 10 & 14 & 11 & 10 \\
     Stoc-CP & 10 & 5 & 5 & 6 & 5 & 5 & 6 & 5 & 5 \\
     Stoc-AMA-SAG & 12 & 8 & 8 & 12 & 7 & 6 & 12 & 7 & 6 \\
     \hline
    \end{tabular}\\[.1in]
    \parbox{1\textwidth}{\small Note: The counts are collected for stochastic algorithms in $10$ repeated runs, each with $80$ epochs (corresponding to $16$ backfitting cycles with 5 batch steps in each block). There are $21$ blocks in total. For example, when $\rho=2^{-18}$ and $\lambda=\|\tilde Y\|_n/2^{6}$, Stoc-CC has a total of $398$ recoveries from $11$ different blocks during the $10$ repeated runs.
    }
    \label{tab: recovery for logistic regression on real data}
\end{table}

\centerline{\bf\Large References}
\begin{description}\addtolength{\itemsep}{-.1in}
\item Condat, L. (2013). A primal–dual splitting method for convex optimization involving Lipschitzian, proximable and linear composite terms. {\em Journal of Optimization Theory and Applications}, 158, 460-479.

\item Friedman, J. (1991). Multivariate adaptive regression splines (with discussion). {\em Annals of
Statistics}, 19 , 79–141.

\item Hunter, D. R., \& Lange, K. (2004). A tutorial on MM algorithms. {\em American Statistician},
58 , 30–37.

\item Hunter, D. R., \& Li, R. (2005). Variable selection using MM algorithms. {\em Annals of Statistics},
33 , 1617.

\item Li, Z., \& Yan, M. (2021). New convergence analysis of a primal-dual algorithm with large stepsizes. {\em Advances in Computational Mathematics}, 47 , 1–20.

\item Parikh, N., \& Boyd, S. (2014). Proximal algorithms. {\em Foundations and Trends in Optimization},
1 , 127–239.

\item Ryu, E. K., \& Yin, W. (2022). \textit{Large-Scale Convex Optimization via Monotone Operators}.
Cambridge University Press.

\item Tseng, P. (1988). Coordinate ascent for maximizing nondifferentiable concave functions.
\textit{Technical Report LIDS-P-1840, MIT}.

\item V\~u, B. C. (2013). A splitting algorithm for dual monotone inclusions involving cocoercive
operators. {\em Advances in Computational Mathematics}, 38 , 667–681.

\end{description}

\end{document}